\documentclass{article}
\usepackage{jfrExamplee_mod,times}
\usepackage{graphicx}
\usepackage{hyperref}
\usepackage{apalike}
\usepackage{setspace}
\usepackage[utf8]{inputenc}
\usepackage{glossaries}
\glsdisablehyper
\usepackage{amssymb}
\usepackage{amsmath}
\usepackage{cleveref}
\usepackage[per-mode = symbol]{siunitx}
\usepackage{subcaption}
\usepackage{algorithm}
\usepackage{algpseudocode}
\usepackage{tabularx}
\usepackage{booktabs}
\usepackage{silence}
\makeglossaries
\usepackage[disable]{todonotes}
\newacronym{asv}{ASV}{autonomous surface vehicle}
\newacronym{auv}{AUV}{autonomous underwater vehicle}
\newacronym{ais}{AIS}{automatic identification system}
\newacronym{colav}{COLAV}{collision avoidance}
\newacronym{colregs}{COLREGs}{International Regulations for Preventing Collisions at Sea}
\newacronym{vo}{VO}{velocity obstacles}
\newacronym{dw}{DW}{dynamic window}
\newacronym{a*}{A*}{A star}
\newacronym{rrt}{RRT}{rapidly exploring random tree}
\newacronym{bc-mpc}{BC-MPC}{branching-course MPC}
\newacronym{sog}{speed}{speed over ground}
\newacronym{rot}{yaw rate}{rate of turn}
\newacronym{mr}{MR}{Maritime Robotics}
\newacronym{fffb}{FF-FB}{feedforward feedback}
\newacronym{pdaf}{PDAF}{probabilistic data association filter}
\newacronym{ros}{ROS}{Robot Operating System}
\newacronym{los}{LOS}{line of sight}
\newacronym{mpc}{MPC}{model predictive control}
\newacronym{ned}{NED}{North-East-Down}
\newacronym{pi}{PI}{proportional-integral}
\newacronym{osd1}{OSD1}{Ocean Space Drone 1}

\title{The Branching-Course MPC Algorithm for Maritime Collision Avoidance}

\newcommand{\authorSpace}{.1cm}
\author{
Bjørn-Olav H. Eriksen \hspace{\authorSpace} Morten Breivik \hspace{\authorSpace} Erik F. Wilthil \hspace{\authorSpace} Andreas L. Fl{\aa}ten  \hspace{\authorSpace} Edmund F. Brekke \\
Centre for Autonomous Marine Operations and Systems \\
Department of Engineering Cybernetics \\
Norwegian University of Science and Technology (NTNU)\\
Trondheim, Norway \\
\texttt{\{bjorn-olav.h.eriksen, morten.breivik\}@ieee.org} \\
\texttt{\{erik.wilthil, andreas.flaten, edmund.brekke\}@ntnu.no} \\
}

\newlength\textwidthcm
\begin{document}
\newcommand{\skewSymThree}[3]{\begin{bmatrix}0 & -#3 & #2 \\ #3 & 0 & -#1 \\ -#2 & #1 & 0\end{bmatrix}}
\newcommand{\bs}{\boldsymbol}
\newcommand{\norm}[1]{\left\lVert#1\right\rVert}
\Crefname{equation}{}{}
\crefname{equation}{}{}
\Crefname{figure}{Fig.}{Figures}
\crefname{figure}{Fig.}{figures}
\Crefname{tabular}{Table}{Tables}
\crefname{tabular}{Table}{tables}
\Crefname{table}{Table}{Tables}
\crefname{table}{Table}{tables}
\crefname{tabular}{Table}{tables}
\crefname{section}{Section}{sections}
\Crefname{section}{Section}{Sections}
\maketitle

\begin{abstract}
This article presents a new algorithm for short-term maritime \gls{colav} named the \gls{bc-mpc} algorithm.
The algorithm is designed to be robust with respect to noise on obstacle estimates, which is a significant source of disturbance when using exteroceptive sensors such as e.g. radars for obstacle detection and tracking.
Exteroceptive sensors do not require vessel-to-vessel communication, which enables \gls{colav} toward vessels not equipped with e.g. \gls{ais} transponders, in addition to increasing the robustness with respect to faulty information which may be provided by other vessels.
The \gls{bc-mpc} algorithm is compliant with rules 8 and 17 of the \gls{colregs}, and favors maneuvers following rules 13--15.
This results in a \gls{colregs}-aware algorithm which can ignore rules 13--15 when necessary.
The algorithm is experimentally validated in several full-scale experiments in the Trondheimsfjord in 2017 using a radar-based system for obstacle detection and tracking.
The \gls{colav} experiments show good performance in compliance with the desired algorithm behavior.
\end{abstract}
\glsresetall
\glsunset{a*}
\glsunset{sog}
\glsunset{rot}

\section{Introduction}
Today's society moves rapidly towards an increased level of automation.
The development of autonomous cars is spearheading this trend, as exemplified by the efforts made by e.g.
Google and Uber.
In recent years, autonomy has also become a hot topic in the maritime domain with research on autonomous passenger and goods transport, seabed surveying and military applications.
An example of this is the Yara Birkeland project in Norway, where an autonomous electrically-powered cargo ship will replace approximately $40000$ diesel-powered truck journeys of fertilizer per year \cite{YaraBirkeland}.
Reduced cost, increased efficiency and reduced environmental impact may be the most obvious benefits of autonomy at sea, but the potential for increased safety is not to be overlooked since reports state that in excess of $75\%$ of maritime accidents are caused by human errors~\cite{Chauvin2011,Levander2017}.
A prerequisite for employing \glspl{asv} in environments where other vessels may be present is, however, that the \glspl{asv} have robust \gls{colav} systems.
Such \gls{colav} systems must make the \glspl{asv}, as other vessels, follow the \gls{colregs} which contains a set of rules on how vessels should behave in situations where there is a risk of collision with another vessel~\cite{Cockcroft2004}.
However, \Gls{colregs} is written for human interpretation with few quantitative rules, which makes it challenging to develop algorithms capturing the intention of \gls{colregs} by machine decision-making.

\Gls{colav} algorithms have typically been divided into reactive and deliberate algorithms.
Reactive algorithms are characterized by considering a limited amount of information, originally only currently available sensor information~\cite{Tan2004a}, and employing little motion planning in a short time frame.
This makes reactive algorithms computationally cheap, and able to react to sudden changes in the environment.
Examples include vessels making sudden unpredicted maneuvers, late detection of obstacles, etc.
However, since reactive algorithms consider a limited amount of information and employ little motion planning, they tend to make suboptimal choices in complex situations which makes them sensitive to local minima.
Examples of reactive algorithms are the \gls{vo}~\cite{Fiorini1998,Kuwata2014} and the \gls{dw}~\cite{Fox1997} algorithms.
Deliberate algorithms consider more information and plan for a longer time frame, which results in more optimal choices at the cost of increased computational requirements.
Examples of deliberate algorithms include the \gls{a*}~\cite{Hart1968} and the \gls{rrt}~\cite{LaValle1998} algorithms.

\begin{figure}
	\centering
	\includegraphics[width=.6\textwidth]{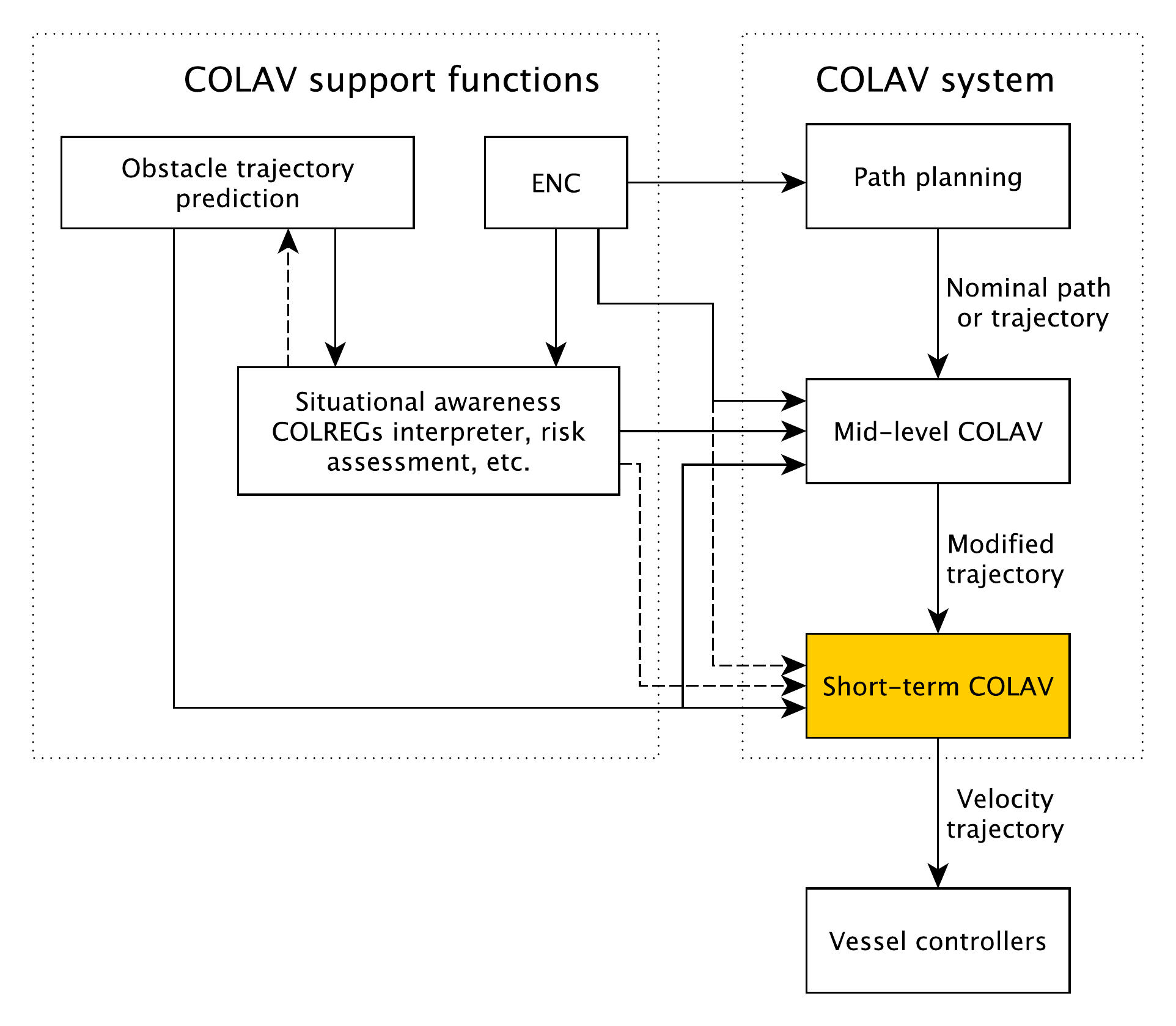}
	\caption{\label{fig:hybrid}A hybrid COLAV architecture with three levels.
The support functions provide relevant information for the COLAV algorithms, including obstacle trajectories, static obstacles from electronic nautical charts (ENC) and situational awareness in the form of COLREGs situations.
The short-term layer does not currently utilize information from ENC or situational awareness.}
\end{figure}

The previously clear border between reactive and deliberate algorithms have become somewhat artificial since few algorithms only utilize currently available sensor information.
However, the idea that the reactive algorithms are capable of responding quickly to changes in the environment and the deliberate algorithms are capable of performing optimal motion planning in a longer time frame is still relevant.
We therefore choose to rather use the terms ``short-term'' and ``long-term'' algorithms to distinguish the algorithms.
In a practical \gls{colav} system, both short-term and long-term algorithms are useful.
For long time frames, all available information should be included, while one may use a less detailed vessel model for planning.
For short-term \gls{colav}, one can include less spatial and temporal information but may need to use a more detailed model of the vessel to ensure dynamically feasible maneuvers.
By combining short-term and long-term algorithms in a hybrid architecture, the benefits of both algorithms can be combined, ensuring both responsiveness, feasibility and optimality.
An example of a hybrid architecture with three \gls{colav} levels is shown in \cref{fig:hybrid}.
The topmost level, named path planning, is intended to produce a nominal path or trajectory from the initial position to the goal.
The spatial and temporal distance between the initial and goal positions may be large, allowing only for a limited complexity in this algorithm.
For instance, moving obstacles could be neglected at this level.
The mid-level \gls{colav} algorithm tries to follow this nominal path or trajectory, while at the same time performing \gls{colav} with respect to all obstacles, characterized as a long-term \gls{colav} algorithm. \Gls{colregs} is a natural part of this level, since it may be complex to decide the appropriate action with respect to \gls{colregs}.
The mid-level algorithm produces a modified trajectory which is passed to the short-term \gls{colav} layer.
This layer perform short-time \gls{colav} making sure to avoid obstacles performing sudden maneuvers or which are detected too late to be handled by the mid-level algorithm, while also ensuring that the maneuvers are feasible with respect to the dynamic constraints of the vessel.
The short-term layer can also act as a backup solution to avoid collisions in cases where the mid-level algorithm fails to produce feasible trajectories, for instance due to time constraints or numerical issues~\cite{Eriksen2017b}.

\Gls{colav} algorithms depend on information about obstacle position, speed and course in order to be able to avoid collisions.
One possible source of such information is using \gls{ais} transponders. \Gls{ais} is a vessel-to-vessel communication system where vessels transmit their current position and velocity to other vessels carrying \gls{ais} transponders \cite{IMOAIS}.
Passenger ships and vessels with a gross tonnage of over $300$ are required to carry \gls{ais} transponders.
This is of course valuable information when it comes to navigation and \gls{colav} at sea.
However, \gls{ais} transponders usually rely on satellite navigation and data inputs from the user, which results in the possibility of transmitting inaccurate or invalid data \cite{Harati-Mokhtari2007}.
Also, vessels or objects not equipped with \gls{ais} transponders will not be detected.
A more robust approach to obtain information about the environment is to employ exteroceptive sensors, which have the advantage of not relying on any infrastructure or collaboration with the obstacles in order to detect them.
A commonly used exteroceptive sensor at sea is radar.
However, the data from a radar usually includes a fair amount of noise, which makes this sensor more complex and difficult to work with than \gls{ais}~\cite{Eriksen2018}.
On-board radars have been used for full-scale \gls{colav} experiments based on the \gls{a*} algorithm in~\cite{Schuster2014}, and using a modified version of the \gls{dw} algorithm in~\cite{Eriksen2018}.
In~\cite{Elkins2010,Kuwata2014}, other exteroceptive sensors such as cameras and lidar are used for \gls{colav}.

\Gls{mpc} has for a long time been a well-known and proven tool for motion planning and \gls{colav} for e.g. ground and automotive robots \cite{Ogren2005,Keller2015,Gray2013}, aerospace applications \cite{Kuwata2011} and underwater vehicles \cite{Caldwell2010}.
In the later years, \gls{mpc} has also been applied for \gls{colav} in the maritime domain, both using sample-based approaches where one considers a finite space of control inputs \cite{Svec2013,Johansen2016,Hagen2018} and conventional gradient-based search algorithms \cite{Abdelaal2016,Eriksen2017b}.
None of these algorithms does, however, consider the amounts of noise which we expect to encounter using a radar-based tracking system.
Gradient-based algorithms have the benefit of exploring the entire control input space, but the complexity of the \gls{colav} problem can make it difficult to guarantee that a feasible solution will be found within the time requirements \cite{Eriksen2017b}.
This makes sample-based approaches well suited for short-term \gls{colav}.
In \cite{Benjamin2006,Benjamin2010}, a protocol-based \gls{colav} algorithm using interval programming is presented.
The algorithm optimizes over multiple functions considering different behaviors, e.g. waypoint following and adherence to different parts of \gls{colregs}, by combining them in an objective function with adaptive weights.
The algorithm does, however, use vessel-to-vessel communication in order to obtain obstacle information, and is not necessarily well suited for use with exteroceptive sensors.

\subsection{The International Regulations for Preventing Collisions at Sea}
\begin{figure}
	\centering
	\includegraphics[width=.4\textwidth]{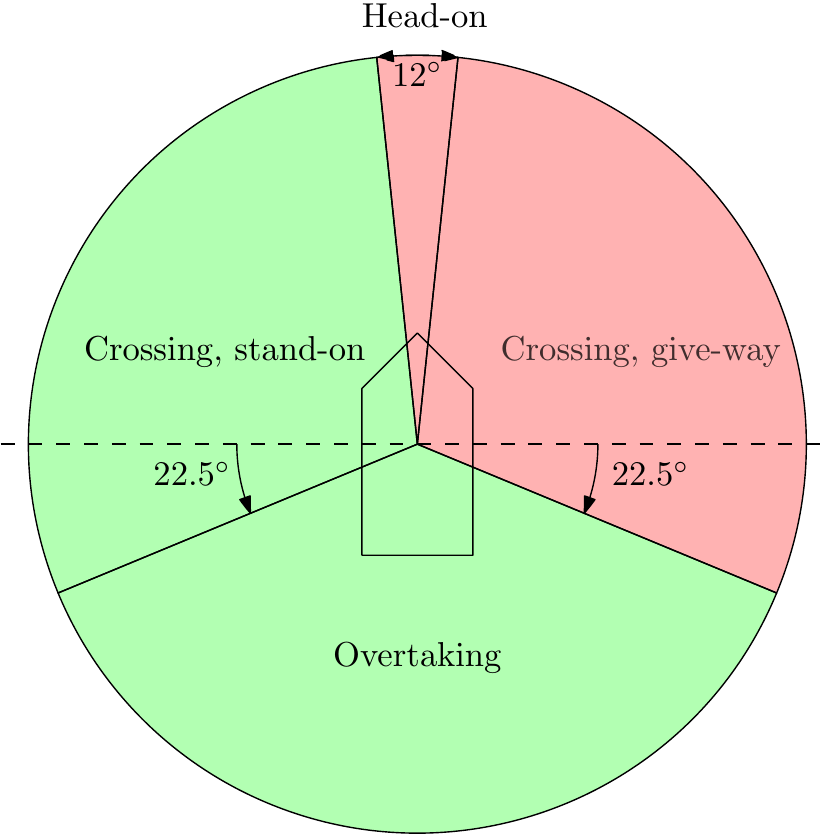}
	\caption{\label{fig:colregsRegions}Graphical illustration of \acrshort{colregs} regions as seen from the ownship.
The light red regions show areas where the ownship is required to maneuver, while the light green region show areas where the ownship should keep the current speed and course.}
\end{figure}

\Gls{colregs} regulate how vessels should behave in situations where there exists a risk of collision.
There is in total 38 rules, where rule 8 and 13--17 are the most relevant ones for designing \gls{colav} algorithms for \glspl{asv}, although the rest must also must be addressed in a \gls{colregs}-compliant system.
Rules 8 and 13--17 can be summarized as:
\begin{description}
	\item[Rule 8:] This rule requires, among other things, that maneuvers applied in situations where a risk of collision exists should be large enough to be readily observable for other vessels.
Small consecutive maneuvers should hence be avoided.
	\item[Rule 13:] In an overtaking situation, where a vessel is approaching another from an angle of more than $22.5\si{\degree}$ abaft the other vessel's beam, the overtaking vessel is deemed the give-way vessel and the overtaken vessel is deemed the stand-on vessel.
The overtaking vessel is allowed to pass on either side.
However, in a case where the overtaken vessel is required to avoid collision with another vessel it may be required to make a starboard maneuver.
To avoid blocking the path of the overtaken vessel in such a situation, we consider it as most suitable to overtake a vessel on her port side.
	\item[Rule 14:] In a head-on situation, where two vessels approaches each other on reciprocal or nearly reciprocal courses (a margin of $\pm 6~\si{\degree}$ is often used), both vessels are required to do starboard maneuvers and pass the other vessel on her port side.
	\item[Rule 15:] This rule handles crossing situations, where a vessel is approaching another vessel from the side, but not in the regions considered as a head on or overtaking situation.
The vessel with the other vessel on her starboard side is deemed the give-way vessel, while the other is deemed the stand-on vessel.
The preferred give-way maneuver is to do a starboard turn and pass behind the stand-on vessel.
	\item[Rule 16:] This rule defines the action for the give-way vessel.
It requires that the give-way vessel performs early and substantial action to avoid collision.
	\item[Rule 17:] This rule defines the action for the stand-on vessel.
It requires that the stand-on vessel keep her current speed and course, while the give-way vessel maneuvers in order to avoid collision.
However, if the give-way vessel fails in her duty of avoiding collision, the stand-on vessel is required to maneuver to avoid collision.
If this occurs in a crossing situation, the stand-on vessel should avoid maneuvering to port if possible.
\end{description}
\Cref{fig:colregsRegions} shows a graphical illustration of the situations given by rules 13--15.
The interested reader is referred to~\cite{Cockcroft2004} for more details on the \gls{colregs} rules.

\subsection{Contributions}
The authors of this article have focused on short-term and reactive \gls{colav} for \glspl{asv} for the last few years, starting with a modified version of the \gls{dw} algorithm designed for use with \glspl{auv}~\cite{Eriksen2016}.
This algorithm was adapted for use with high-speed \glspl{asv}, and tested in conjunction with a radar-based tracking system~\cite{Wilthil2017} successfully demonstrating closed-loop radar-based \gls{colav} in full scale experiments~\cite{Eriksen2018}.
However, the experiments revealed challenges with using radar-based tracking systems for \gls{colav}, especially noisy estimates of obstacle speed and course caused problems.
The \gls{dw} algorithm is not particularly robust with respect to such noise, causing the vessel to repeatedly change the planned maneuver.
In addition, the \gls{dw} algorithm assumes the \gls{asv} to keep a constant turn rate for the entire prediction horizon.
This does not resemble the way vessels usually maneuver at sea, where one usually performs a corrective maneuver by changing the course and/or speed, followed by keeping the speed and course constant.
These issues motivate us to develop a new short-term \gls{colav} algorithm which is less sensitive to noisy obstacle estimates while also producing more ``maritime-like'' maneuvers.

In this article, we therefore present a new algorithm for short-term \gls{colav} named the \gls{bc-mpc} algorithm.
This algorithm is based on sample-based \gls{mpc} and is designed to be robust with respect to noisy obstacle estimates, which is an important consideration when using radar-based tracking systems for providing obstacle estimates.
In contrast to sample-based \gls{mpc} algorithms previously applied to \glspl{asv}, the \gls{bc-mpc} algorithm considers a sequence of maneuvers, enabling the algorithm to plan more complex trajectories than just a single avoidance maneuver.
Furthermore, the \gls{bc-mpc} algorithm complies with rules 8 and 17 of \gls{colregs}, while favoring maneuvers complying with rules 13--15.
In cases where the algorithm chooses to ignore rules 13--15, which can be required by rule 17, the maneuvers have increased clearance to obstacles.
The term ``\gls{colregs}-compliance'' is often abused in the literature by using it for algorithms only complying with parts of \gls{colregs}.
With this in mind, we consider the algorithm as being ``\gls{colregs}-aware''.
The algorithm is implemented on an under-actuatuated \gls{asv} and validated through several full-scale closed-loop \gls{colav} experiments using a radar-based tracking system for providing estimates of obstacle course, speed and position.

\subsection{Outline}
The rest of the article is structured as follows: \Cref{sec:ASVmodelingControl} describes modeling and control of \glspl{asv}, \cref{sec:BC-MPC} presents the \gls{bc-mpc} algorithm, while \cref{sec:expres} contains results from the full-scale closed-loop \gls{colav} experiments.
Finally, \cref{sec:conclusions} concludes the article and presents possibilities for further work.

\section{ASV modeling and control}\label{sec:ASVmodelingControl}
\begin{figure}
	\centering
	\includegraphics[width=.8\textwidth]{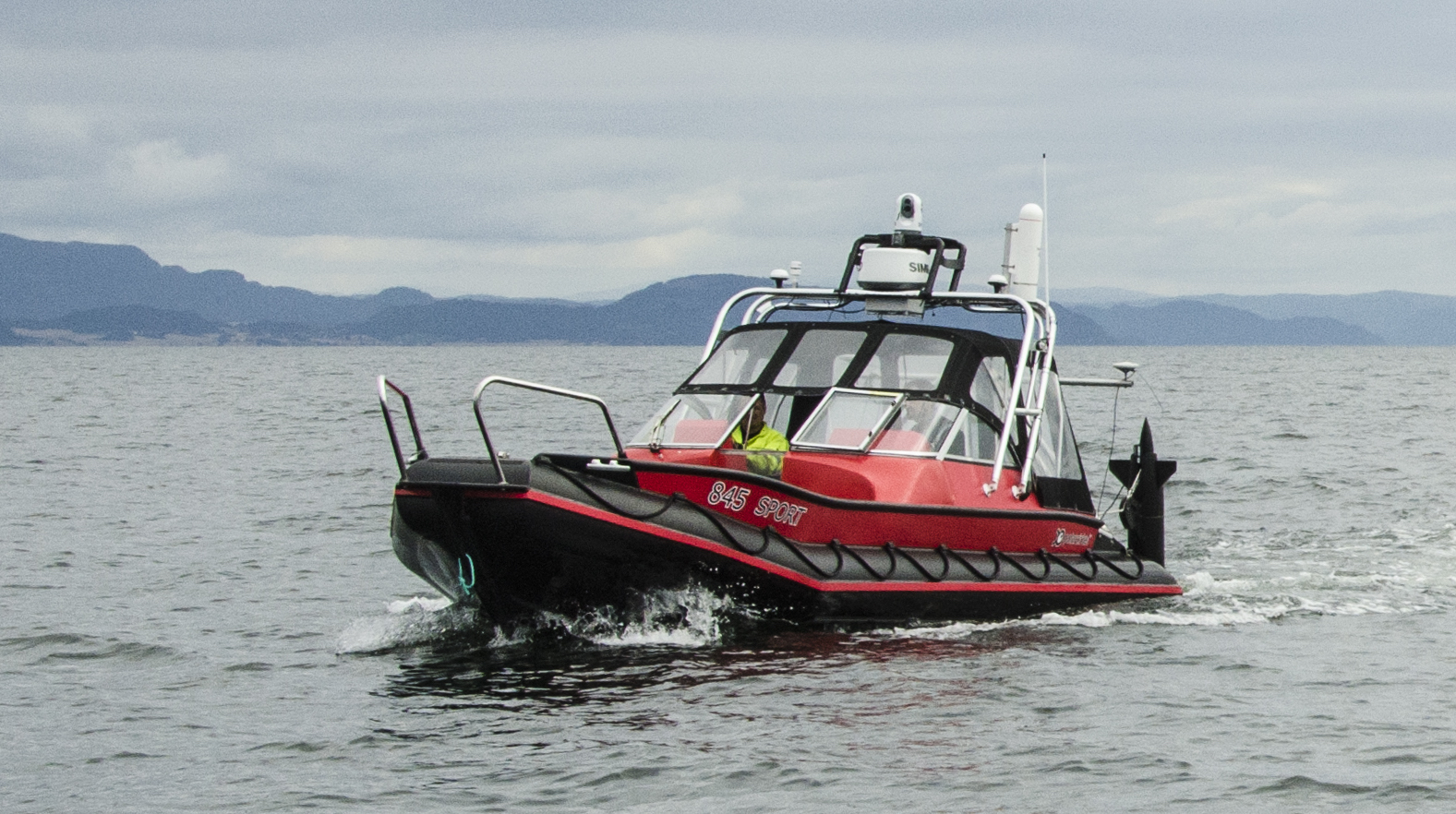}
	\caption{\label{fig:telemetron}The Telemetron ASV, designed for both manned and unmanned operations.
Courtesy of Maritime Robotics.}
\end{figure}
The vessel of interest in this work is the Telemetron \gls{asv} shown in \cref{fig:telemetron}, which is owned and operated by \gls{mr}.
The vessel is $8.45~\si{\meter}$ long, and uses a single steerable outboard engine for propulsion, which makes the vessel underactuated.
\subsection{ASV modeling}
\Glspl{asv} are in general small and agile vessels, capable of operating at high speeds.
At low speeds the hydrostatic pressure mainly carries the weight of the vessel, and it operates in the displacement region.
When the vessel speed increases, the hydrodynamic pressure increases, eventually dominating over the hydrostatic pressure.
At this point, we are in the planing region.
In between the displacement and planing region we have the semi-displacement region.
The Telemetron \gls{asv} is a high-speed vessel, capable of speeds up to $18~\si{\meter\per\second}$, which combined with the vessel length of $8.45~\si{\meter}$ makes for a vessel operating in the displacement, semi-displacement and planing regions~\cite{Fossen2011,Faltinsen2005}.

\begin{figure}
	\centering
	\includegraphics[width=.4\textwidth]{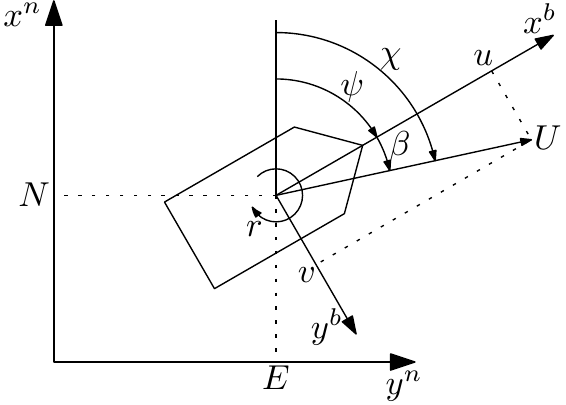}
	\caption{\label{fig:SurgeSwayYaw}Vessel variables.
The superscripts $(\cdot)^n$ and $(\cdot)^b$ denote the NED and body reference frames~\cite{Fossen2011}, respectively.
The variables $N$, $E$ and $\psi$ represent the vessel pose, $u$, $v$ and $r$ represent the body-fixed vessel velocity and $U$ is the vessel speed over ground.
The course $\chi$ is the sum of the heading $\psi$ and the sideslip $\beta$.}
\end{figure}

The conventional approach to modeling \glspl{asv} is by using the 3DOF model~\cite{Fossen2011}:
\begin{subequations}\label{eq:Fossen3DOF}
	\begin{equation}\label{eq:Fossen3DOFKinematics}
		\dot{\bs \eta} 	= \bs R(\psi) \bs \nu
	\end{equation}
	\begin{equation}\label{eq:Fossen3DOFDynamics}
		\bs M \dot{\bs \nu} + \bs C(\bs\nu) \bs \nu + \bs D(\bs \nu) \bs \nu	= \bs \tau,
	\end{equation}
\end{subequations}
where $\bs \eta = \begin{bmatrix} N & E & \psi \end{bmatrix}^T$ is the vessel pose in an earth-fixed \gls{ned} reference frame, $\bs \nu = \begin{bmatrix} u & v & r \end{bmatrix}^T$ is the vessel velocity and $\bs \tau = \begin{bmatrix} X & Y & N \end{bmatrix}^T$ is a vector of forces and torque, both given in the body-fixed reference frame.
See \cref{fig:SurgeSwayYaw} for an illustration of the variables.
The matrix $\bs R(\psi)$ is a rotation matrix, while $\bs M$, $\bs C(\bs \nu)$ and $\bs D(\bs \nu)$ are the mass, Coriolis and centripetal and damping matrices, respectively.

There exist many versions of the model \eqref{eq:Fossen3DOF}~\cite{Fossen2011}, but they require that the vessel operates in the displacement region.
For the Telemetron \gls{asv}, this would require a maximum operating speed of approximately $3.5~\si{\meter\per\second}$~\cite{Eriksen2017}.
This is quite a big limitation, and we therefore rather use a control-oriented non-first principles model developed for high-speed \glspl{asv}~\cite{Eriksen2017}, valid for the displacement, semi-displacement and planing regions:
\begin{equation}\label{eq:Model}
    \bs M(\bs x) \dot{\bs x} + \bs \sigma(\bs x) = \bs \tau,
\end{equation}
where $\bs x = \begin{bmatrix} U & r \end{bmatrix}^T$ is the vessel state, with $U = \sqrt{u^2 + v^2}$ being the vessel speed over ground and $r$ being the vessel \gls{rot}, while $\bs \tau = \begin{bmatrix} \tau_m & \tau_\delta \end{bmatrix}^T$ is a normalized control input.
In this article, we also refer to the vessel speed over ground as the vessel \gls{sog}.
The matrix $\bs M(\bs x)$ is a diagonal state-dependent inertia matrix with nonlinear terms and $\bs \sigma(\bs x) = \begin{bmatrix} \sigma_U(\bs x) & \sigma_r(\bs x) \end{bmatrix}^T$ is a vector of nonlinear damping terms.
Notice that the model is in 2DOF, designed for underactuated \glspl{asv}, where the speed and course are usually controlled.
Using the state variable from \eqref{eq:Model}, the kinematics can be defined as:
\begin{equation}\label{eq:KinematicsU}
	\begin{aligned}
		\dot{\bs \eta} 	&= \begin{bmatrix} \cos(\chi) & 0 \\ \sin(\chi) & 0 \\ 0 & 1 \end{bmatrix} \begin{bmatrix} U \\ r \end{bmatrix} \\
		\dot{\chi}		&= r + \dot \beta,
	\end{aligned}
\end{equation}
where $\chi$ is the vessel course and $\beta$ is the sideslip.
For more details on the model, see~\cite{Eriksen2017}.

\subsection{ASV control design}\label{sec:ASVcontrol}
As shown in \Cref{fig:hybrid}, the \gls{colav} system is built on top of the vessel controllers.
Hence, the performance of the \gls{colav} system can be limited by the performance of the vessel controllers.
It is therefore beneficial to use high-performance vessel controllers ensuring that the maneuvers that the \gls{colav} system specifies are properly executed, not limiting the performance of the \gls{colav} system.

The model \eqref{eq:Model} can be used in control design, particularly using it for model-based feedforward in \gls{sog} and \gls{rot} is shown to provide good performance~\cite{Eriksen2017}.
A controller named the \gls{fffb} controller is presented in~\cite{Eriksen2017}, which combines model-based feedforward terms with a gain-scheduled \gls{pi} feedback controller for controlling the vessel \gls{sog} and \gls{rot}.
For the \gls{bc-mpc} algorithm, we need a controller capable of following a \gls{sog} and course trajectory.
The \gls{fffb} controller has proven to have high performance in experiments~\cite{Eriksen2018,Eriksen2017}, so we therefore extend the \gls{fffb} controller to include course control:
\begin{equation}\label{eq:vesselController}
	\bs \tau = \bs M(\bs x) \dot{\bs x}_d + \bs \sigma(\bs x_d) - \bs M(\bs x)\bs K_p \tilde{\bs \zeta} - \bs K_i \int_{t_0}^t \tilde{\bs \zeta}_1(\gamma)\mathrm{d}\gamma,
\end{equation}
where $\bs x_d = \begin{bmatrix} U_d & r_d \end{bmatrix}^T$, $\bs K_p>0$ is a matrix of proportional gains, $\bs K_i>0$ is a diagonal matrix of integral gains, and:
\begin{equation}
	\tilde{\bs \zeta} = \begin{bmatrix} \tilde U \\ \tilde r \\ \tilde \chi \end{bmatrix}, \quad \tilde{\bs \zeta}_1 = \begin{bmatrix} \tilde U \\ \tilde \chi \end{bmatrix},
\end{equation}
where $\tilde U = U - U_d$, $\tilde r = r - r_d$ and $\tilde \chi = \Upsilon(\chi - \chi_d)$ are the \gls{sog}, \gls{rot} and course errors, respectively.
The function $\Upsilon: \mathbb{R} \to S^1$ maps an angle to the domain $[-\pi,\pi)$.

In the control law \eqref{eq:vesselController}, we use the desired \gls{rot} $r_d$ and its derivative $\dot r_d$.
Through \eqref{eq:KinematicsU}, the relation between the course and \gls{rot} is stated as $r = \dot \chi - \dot \beta$, where the derivative of the sideslip enters the equation.
At this stage, we do not have a sideslip model of the Telemetron \gls{asv}.
However, we have seen in experiments that at moderate speeds the sideslip is sufficiently constant to be neglected without major implications.
We therefore simplify the relation by assuming constant sideslip and defining the desired \gls{rot} and its derivative as:
\begin{equation}
	\begin{aligned}
		r_d &= \dot \chi_d \\
		\dot r_d &= \ddot \chi_d.
	\end{aligned}
\end{equation}

The interested reader is referred to~\cite{Eriksen2018b} for more details on the speed and course controller.

\section{The BC-MPC algorithm}\label{sec:BC-MPC}
The \gls{bc-mpc} algorithm is intended to avoid collisions with moving obstacles while respecting the dynamic constraints of the vessel in order to ensure feasible maneuvers, which is ideal for short-term \gls{colav}.
The algorithm is based on \acrfull{mpc}, and plans vessel-feasible trajectories with multiple maneuvers where only the first maneuver is executed.
The trajectories have continuous acceleration, which is beneficial for vessel controllers utilizing model-based feedforward terms, such as \eqref{eq:vesselController}.
To fit well with tracking systems based on exteroceptive sensors, such as e.g. radars, the algorithm is designed to be robust with respect to noisy obstacle estimates.
Furthermore, the algorithm is designed with the short-term perspective of \gls{colregs} in mind.
The algorithm is also modular, so it can easily be tailored for different applications.

The \gls{bc-mpc} algorithm can be described by two steps, which will be explained in detail in the following sections:
\begin{enumerate}
	\item Generate a search space consisting of feasible trajectories with respect to the dynamic constraints of the vessel.
	\item Discretize the search space and compute an objective function value on the trajectories.
The optimal trajectory is then selected as the one with the lowest objective function value.
\end{enumerate}

\begin{figure}
	\centering
	\includegraphics[width=\textwidth]{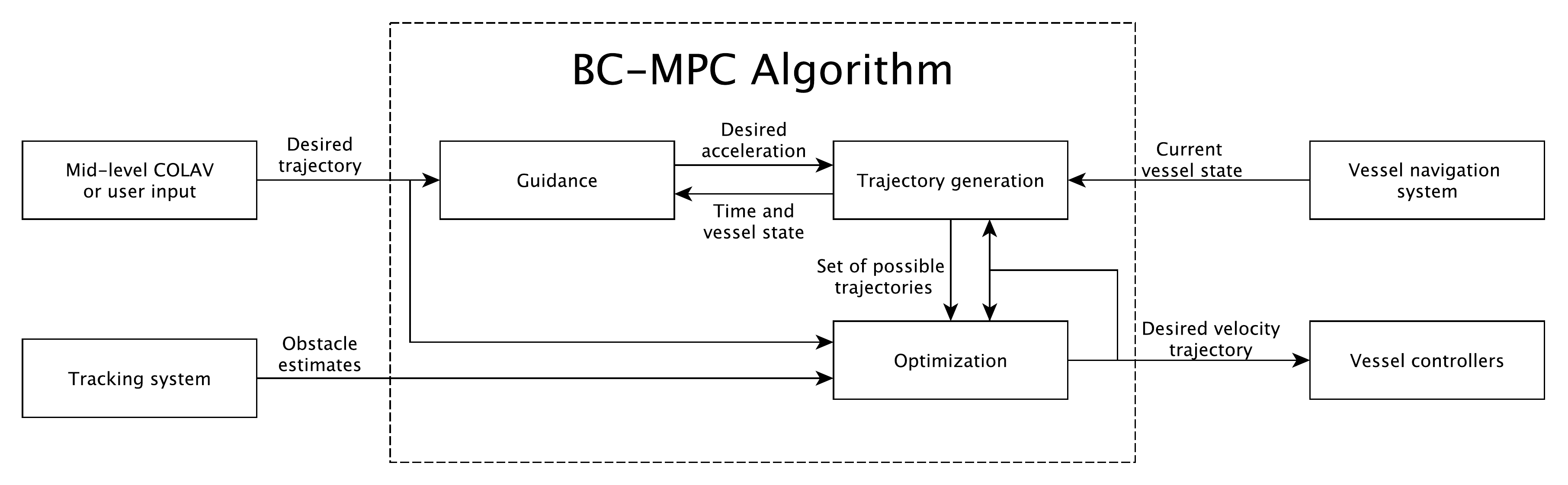}
	\caption{\label{fig:bc-mpc_architecture}BC-MPC algorithm overview.}
\end{figure}

The \gls{bc-mpc} algorithm architecture is shown in \cref{fig:bc-mpc_architecture}.
The algorithm inputs a desired trajectory, which can originate from either another \gls{colav} algorithm or directly from a user.
The guidance function receives the desired trajectory, and computes a desired acceleration given a vessel state and time specified by the trajectory generation.
The trajectory generation block creates a set of possible vessel trajectories, given an initial vessel state, initial desired velocity and a desired acceleration from the guidance function.
A tracking system provides obstacle estimates, which are used to calculate a part of the objective function.
The optimization block computes the optimal trajectory based on an objective function, and outputs this as a desired velocity trajectory to the vessel controller \eqref{eq:vesselController}.

\subsection{Trajectory generation}
The search space consists of a number of trajectories, each consisting of a sequence of sub-trajectories each containing one maneuver.
This section describes how the trajectories are generated.
Each trajectory is defined by a desired velocity trajectory containing a \gls{sog} and course trajectory with continuous acceleration, and feedback-corrected predicted pose and velocity trajectories.

\subsubsection{Trajectory generation: A single step}
As mentioned, each trajectory consists of a sequence of maneuvers, resulting in trajectories that branches out from each other.
Hence, the trajectory generation can be divided in repeatable steps.
At each step, a set of sub-trajectories, each containing one maneuver, are computed given an initial vessel configuration, initial time and some step-specific parameters:
\begin{itemize}
	\item The number of \gls{sog} maneuvers $N_U$
	\item The number of course maneuvers $N_\chi$
	\item The time allowed for changing the actuator input, named the ramp time $T_{\text{ramp}}$
	\item The maneuver time length in \gls{sog} $T_U$ and course $T_\chi$
	\item The total step time length $T$
\end{itemize}

\begin{figure}
	\centering
	\includegraphics[width=.7\textwidth]{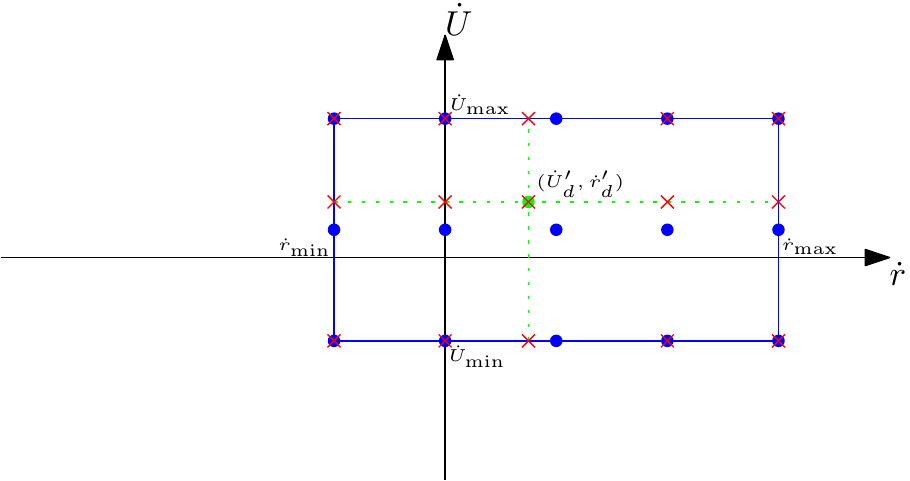}
	\caption{\label{fig:AccSamples}Set of possible accelerations shown with the blue line, with initial samples shown as blue circles.
The desired acceleration $(\dot U'_d, \dot r'_d)$ is shown as a green circle, while the final samples are shown as red crosses.}
\end{figure}

We start by generating the desired velocity trajectories, which should be feasible with respect to actuator rate and magnitude saturations.
To ensure feasibility with respect to the actuator rate saturations, we start from the model \eqref{eq:Model} by calculating the possible \gls{sog} and course accelerations given our current configuration as:
\begin{equation}\label{eq:possibleAcc}
	\begin{aligned}
		\dot{\bs X}_{\max} &= \bs M^{-1}\left( \bs\tau_{\max} - \bs\sigma(\bs X_0) \right) \\
		\dot{\bs X}_{\min} &= \bs M^{-1}\left( \bs\tau_{\min} - \bs\sigma(\bs X_0) \right),
	\end{aligned}
\end{equation}
where $\dot{\bs X}_{\max} = \begin{bmatrix} \dot U_{\max} & \dot r_{\max} \end{bmatrix}^T$, $\dot{\bs X}_{\min} = \begin{bmatrix} \dot U_{\min} & \dot r_{\min} \end{bmatrix}^T$, $\bs X_0$ is the current vessel velocity and:
\begin{equation}\label{eq:possibleActuator}
	\begin{aligned}
		\bs\tau_{\max} &= \text{sat}\left(\bs \tau_0 + T_{\text{ramp}} \dot{\bs \tau}_{\max}, \bs \tau_{\min},  \bs \tau_{\max}\right) \\
		\bs\tau_{\min} &= \text{sat}\left(\bs \tau_0 + T_{\text{ramp}} \dot{\bs \tau}_{\min}, \bs \tau_{\min},  \bs \tau_{\max}\right),
	\end{aligned}
\end{equation}
where $T_{\text{ramp}}>0$ is the ramp time, $\bs \tau_0$ is the current control input, $\bs \tau_{\max}$ and $\bs \tau_{\min}$ are the maximum and minimum control input, respectively, and $\dot{\bs \tau}_{\max}$ and $\dot{\bs \tau}_{\min}$ are the maximum and minimum control input rate of change, respectively.
The saturation function $\text{sat}(\bs a,\bs a_{\min},\bs a_{\max})$ is defined as $\text{sat}:\mathbb{R}^K\times\mathbb{R}^K\times\mathbb{R}^K \rightarrow \mathbb{R}^K$ with:
\begin{equation}
	a^*_i = \begin{cases}
  	 a_{\min,i}     &, a_i < a_{\min,i} \\
  	 a_{\max,i}     &, a_i > a_{\max,i} \\
  	 a_i			&, \text{otherwise},
  	\end{cases}
\end{equation}
for $\bs a^* = \text{sat}(\bs a,\bs a_{\min},\bs a_{\max})$, $i\in\{1,2,\ldots,K\}$ and $(\cdot)_i$ denoting element $i$ of a vector.
Following this, we create a set of possible accelerations as:
\begin{equation}
	A_d = \left\{ (\dot U, \dot r) \in \mathbb{R}\times\mathbb{R} \big | \dot U \in [\dot U_{\min}, \dot U_{\max}], \dot r \in [\dot r_{\min}, \dot r_{\max}] \right\}.
\end{equation}
The set of possible accelerations is then sampled uniformly to create a discrete set of candidate maneuver accelerations:
\begin{equation}
	\begin{aligned}
		\dot{\bs U}_{\text{samples}} &= \left\{ \dot U_1, \dot U_2, \ldots, \dot U_{N_U} \right\} \\
		\dot{\bs r}_{\text{samples}} &= \left\{ \dot r_1, \dot r_2, \ldots, \dot r_{N_\chi} \right\},
	\end{aligned}
\end{equation}
where $\dot U_i$, $i\in [1,N_U]$ are \gls{sog} acceleration samples and $\dot r_i$, $i\in [1,N_\chi]$ are course acceleration samples.
To be able to include a specific maneuver in the search space, which can be beneficial e.g. to converge to a specific desired trajectory, we allow to modify some of the sampled accelerations if a desired acceleration $(\dot U'_d, \dot r'_d)$ is inside the set of possible accelerations as follows: If $\dot U'_d \in A_d$, we change the closest \gls{sog} acceleration sample in $\dot{\bs U}_{\text{samples}}$ to $\dot U'_d$.
Similarly for course, if $\dot r'_d \in A_d$, we change the closest course acceleration sample in $\dot{\bs r}_{\text{samples}}$ to $\dot r'_d$.
Following this, we create a set of candidate maneuver accelerations by combining the \gls{sog} and course candidate maneuvers as $\dot{\bs U}_{\text{samples}} \times \dot{\bs r}_{\text{samples}}$.
This concept is illustrated in \cref{fig:AccSamples}, where $A_d$ is sampled with $N_U = 3$ \gls{sog} samples and $N_\chi = 5$ course samples.

\begin{figure}
  	\centering
  	\begin{subfigure}[b]{.7\textwidth}
    	\includegraphics[width=\textwidth]{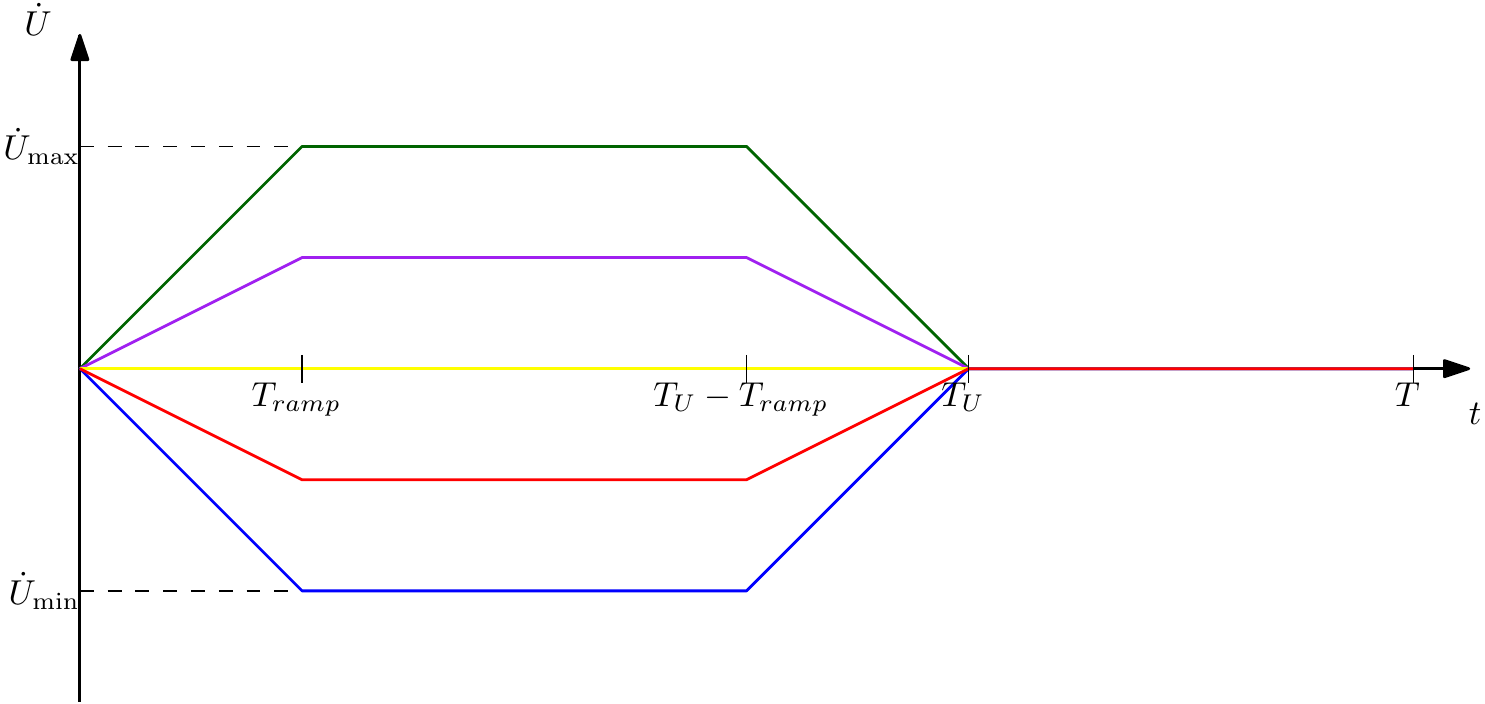}
    	\caption{Speed acceleration motion primitives}
  	\end{subfigure} \\
  	\begin{subfigure}[b]{.7\textwidth}
    	\includegraphics[width=\textwidth]{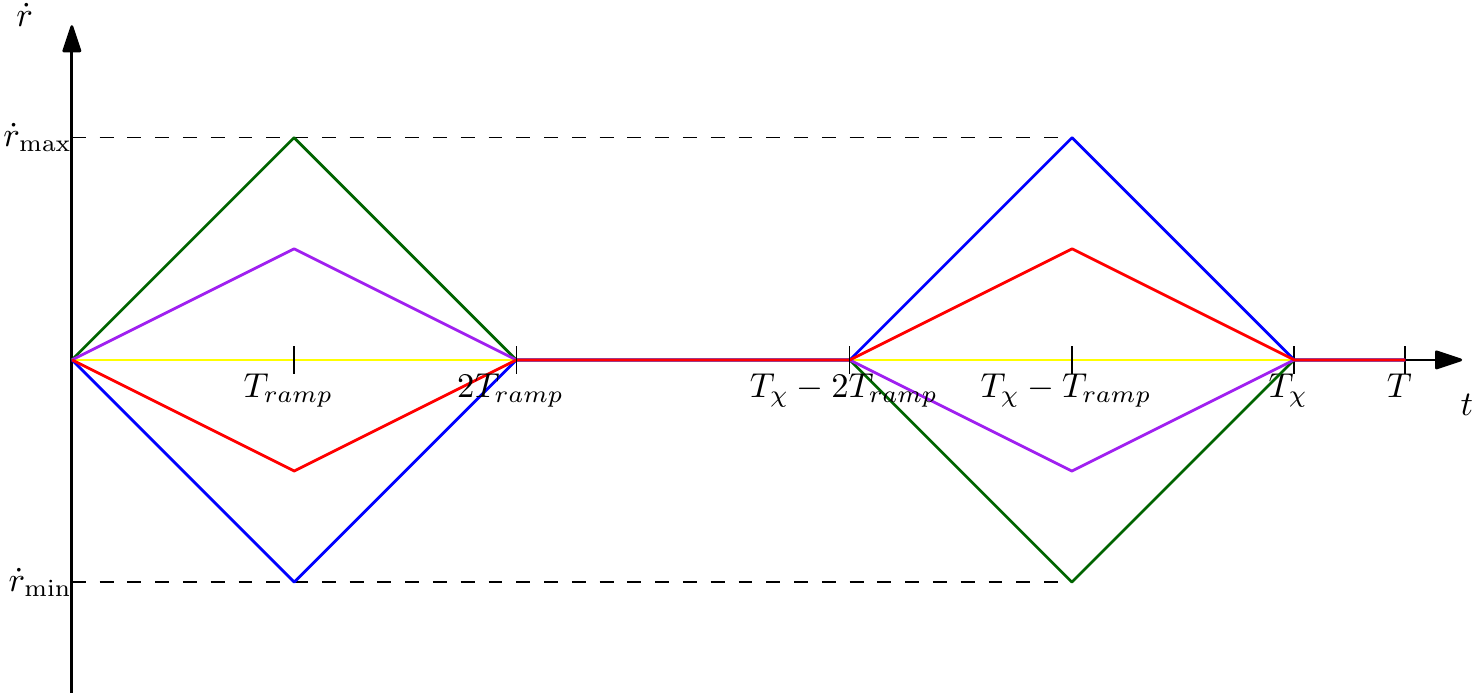}
    	\caption{Course acceleration motion primitives.
Note that the integral of each course acceleration trajectory is zero.}
  	\end{subfigure}
  	\caption{\label{fig:motionPrimitives}Acceleration motion primitives, where $T$ is the step time, $T_{ramp}$ denotes the ramp time while $T_U$ and $T_\chi$ are the \gls{sog} and course maneuver time lengths, respectively.}
\end{figure}

Given the acceleration samples, we create a set of $N_U$ motion primitives for \gls{sog} based on the piecewise-linear \gls{sog} acceleration trajectories:
\begin{equation}\label{eq:sog_acc}
  	\dot U_{d,i}(t) = \begin{cases}
  	 k_{U,i} t                                      &, 0 \leq t < T_{\text{ramp}}  \\
  	 \dot U_i                              &, T_{\text{ramp}} \leq t < T_U - T_{\text{ramp}} \\
  	 \dot U_i - k_{U,i} (t - (T_U - T_{\text{ramp}})) &, T_U - T_{\text{ramp}} \leq t < T_U \\
  	 0,                                             &, T_U \leq t \leq T,
  	\end{cases}
\end{equation}
where $k_{U,i} = \frac{\dot U_i}{T_{\text{ramp}}}$, $\dot U_i$ is the sampled acceleration for \gls{sog} motion primitive $i\in[1,N_U]$, $T_U>0$ is the \gls{sog} maneuver length and $T>0$ is the total trajectory length.
Similarly, we define $N_\chi$ course motion primitives by the piecewise-linear course acceleration trajectories:
\begin{equation}\label{eq:course_acc}
  	\dot r_{d,i}(t) = \begin{cases}
  	 k_{r,i} t                                        			&, 0 \leq t < T_{\text{ramp}}  \\
  	 2 \dot r_i - k_{r,i} t                      		&, T_{\text{ramp}} \leq t < 2 T_{\text{ramp}} \\
  	 0                                                			&, 2 T_{\text{ramp}} \leq t < T_\chi - 2 T_{\text{ramp}} \\
  	 -k_{r,i} (t - (T_\chi - 2 T_{\text{ramp}}))                     	&, T_\chi - 2 T_{\text{ramp}} \leq t < T_\chi - T_{\text{ramp}}  \\
  	 -2 \dot r_i + k_{r,i} (t - (T_\chi - T_{\text{ramp}})) 	&, T_\chi - T_{\text{ramp}} \leq t < T_\chi \\
  	 0                                                			&, 2 T_\chi \leq t < T,
  	\end{cases}
\end{equation}
where $k_{r,i} = \frac{\dot r_i}{T_{\text{ramp}}}$, $\dot r_i$ is the sampled acceleration for course motion primitive $i\in[1,N_\chi]$ and $T_\chi>0$ is the course maneuver length.
For notational simplicity and without loss of generality, we assumed zero initial time $t_0=0$ in \eqref{eq:sog_acc} and \eqref{eq:course_acc}.
The acceleration trajectories and parameters for $N_U = 5$ \gls{sog} motion primitives and $N_\chi = 5$ course motion primitives are illustrated in \cref{fig:motionPrimitives}.
Notice that the integral of the course acceleration maneuvers are zero, hence if the maneuver is initialized with zero \gls{rot} the maneuver will end with zero \gls{rot}.
\todo{Burde være tekst i ramp.
Ny figur i CAMS artikkel}

\begin{figure}
	\centering
	\begin{subfigure}[b]{.7\textwidth}
    	\includegraphics[width=\textwidth]{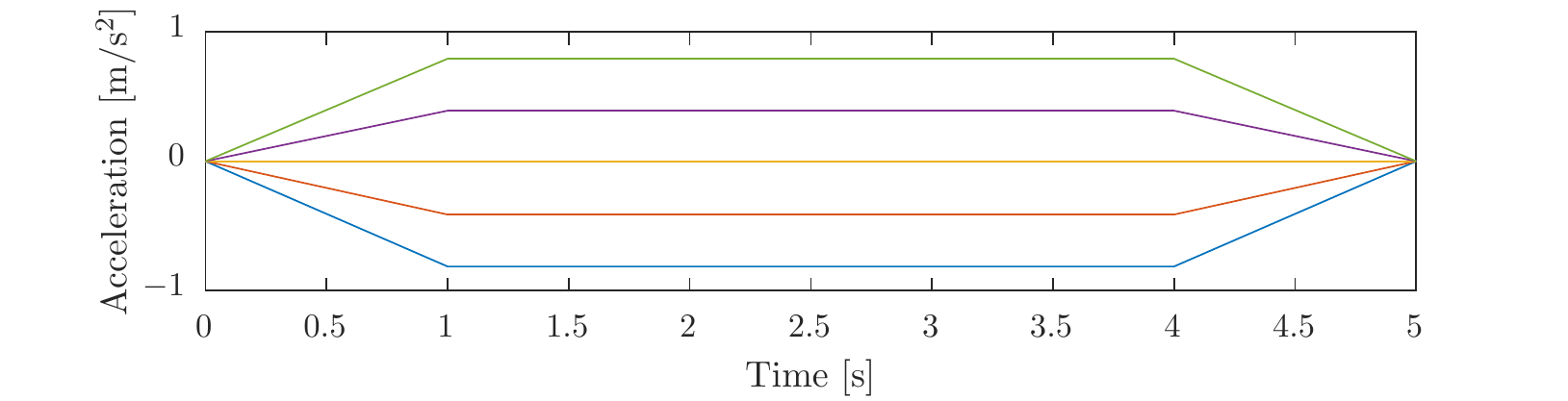}
  	\end{subfigure} \\
	\begin{subfigure}[b]{.7\textwidth}
    	\includegraphics[width=\textwidth]{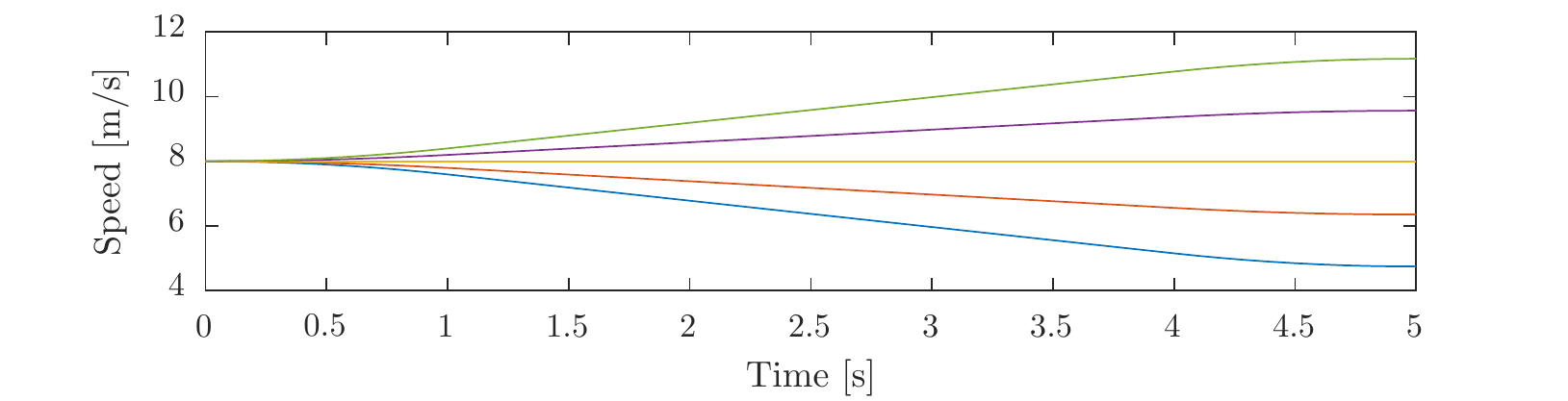}
  	\end{subfigure}
  	\caption{Example of $N_U=5$ \gls{sog} trajectories with $T_{\text{ramp}}=1~\si{\second}$ and $T=T_U=5~\si{\second}$.
Acceleration in the top plot and speed in the bottom plot.\label{fig:SOGtraj}}
\end{figure}

\begin{figure}
	\centering
	\begin{subfigure}[b]{.7\textwidth}
    	\includegraphics[width=\textwidth]{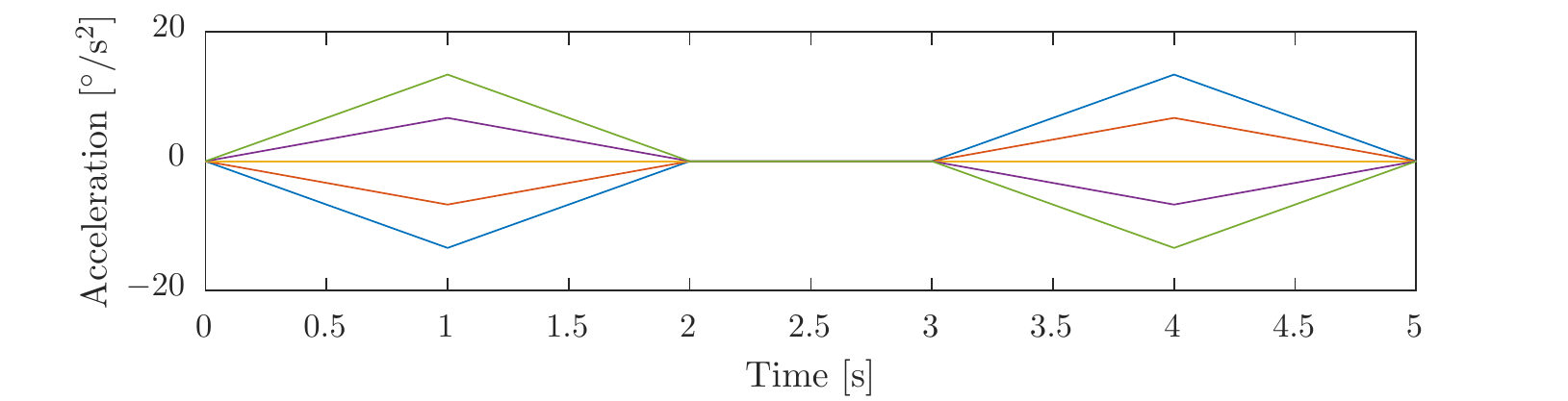}
  	\end{subfigure} \\
	\begin{subfigure}[b]{.7\textwidth}
    	\includegraphics[width=\textwidth]{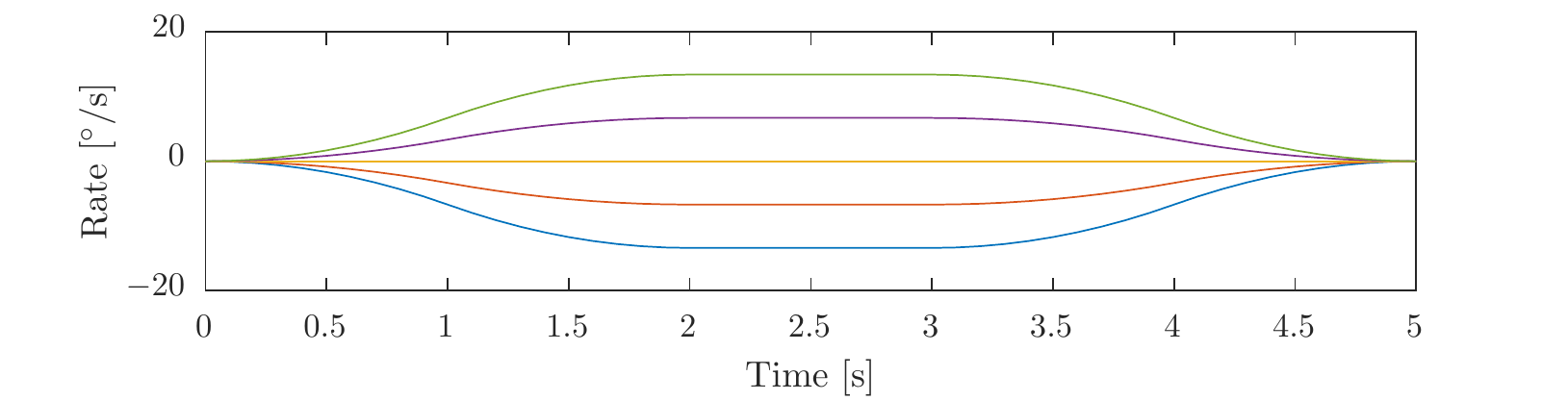}
  	\end{subfigure} \\
  	\begin{subfigure}[b]{.7\textwidth}
    	\includegraphics[width=\textwidth]{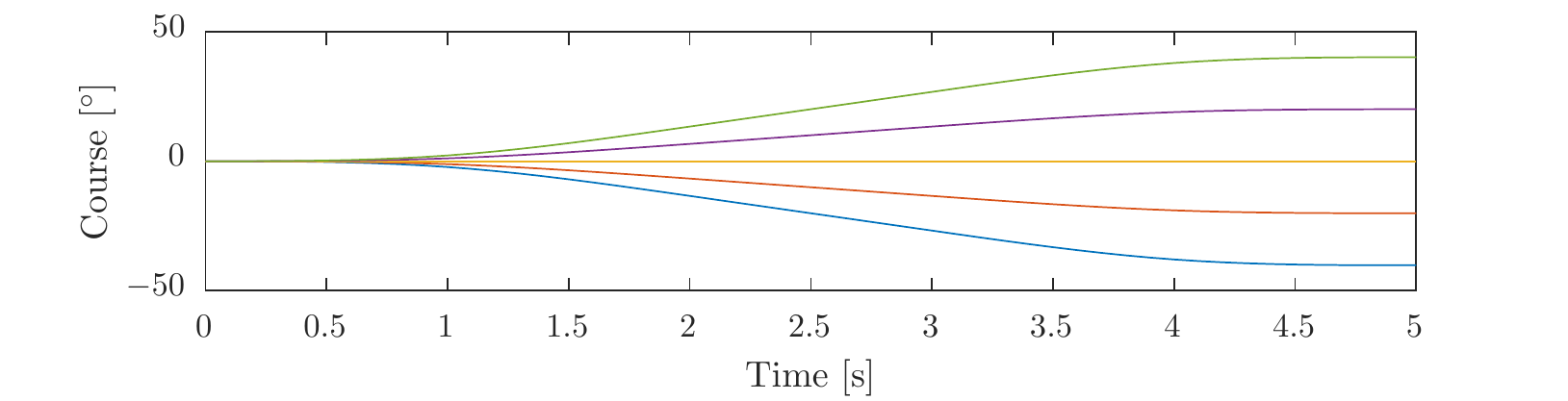}
  	\end{subfigure}
  	\caption{Example of $U_\chi=5$ course trajectories with $T_{\text{ramp}}=1~\si{\second}$ and $T=T_\chi=5~\si{\second}$.
Acceleration in the top plot, rate in the middle and course in the bottom plot.\label{fig:Coursetraj}}
\end{figure}

Based on the acceleration trajectories, we create trajectories for the desired \gls{sog}, \gls{rot} and course by integrating the expressions \eqref{eq:sog_acc} and \eqref{eq:course_acc} as:
\begin{equation}\label{eq:velocity_trajs}
	\begin{aligned}
		U_{d,i}(t) &= U_{d,0} + \int_{t_0}^t \dot U_{d,i}(\gamma) \mathrm{d}\gamma, \quad i \in [1,N_U] \\
		r_{d,i}(t) &= r_{d,0} + \int_{t_0}^t \dot r_{d,i}(\gamma) \mathrm{d}\gamma, \quad i \in [1,N_\chi] \\
		\chi_{d,i}(t) &= \chi_{d,0} + \int_{t_0}^t r_{d,i}(\gamma) \mathrm{d}\gamma, \quad i \in [1,N_\chi]. \\
	\end{aligned}
\end{equation}
The initial values $U_{d,0}$, $r_{d,0}$ and $\chi_{d,0}$ are taken as the corresponding desired values from the last \gls{bc-mpc} iteration (or sub-trajectory, if computing trajectories for subsequent maneuvers), such that the desired trajectories passed to the vessel controllers are continuous.
This implies that we do not include feedback in the desired trajectories.
Furthermore, as in \cref{sec:ASVmodelingControl}, the vessel sideslip is neglected.
This could, however, be included by using a vessel model including sideslip.
A numerical example of $5$ \gls{sog} and $5$ course trajectories is shown in \cref{fig:SOGtraj,fig:Coursetraj}, where a maneuver length of $5~\si{\second}$ is used for both \gls{sog} and course.
Vessels at sea usually maneuver by either keeping a constant speed and course or by performing a speed and/or course change and continuing with this new speed and course for some time.
By selecting the initial yaw rate in \eqref{eq:velocity_trajs} as $r_{d,0} = 0$ we ensure that maneuvers start and end with constant-course motion, which mimics this behavior while also producing maneuvers that should be readily observable for other vessels, as required by rule 8 of \gls{colregs}.

Following this, we create a union set of the desired velocity trajectories as:
\begin{equation}\label{eq:desiredVelocities}
	\mathcal{U}_d = \{U_{d,1}(t), U_{d,2}(t), \ldots, U_{d,{N_U}}(t)\} \times \{ \chi_{d,1}(t), \chi_{d,2}(t), \ldots , \chi_{d,{N_\chi}}(t)\},
\end{equation}
resulting in a total of $N_U\!\cdot\! N_\chi$ desired velocity trajectories.
Notice that the \gls{sog} trajectories in $\mathcal{U}_d $ are continuously differentiable, while the course trajectories are twice continuously differentiable.
Velocity trajectories containing infeasible steady-state vessel velocities are removed from $\mathcal{U}_d $ by checking the feasibility using the vessel model \eqref{eq:Model} together with the actuator saturation constraints.

Given the desired velocity trajectories, we calculate the feedback-corrected pose trajectories.
To do this, we first predict the resulting \gls{sog} and course trajectories, $\bar U_i(t)$, $i\in [1,N_U]$ and $\bar \chi_i(t)$, $i\in[1,N_\chi]$, respectively.
This is done by simulating the closed-loop error dynamics of the vessel and vessel controllers using the desired velocity trajectories as the input.
In this article, we approximate the error dynamics using first order linear models, which may seem as quite rough approximations.
However, this is justified by noting that the model-based speed and course controller demonstrates very good control performance for the Telemetron \gls{asv}, resulting in small control errors~\cite{Eriksen2018b}.
Furthermore, the control errors are dominated by environmental disturbances, which is difficult to model without increasing the complexity to an unnecessarily high level.
The closed-loop error models are given as:
\begin{equation}\label{eq:errorModel}
  	\begin{aligned}
    	\dot{\tilde U} &= \frac{1}{T_{\tilde U}} \tilde U \\
    	\dot{\tilde \chi} &= \frac{1}{T_{\tilde \chi}} \tilde \chi,
  	\end{aligned}
\end{equation}
where $\tilde U = \bar U - U_d$, $\tilde \chi = \bar \chi - \chi_d$, and $T_{\tilde U}>0$ and $T_{\tilde \chi}>0$ are time constants.
The time constants are heuristically determined through simulations and experiments.
Using the error model, the predicted \gls{sog} and course trajectories are found as:
\begin{equation}\label{eq:velocityPrediction}
  	\begin{aligned}
    	\bar U_i(t) &=  \tilde U_0 \mathrm{e}^{-\frac{1}{T_{\tilde U}}(t - t_0) } + U_{d,i}(t), & i \in [1,N_U]\\
    	\bar \chi_i(t) &=  \tilde \chi_0 \mathrm{e}^{-\frac{1}{T_{\tilde \chi}}(t - t_0) } + \chi_{d,i}(t), & i \in [1,N_\chi],
  	\end{aligned}
\end{equation}
where $\tilde U_0= U_0 - U_{d,0}$ and $\tilde \chi_0 = \chi_0- \chi_{d,0}$ introduces feedback in the prediction through the current vessel \gls{sog} and course, $U_0$ and $\chi_0$, respectively.
Similarly as \eqref{eq:desiredVelocities}, we construct a set of predicted velocity trajectories:
\begin{equation}\label{eq:predictedVelocities}
	\bar{\mathcal{U}} = \{\bar U_1(t), \bar U_2(t), \ldots, \bar U_{N_U}(t)\} \times \{ \bar \chi_1(t), \bar \chi_2(t), \ldots , \bar \chi_{N_\chi}(t)\}.
\end{equation}
Combinations of \gls{sog} and course trajectories that was considered infeasible when forming $\mathcal{U}_d$ are also removed from $\bar{\mathcal{U}}$.
Following this, vessel position trajectories $\bar{\bs p}(t) = \begin{bmatrix} \bar N(t) & \bar E(t) \end{bmatrix}^T$ are calculated from the predicted velocity trajectories using a kinematic model:
\begin{equation}\label{eq:kinematicPositionIntegral}
  	\begin{aligned}
    	\dot{\bar{\bs p}} &= \begin{bmatrix} \cos(\bar\chi) \\ \sin(\bar\chi) \end{bmatrix} \bar U,
  	\end{aligned}
\end{equation}
which is integrated using the current vessel position as the initial condition.
The feedback-corrected predicted vessel pose trajectories are finally combined in the set $\bar{\mathcal{H}}$ as:
\begin{equation}\label{eq:predictedPosition}
	\bar{\mathcal{H}} = \left\{\bar {\bs\eta}(t; \bar U(t), \bar \chi(t)) \big | (\bar U(t), \bar \chi(t)) \in \bar{\mathcal{U}} \right\},
\end{equation}
where $\bar{\bs \eta} = \begin{bmatrix} \bar N(t) & \bar E(t) & \bar \chi(t) \end{bmatrix}^T$.\todo{Dette er jo ikke en pose? Men kan kanskje skrive om til $\bar{\bs \eta} = [ \bar N(t), \bar E(t), \bar \psi(t) ]^T$, $\bar\chi(t)$ ligger jo implisitt inne i $\bar{\bs \eta}(t)$ som $atan2(\dot E(t), \dot N(t))$.}

To summarize, a single step of a trajectory is defined by the set of desired velocity trajectories $\mathcal{U}_d$, the set of predicted velocity trajectories $\bar{\mathcal{U}}$ and the set of set of predicted pose trajectories $\bar{\mathcal{H}}$.

\subsubsection{Trajectory generation: The full trajectory generation}
A full trajectory consists of multiple sub-trajectories, each containing one maneuver and constructed using the single-step procedure.
This naturally forms a tree structure, with nodes representing vessel states and edges representing sub-trajectories.
The depth of the tree will be equal to the desired number of maneuvers in each trajectory.
The tree is is initialized with the initial state as the root node, which the single-step procedure is performed on, generating a number of sub-trajectories and leaf nodes.
Following this, the the single-step procedure is performed on each of the leaf nodes, adding the next sub-trajectory and leaf nodes to the existing trajectories and expanding the tree depth.
This procedure is repeated until the tree has the desired depth, resulting in each trajectory having the desired number of maneuvers.
Using the same number of \gls{sog} and course maneuvers at each level would result in the tree growing exponentially with the number of levels.
To limit the growth, we therefore allow for choosing a different number of \gls{sog} and course maneuvers at each level, for instance keeping the \gls{sog} constant in all levels except the first, only allowing the \gls{sog} to be changed during the first maneuver of a trajectory.

The remaining parameters can also be chosen differently for each level, and in principle the acceleration trajectories \eqref{eq:sog_acc} and \eqref{eq:course_acc} can also be designed using different structures.
However, we choose to use the same acceleration trajectory structure for each level, while also keeping the ramp time and maneuver time lengths constant.
This leaves only the step time length and number of \gls{sog} and course maneuvers as parameters that can change throughout the tree depth.
Choosing different step time lengths can be considered as an \gls{mpc} input blocking scheme, requiring that the step lengths are integer dividable by the algorithm sample time.

A full trajectory generation can hence be defined by the following parameters:
\begin{itemize}
  	\item An initial vessel state including the current desired velocity.
  	\item The number of maneuvers in each trajectory, or levels, defined as $B>0$.
  	\item The step time at each level $\bs T = \begin{bmatrix} T_1 & T_2 & \ldots & T_{B} \end{bmatrix}$, the ramp time $T_{\text{ramp}}$ and the \gls{sog} and course maneuver lengths $T_U$ and $T_\chi$, respectively.
  	\item The number of \gls{sog} maneuvers at each level $\bs N_U = \begin{bmatrix} N_{U,1} & N_{U,2} & \ldots & N_{U,B} \end{bmatrix}$.
  	\item The number of course maneuvers at each level $\bs N_\chi = \begin{bmatrix} N_{\chi,1} & N_{\chi,2} & \ldots & N_{\chi,B} \end{bmatrix}$.
\end{itemize}

\begin{figure}
	\centering
	\includegraphics[width=.8\textwidth]{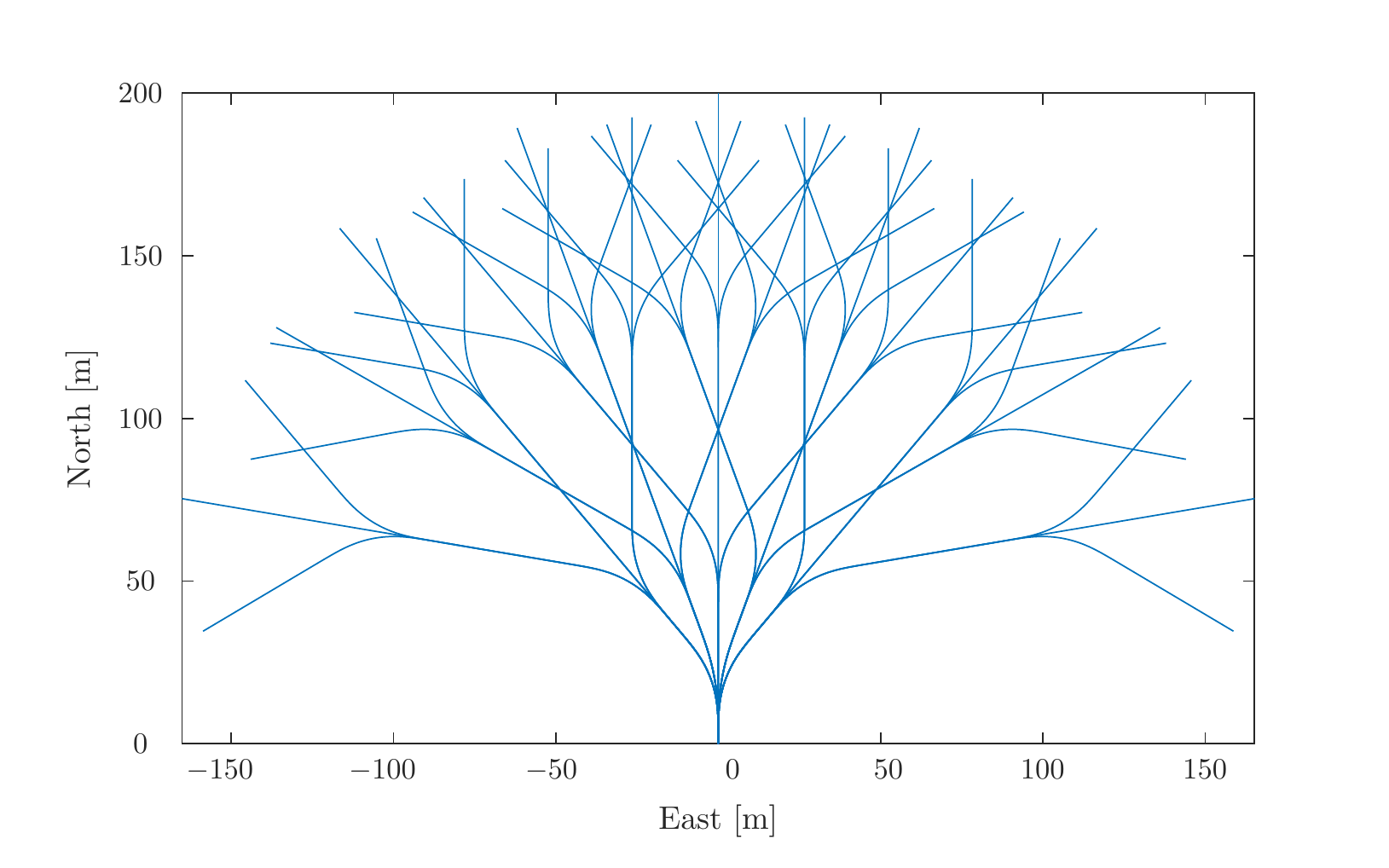}
  	\caption{A set of predicted pose trajectories with 3 levels.\label{fig:postraj}}
\end{figure}

A set of predicted vessel pose trajectories with $B=3$ levels is shown in \Cref{fig:postraj}.
The step time is chosen as $\bs T = \begin{bmatrix} 5~\si{\second} & 10~\si{\second} & 10~\si{\second} \end{bmatrix}$, making the trajectories $25~\si{\second}$ long in total.
The trajectories have $5$ course maneuvers at the first level and three in the later levels, while there for illustrational purposes are only one \gls{sog} maneuver at each step.
Hence, $\bs N_U = \begin{bmatrix} 1 & 1 & 1 \end{bmatrix}$ and $\bs N_\chi = \begin{bmatrix} 5 & 3 & 3 \end{bmatrix}$.
The ramp time and maneuver lengths are chosen as $T_{\text{ramp}} = 1~\si{\second}$ and $T_U = T_\chi = 5~\si{\second}$, respectively.
Notice that the maneuver length of $5$ seconds results in the second and third maneuver having a straight-course segment after the turn, which increases the prediction horizon without increasing the computational load while also increasing the maneuver observability.

\subsubsection{Calculating a desired acceleration}
In the single-step trajectory generation, a desired acceleration $(\dot U_d', \dot r_d')$ can be used to include a desired maneuver in the search space.
We therefore use a guidance algorithm to ensure that there exists a trajectory in the search space that converges towards the desired trajectory inputted to the \gls{bc-mpc} algorithm.
To achieve this, we use a modified version of a path tracking algorithm ensuring vessel convergence to a curved path~\cite{Breivik2004}.
The control law is based on \gls{los} guidance~\cite{Fossen2011}, together with defining a desired point on the path which the velocity of is controlled, named the path particle (PP).
The desired course is stated as:
\begin{equation}\label{eq:trackingChi}
	\chi_{d,LOS} = \chi_{path} + \arctan\left(- \frac{e}{\Delta}\right),
\end{equation}
where $\chi_{path}$ is the path angle at the desired point, $e$ is the cross-track error and $\Delta > 0$ is the lookahead distance.
The path particle velocity along the path is stated as:
\begin{equation}\label{eq:trackingUpp}
	U_{PP} = U \cos(\chi - \chi_{path}) + \gamma_s s,
\end{equation}
where $U$ is the vessel speed, $\gamma_s > 0$ is a tuning parameter and $s$ is the along-track distance.
This control law controls the speed along the path $U_{PP}$ as a function of the vessel speed, course and the along-track distance to the path particle, letting the vessel converge towards the path with a constant speed.
We rather want to be able to follow a desired trajectory by controlling the vessel speed and course based on the desired trajectory.
We therefore fix the path particle at the desired position on the trajectory, given the current time, and by reformulating \eqref{eq:trackingUpp} we obtain a desired vessel speed given the trajectory velocity:
\begin{equation}\label{eq:LOSSpeed}
	U_{d,LOS} = \begin{cases}	\text{sat}\left(\frac{U_{t} - \gamma_s s}{\cos(\chi - \chi_{path})},0,U_{\max,LOS}\right)  & \text{if } \lvert \cos(\chi - \chi_{path}) \rvert > \epsilon \\
						\text{sat}\left(\frac{U_{pp} - \gamma_ss}{\epsilon},0,U_{\max,LOS}\right)  & \text{else,} \end{cases}
\end{equation}
where $U_t$ is the trajectory velocity and $\epsilon > 0$ is a small constant to avoid division by zero.
The saturation function ensures that the desired vessel speed is in the interval $[0,U_{\max}]$, where $U_{\max} >0$ is the maximum vessel operating speed.
Given a desired speed and course, we compute the desired \gls{sog} and course acceleration:
\begin{equation}\label{eq:LOS_acceleration}
	\begin{aligned}
		\dot U_d' &= \frac{U_{d,LOS} - U_{d,0}}{T_U - T_{\text{ramp}}} \\
		\dot r_d' &= \frac{\chi_{d,LOS} - \chi_{d,0}}{T_{\text{ramp}}\left(T_\chi - 2 T_{\text{ramp}}\right)},
	\end{aligned}
\end{equation}
which are found by solving \eqref{eq:velocity_trajs} for the final desired \gls{sog} and course.
Notice that in cases where there is only one \gls{sog} and/or course maneuver, the corresponding desired acceleration should be selected as zero to keep a constant speed and/or course.

The obvious singularity in \eqref{eq:LOSSpeed} when the vessel course is perpendicular to the desired path (and hence $\cos(\chi - \chi_{path} = 0$) is handled by avoiding division by {zero\todo{MB:Vi får oppdatere denne i neste iterasjon :)}} and ensuring that the desired speed is inside the possible operating speed of the vessel, which makes it difficult to guarantee stability and convergence of this guidance scheme.
However, the desired acceleration is only used to modify some trajectories in the \gls{bc-mpc} search space, and will hence not constrain the algorithm to choose a trajectory based on \eqref{eq:LOS_acceleration}.
One could employ other schemes, e.g. \cite{Paliotta2017} which guarantees convergence to curved trajectories.
This does, however, increase the complexity by depending on a detailed 3DOF model of the vessel while also employing a feedback-linearizing controller to control the vessel.
It is in general difficult to obtain detailed models of high-speed \glspl{asv}, while time delays, sensor noise and modeling uncertainties are shown to cause robustness issues when using feedback-linearizing controllers \cite{Eriksen2017}.
Hence, the simplicity of \eqref{eq:trackingChi}--\eqref{eq:LOSSpeed} is appealing when a guarantee of stability and convergence is not required.

\subsection{Selecting the optimal trajectory}
Given the set of feasible trajectories, we solve an optimization problem to select the optimal trajectory.
We start by defining a cost function to assign a cost to each trajectory:
\begin{equation}\label{eq:objFunc}
 	G(\bar{\bs \eta}(t), \bs u_d(t);\bs p_d(t)) = w_{\text{al}}\text{align}(\bar{\bs \eta}(t); \bs p_d(t)) + w_{\text{av}}\text{avoid}(\bar{\bs \eta}(t)) + w_{\text{t}}\text{tran}(\bs u_d(t)),
\end{equation}
where $(\bar{\bs \eta}(t), \bs u_d(t))$ is the predicted vessel pose and desired velocity of a candidate trajectory, $\text{align}(\bar{\bs \eta}(t); \bs p_d(t))$ measures the alignment between the predicted pose trajectory and a desired trajectory $\bs p_d(t)$, $\text{avoid}(\bar{\bs \eta}(t))$ assigns cost to trajectories traversing close to obstacles, while $\text{tran}(\bs u(t))$ introduces transitional cost in the objective function to avoid wobbly behavior.
The parameters $w_{\text{al}}, w_{\text{av}}, w_{\text{t}} \geq 0$ are tuning parameters to control the weighting of the different objective terms.

Using \eqref{eq:objFunc}, we define the optimization problem:
\begin{equation}\label{eq:ocp}
	\bs u_d^*(t) = \underset{(\bar{\bs \eta}_k(t), \bs u_{d,k}(t)) \in (\bar{\mathcal{H}}, \mathcal{U}_d)}{\text{argmin}} G(\bar{\bs \eta}_k(t), \bs u_{d,k}(t); \bs p_d(t)),
\end{equation}
where $\bs u_d^*(t)$ is the optimal desired velocity trajectory to be used as the reference for the vessel controllers.
The optimization problem is solved by simply calculating the cost over the finite discrete set of trajectories and choosing the one with the lowest cost.

The next sections describe the different terms of the objective function \eqref{eq:objFunc}.
Notice that we strive to avoid using discontinuities and logic in order to improve the robustness with respect to obstacle estimate noise.

\subsubsection{Trajectory alignment}
The alignment between the desired trajectory and a candidate trajectory is used in the objective function \eqref{eq:objFunc} to motivate the algorithm to follow the desired trajectory.
Given a desired trajectory $\bs p_d(t) : \mathrm{R}^+ \to \mathrm{R}^2$, required to be $C^1$, we obtain a desired course as:
\begin{equation}
  	\chi_d(t) = \text{atan2}(\dot E_d(t), \dot N_d(t)),
\end{equation}
with $\bs p_d(t) = \begin{bmatrix} N_d(t) & E_d(t) \end{bmatrix}^T$.
Given this, we define a weighted metric of Euclidean distance and orientation error as:
\begin{equation}
  	\text{align}(\bar{\bs \eta}(t); \bs p_d(t)) = \int_{t_0}^{t_0+T_{\text{full}}} \left( w_p\norm{\begin{bmatrix} \bar N(\gamma) \\ \bar E(\gamma) \end{bmatrix} - \bs p_d(\gamma)}_2 + w_\chi |\Upsilon\left(\bar\chi(\gamma) - \chi_d(\gamma)\right)|\right)\mathrm{d}\gamma,
\end{equation}
where $w_p, w_\chi>0$ are weights controlling the influence of the Euclidean and angular error, respectively, $T_{\text{full}} = \sum_{i=1}^B T_i$ denotes the entire trajectory prediction horizon, while $\Upsilon: \mathbb{R} \to S^1$ maps an angle to the domain $[-\pi,\pi)$.
For simplicity, we fix $w_p = 1$ and leave $w_\chi$ and $w_{\text{al}}$ to control the weighting.

\subsubsection{Obstacle avoidance}
Obstacle avoidance is achieved by penalizing candidate trajectories with small distances to obstacles.
We define three regions around the obstacles, named the collision, safety and margin regions, respectively.
The idea behind this is to make it possible use different gradients on the penalty depending on how close the ownship is to the obstacle, which should improve the robustness with respect to noise on the obstacle estimates.

\begin{figure}
  	\centering
  	\includegraphics[width=.5\textwidth]{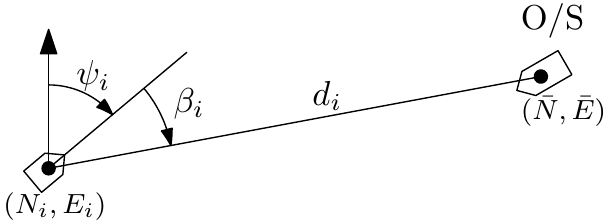}
  	\caption{\label{fig:obsDistPose}Distance $d_i$ and relative bearing $\beta_i$ to obstacle $i$.
The ownship is marked O/S.}
\end{figure}

We define a time-varying vector between obstacle $i$ and a predicted vessel trajectory as:
\begin{equation}\label{eq:obstacleVector}
   	\bs r_i(\bar{\bs \eta}(t); \bs p_i(t)) = \bs p_i(t) - \begin{bmatrix} \bar N(t) \\ \bar E(t) \end{bmatrix},
\end{equation}
where $\bs r_i = \begin{bmatrix} r_{N,i} & r_{E,i} \end{bmatrix}^T$ and $\bs p_i(t)$ is the position of obstacle $i$ at time $t$.
The obstacle position in future time is computed under the common assumption that obstacles will keep their current speed and course~\cite{Johansen2016,Kuwata2014,Eriksen2018}, which is a reasonable assumption for relatively short time periods.
More complex techniques can also be applied for predicting the future position of obstacles, for instance based on historic \gls{ais} data~\cite{Dalsnes2018} or by estimating the turn rate of the obstacles~\cite{Flaaten2017}.
Using \eqref{eq:obstacleVector}, we define the distance and relative bearing to obstacle $i$ given a predicted vessel trajectory $\bar{\bs \eta}(t)$ as:
\begin{equation}
  	\begin{aligned}
    	d_i(\bar{\bs \eta}(t);\bs p_i(t)) &= \norm{\bs r_i(\bar{\bs \eta}(t);\bs p_i(t))}_2 \\
    	\beta_i(\bar{\bs \eta}(t);\bs p_i(t)) &= \Upsilon\left(\text{atan2}\big(r_{E,i}(\bar{\bs \eta}(t)), r_{N,i}(\bar{\bs \eta}(t))\big) - \chi_i(t)\right),
  	\end{aligned}
\end{equation}
where $\chi_i(t)$ is the course of obstacle $i$, calculated as $\chi_i(t) = \text{atan2}\left(\dot E_i(t), \dot N_i(t) \right)$ with $\bs p_i(t) = \begin{bmatrix} N_i(t) & E_i(t) \end{bmatrix}^T$.
The distance $d_i$ and relative bearing $\beta_i$ are illustrated in \Cref{fig:obsDistPose}.

The obstacle distance and relative bearing is used to calculate a penalty function, which we use to define the avoidance function as:
\begin{equation}
  	\text{avoid}(\bar{\bs \eta}(t)) = \sum_{i=1}^M \int_{t=t_0}^{t_0+T_{\text{full}}} w_{i}(\gamma) \text{penalty}_i(\bar{\bs \eta}(\gamma))\mathrm{d}\gamma,
\end{equation}
where $M$ is the number of obstacles, $\text{penalty}_i(\bar{\bs \eta}(t))$ assigns a penalty to the predicted vessel trajectory $\bar{\bs \eta}(t)$ at time $t$ with respect to obstacle $i$, while $w_{i}(t)$ are time and obstacle dependent weights.
The weights can be useful for prioritizing vessels in multi-encounter situations where properties like vessel type, size, speed, etc. can be used for differentiating the importance of avoiding the given vessels in severe situations.
The weights can also facilitate time-dependent weighting, for instance as a heuristic method to incorporate uncertainty on obstacle estimates, combined with obstacle and time-dependent scaling of the obstacle region sizes.
For simplicity, we keep the weights constant at $w_{i}(t) = 1 \, \forall \,i$.

The penalty function can be designed in a variety of ways, with the simplest possibly being a circular penalty function.
When using a circular penalty function, the relative bearing to the obstacle does not matter, and the function can be defined as:
\begin{equation}\label{eq:circularPenaltyEq}
  	\text{penalty}_{i,\text{circular}}(\bar{\bs \eta}(t)) =
  	\begin{cases}
  	  	1    & \text{if } d_i < D_0 \\
  	  	1 + \frac{\gamma_1 - 1}{D_1 - D_0} (d_i - D_0)    & \text{if } D_0 \leq d_i < D_1 \\
  	  	\gamma_1 - \frac{\gamma_1}{D_2 - D_1} (d_i - D_1)    & \text{if } D_1 \leq d_i < D_2 \\
  	  	0       & \text{else},
  	\end{cases}
\end{equation}
where the parameters of $d_i(\bar{\bs \eta}(t);\bs p_i(t))$ are omitted for notational simplicity.
The variables $D_2 > D_1 > D_0 > 0$ are the margin, safety and collision region sizes, respectively, while $\gamma_1 \in (0,1)$ is a tuning parameter controlling the cost gradient inside the margin and safety regions.
\begin{figure}
  	\centering
  	\begin{subfigure}[b]{.48\textwidth}
  	  	\includegraphics[width=\textwidth]{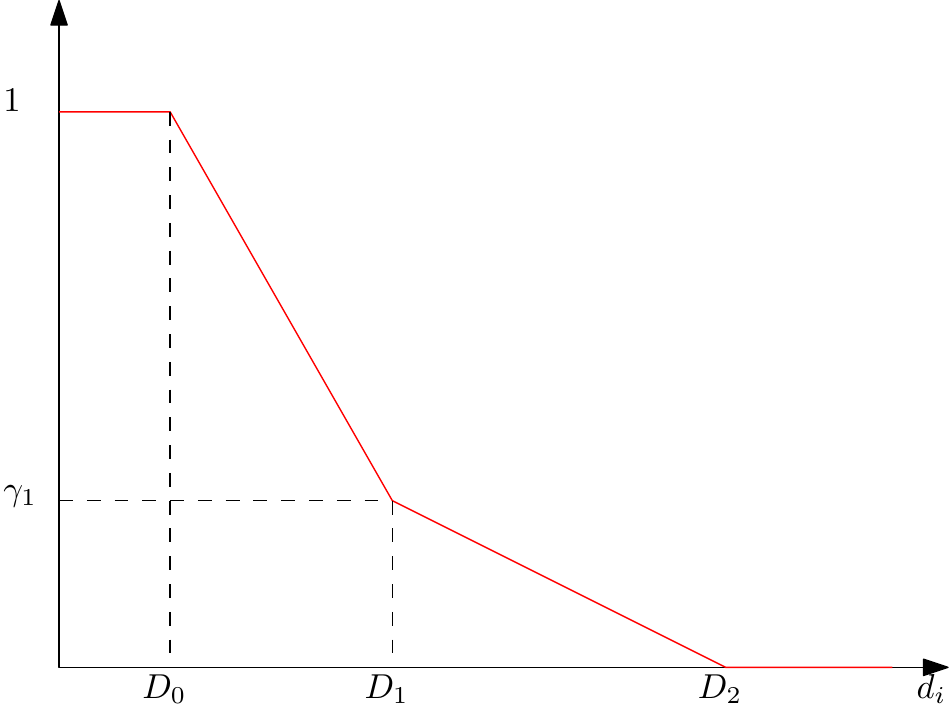}
  	  	\caption{\label{fig:circularObsCostFunc}Function value.}
  	\end{subfigure}\quad
\begin{subfigure}[b]{.48\textwidth}
  	  	\includegraphics[width=\textwidth]{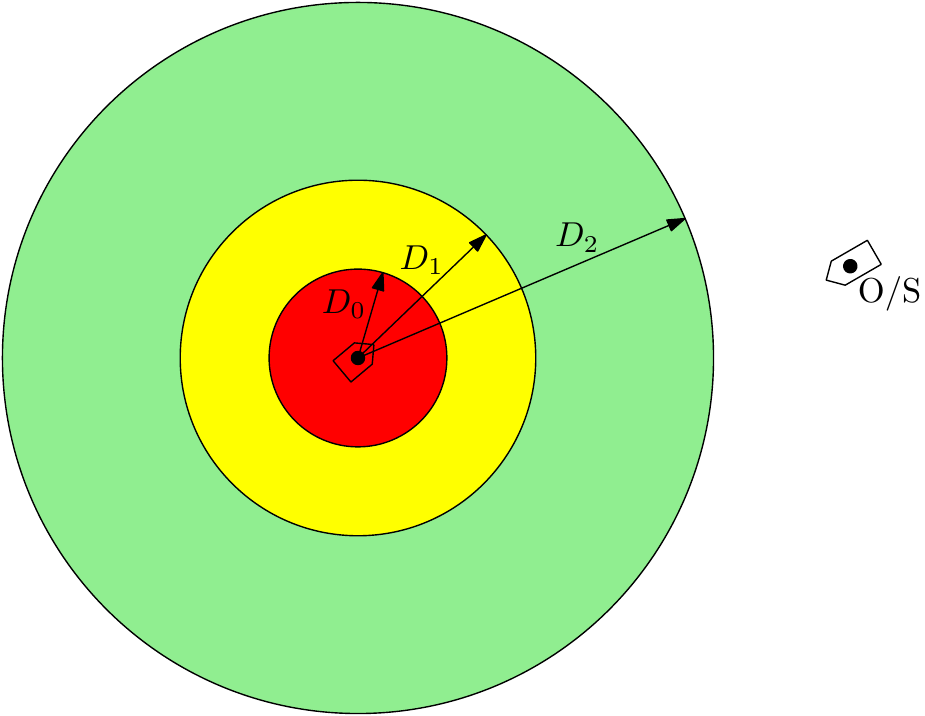}
  	  	\caption{\label{fig:circularObsCostRegions}Function regions.
The red region is the collision region, yellow is the safety region and green is the margin region, given by the radiuses $D_0$, $D_1$ and $D_2$, respectively.}
  	\end{subfigure}
  	\caption{\label{fig:CicPenalty}Circular penalty function.}
\end{figure}The circular penalty function is illustrated in \Cref{fig:CicPenalty}.

\begin{figure}
  	\centering
  	\begin{subfigure}[b]{.54\textwidth}
	  	\includegraphics[width=\textwidth]{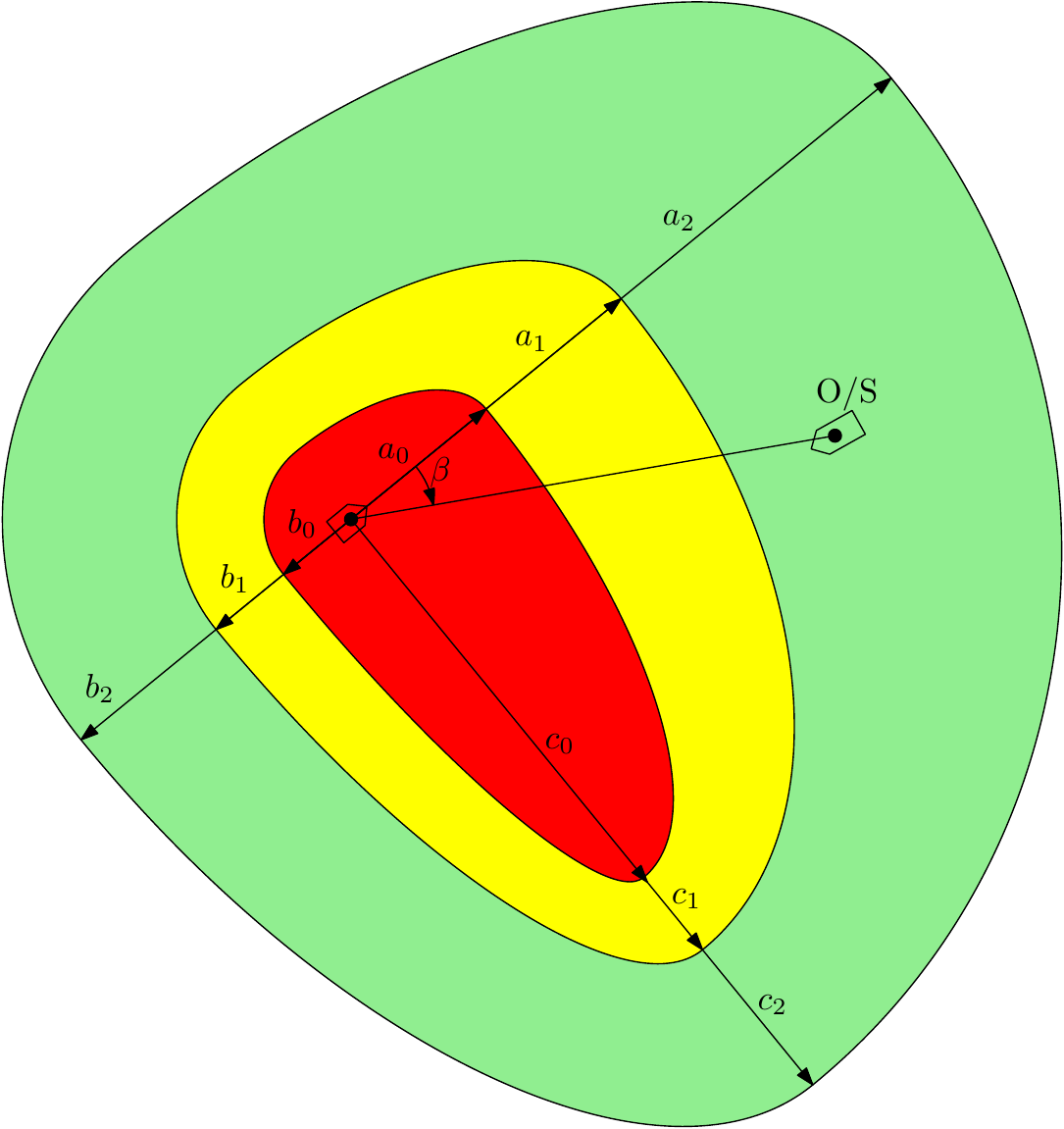}
	  	\caption{\label{fig:ellipticalColregsObsCostRegions}Function regions, each constructed by one circle and three ellipses.}
  	\end{subfigure}
  	\hfill
  	\begin{subfigure}[b]{.4\textwidth}
	  	\includegraphics[width=\textwidth]{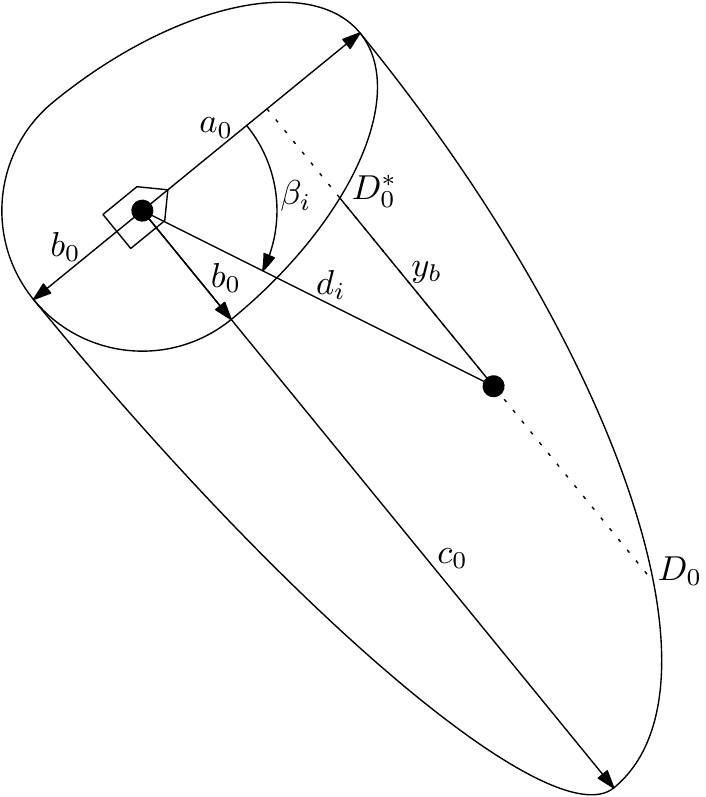}
  		\caption{\label{fig:ellipticalColregsDistFromDinner}Variables for the inner penalty function for the elliptical COLREGs penalty function given a point $(d_i, \beta_i)$.}
  	\end{subfigure}
  	\caption{\label{fig:ellipticalColregsFigure}Elliptical \gls{colregs} penalty function.}
\end{figure}

A circular penalty function is useful for static objects where there is no preference on which side of the object one should pass.
For moving vessels, it should be considered to be more dangerous to be in front of the vessel than on the side or behind it, and \gls{colregs} also introduce preferences on which side one should pass an obstacle.
For short-term \gls{colav}, it is not beneficial to constrain the algorithm to strictly follow the overtaking, head on and crossing rules (rules 13--15), since maneuvers ignoring these rules may be required to fulfill rule 17.
However, the algorithm should choose maneuvers compliant with rules 13--15 when this is possible.
We therefore motivate the algorithm to choose \gls{colregs}-aware maneuvers by defining an elliptical \gls{colregs} penalty function by letting the region sizes $D_0, D_1$ and $D_2$ be dependent on the relative bearing.
Each region is defined by a combination of three ellipses and one circle as:
\begin{equation}\label{eq:COLREGs_regions}
  	D_k(\beta_i) = \begin{cases}
     	b_k & \text{if } \beta_i < -\frac{\pi}{2} \\
     	\frac{a_k b_k}{\sqrt{(b_k \cos \beta_i )^{2}+(a_k\sin \beta_i )^{2}}} & \text{if } -\frac{\pi}{2} \leq \beta_i < 0 \\
     	\frac{a_k c_k}{\sqrt{(c_k \cos \beta_i )^{2}+(a_k\sin \beta_i )^{2}}} & \text{if } 0 \leq \beta_i < \frac{\pi}{2} \\
     	\frac{b_k c_k}{\sqrt{(c_k \cos \beta_i )^{2}+(b_k\sin \beta_i )^{2}}} & \text{if } \frac{\pi}{2} \leq \beta_i,
  	\end{cases}
\end{equation}
where $a_k$, $b_k$ and $c_k = b_k + d_{\text{COLREGs}}$ with $k\in\{0,1,2\}$ defines the major and minor ellipsis axis.
The parameter $d_{\text{COLREGs}}$ controls the region expansion of the starboard side of the obstacle.
The regions are illustrated in \Cref{fig:ellipticalColregsObsCostRegions}.

If we were to use \eqref{eq:circularPenaltyEq} with $D_k$ from \eqref{eq:COLREGs_regions} as the elliptical \gls{colregs} penalty function, the entire collision region would have a constant penalty.
This poses a potential problem since all points inside the region is considered to be equally costly.
For the circular penalty function, this region is so small that the impact is quite low.
For the elliptical \gls{colregs} penalty function, however, it is natural to have a non-constant cost inside the collision region since this is rather large.
We therefore define the elliptical \gls{colregs} penalty function as:
\begin{equation}\label{eq:ellipticalColregsPenaltyEq}
  	\text{penalty}_{i, \text{COLREGs}}(\bar{\bs \eta}(t)) = \text{inner\_penalty}_i(\bar{\bs \eta}(t)) +
  	\begin{cases}
  	  	1    & \text{if } d_i < D_0 \\
  	  	1 + \frac{\gamma_1 - 1}{D_1 - D_0} (d_i - D_0)    & \text{if } D_0 \leq d_i < D_1 \\
  	  	\gamma_1 - \frac{\gamma_1}{D_2 - D_1} (d_i - D_1)    & \text{if } D_1 \leq d_i < D_2 \\
  	  	0       & \text{else},
  	\end{cases}
\end{equation}
where $D_k$, $k\in\{0,1,2\}$ are given by \eqref{eq:COLREGs_regions} and $\text{inner\_penalty}_i(\bar{\bs \eta}(t))$ is an additional cost inside the collision region.
This additional cost is given as:
\begin{equation}
	\text{inner\_penalty}_i(\bar{\bs \eta}(t)) =
	\begin{cases}
  	  	1    & \text{if } d_i < D_{0}^* \\
  	  	1 - \frac{y_b(d_i, \beta_i)}{d_{\text{COLREGs}}}    & \text{if } D_{0}^* \leq d_i < D_0, \\
  	  	0 													& \text{else},
  	\end{cases}
\end{equation}
where $D_{0}^*$ given as:
\begin{equation}
  	D_0^*(\beta_i) = \begin{cases}
     	\frac{a_0 b_0}{\sqrt{(b_0 \cos \beta_i )^{2}+(a_0\sin \beta_i )^{2}}} & \text{if } |\beta_i| < \frac{\pi}{2} \\
     	b_0 & \text{else},
  	\end{cases}
\end{equation}
and $y_b(d_i, \beta_i)$ is the distance from the $D_{0}^*$ region to the point $(d_i, \beta_i)$ along the y-direction of the obstacle body frame, as illustrated in \Cref{fig:ellipticalColregsDistFromDinner}.

\subsubsection{Transitional cost}
An important design criteria for the algorithm is that it should be robust with respect to noise on the obstacle estimates, making it well suited for use with tracking systems based on exteroceptive sensors.
By introducing transitional cost in the objective function, a certain level of cost reduction will be required to make the algorithm change the current planned maneuver.
This should increase the robustness to noise on the obstacle estimates, while also making the algorithm less affected by noise in the vessel state estimates and external disturbances, for instance wave induced motion.

Denoting the desired velocity trajectory from the previous iteration as $\bs u_d^-(t)$, which is currently being tracked by the vessel controllers, the transitional cost is computed as:
\begin{equation}
  \text{tran}(\bs u_d(t)) = \begin{cases}1 & \text{if } \int_{t_0}^{t_0 + T_1} \left|U_d(\gamma) - U_d^-(\gamma)\right|\mathrm{d}\gamma > e_{U,\min} \text{ or } \int_{t_0}^{t_0 + T_1} \left|\chi_d(\gamma) - \chi_d^-(\gamma)\right|\mathrm{d}\gamma > e_{\chi,\min}\\ 0 & \text{else},\end{cases}
\end{equation}
with $\bs u_d(t) = \begin{bmatrix} U_d(t) & \chi_d(t) \end{bmatrix}^T$, $\bs u_d^-(t) = \begin{bmatrix} U_d^-(t) & \chi_d^-(t) \end{bmatrix}^T$ and where $T_1$ is the step time of the first trajectory maneuver.
The variables $e_{U,\min}$ and $e_{\chi,\min}$ denote the minimum difference between the previous desired velocity trajectory and the candidates:
\begin{equation}
  \begin{aligned}
    e_{U,\min} &= \underset{\bs u_d(t) \in \mathcal{U}_d}{\text{min}} \int_{t_0}^{t_0+T_1} \left|U_d(\gamma) - U_d^-(\gamma)\right|\mathrm{d}\gamma \\
    e_{\chi,\min} &= \underset{\bs u_d(t) \in \mathcal{U}_d}{\text{min}} \int_{t_0}^{t_0 + T_1} \left|\chi_d(\gamma) - \chi_d^-(\gamma)\right|\mathrm{d}\gamma.
  \end{aligned}
\end{equation}
The transitional cost term is zero if the first maneuver of the candidate desired velocity trajectory $\bs u_d(t)$ is the one closest to the desired velocity trajectory from the previous iteration $\bs u_d^-(t)$, and one otherwise.
Notice that the transitional cost term introduces discontinuities, which we previously stated that we would like to avoid in order to improve the robustness with respect to noise on obstacle estimates.
The transitional cost term does, however, not rely on obstacle estimates, making the term insensitive to noise on the obstacle estimates and justifying the use of a discontinuous transitional cost function.


\section{Experimental results}\label{sec:expres}
Full scale experiments were conducted in the Trondheimsfjord, Norway, on the 12\textsuperscript{th} of October $2017$.
This chapter describes the experimental setup and results.

\subsection{Experimental setup}
The Telemetron \gls{asv}, briefly introduced in \cref{sec:ASVmodelingControl}, was used as the ownship.
The vessel is fitted with a SIMRAD Broadband 4G™ Radar, and a Kongsberg Seatex Seapath 330+ GNSS-aided inertial navigation system was used during the experiments.
See \cref{tab:HW} for more details on the vessel specifications.
The \gls{bc-mpc} algorithm was implemented in discrete time using the Euler method to discretize the algorithm, see \cref{tab:BCMPC_param} for the algorithm parameters.
We inputted a user-specified straight line trajectory with constant speed as the desired trajectory, and used the elliptical \gls{colregs} penalty function for obstacle avoidance.
The \gls{bc-mpc} algorithm was run at a rate of $0.2~\si{\hertz}$.
\begin{table}
\caption{\label{tab:HW}Telemetron ASV specifications.}
\begin{tabularx}{\textwidth}{lrX}
\toprule
\textbf{Component} & & \textbf{Description} \\
\midrule
Vessel hull                   &   & Polarcirkel Sport 845 \\
  \hspace{1cm}Length              & & $8.45~\si{\meter}$  \\
  \hspace{1cm}Width             & & $2.71~\si{\meter}$  \\
  \hspace{1cm}Weight              & & $1675~\si{\kilogram}$ \\
Propulsion system               &   & Yamaha $225$~HP outboard engine \\
  \hspace{1cm}Motor control           &   & Electro-mechanical actuation of throttle valve \\
  \hspace{1cm}Rudder control          &   & Hydraulic actuation of outboard engine angle with proportional-derivative (PD) feedback control \\
Navigation system                 &   & Kongsberg Seatex Seapath 330+ \\
Radar                                           & & Simrad Broadband 4G™ Radar \\
Processing platform                             & & Intel® i7 $3.4~\si{\giga\hertz}$ CPU, running Ubuntu 16.04 Linux \\
\bottomrule
\end{tabularx}
\end{table}

\begin{table}
  \centering
  \caption{\label{tab:BCMPC_param}\Gls{bc-mpc} algorithm parameters.}
  \begin{tabular}{lll}
  \toprule
    \textbf{Parameter}        & \textbf{Value}    & \textbf{Description} \\
  \midrule
    $\bs T$                   & $\begin{bmatrix} 5 & 20 & 30 \end{bmatrix}~\si{\second}$  & Prediction horizon  \\
    $\bs N_U$                 & $\begin{bmatrix} 5 & 1 & 1 \end{bmatrix}$                 & Number of \gls{sog} maneuvers  \\
    $\bs N_\chi$              & $\begin{bmatrix} 5 & 3 & 3 \end{bmatrix}$                 & Number of course maneuvers  \\
    $T_{\text{ramp}}$         & $1~\si{\second}$                                          & Ramp time  \\
    $T_U$                     & $5~\si{\second}$                                          & \Gls{sog} maneuver length  \\
    $T_\chi$                  & $5~\si{\second}$                                          & Course maneuver length  \\
    $T_{\tilde U}$            & $5~\si{\second}$                                          & \Gls{sog} error model time constant  \\
    $T_{\tilde \chi}$         & $5~\si{\second}$                                          & Course error model time constant  \\
  \cmidrule{1-3}
    $\Delta$                  & $500~\si{\meter}$                                         & \Gls{los} lookahead distance  \\
    $\gamma_s$                  & $0.005~\si{\per\second}$                                   & \Gls{los} along track distance gain  \\
  \cmidrule{1-3}
    $w_{\text{al}}$                  & $1$                                                       & Align weight  \\
    $w_{\text{av}}$                  & $6000$                                                    & Avoid weight  \\
    $w_{\text{t}}$                   & $4200$                                                    & Transitional cost weight  \\
    $w_{\chi}$                & $100$                                                     & Angular error scaling weight  \\
  \cmidrule{1-3}
    $a_0$               & $50~\si{\meter}$    & Collision region major axis \\
    $a_1$               & $150~\si{\meter}$   & Safety region major axis\\
    $a_2$               & $250~\si{\meter}$   & Margin region major axis\\
    $b_0$               & $25~\si{\meter}$    & Collision region minor axis\\
    $b_1$               & $75~\si{\meter}$    & Safety region minor axis\\
    $b_2$               & $125~\si{\meter}$   & Margin region minor axis\\
    $d_{\text{COLREGs}}$       & $100~\si{\meter}$   & \Gls{colregs} distance\\
    $\gamma_1$          & $0.1$               & Obstacle cost gradient parameter\\
  \bottomrule
  \end{tabular}
\end{table}

\begin{figure}
  \centering
  \includegraphics[width=.9\textwidth]{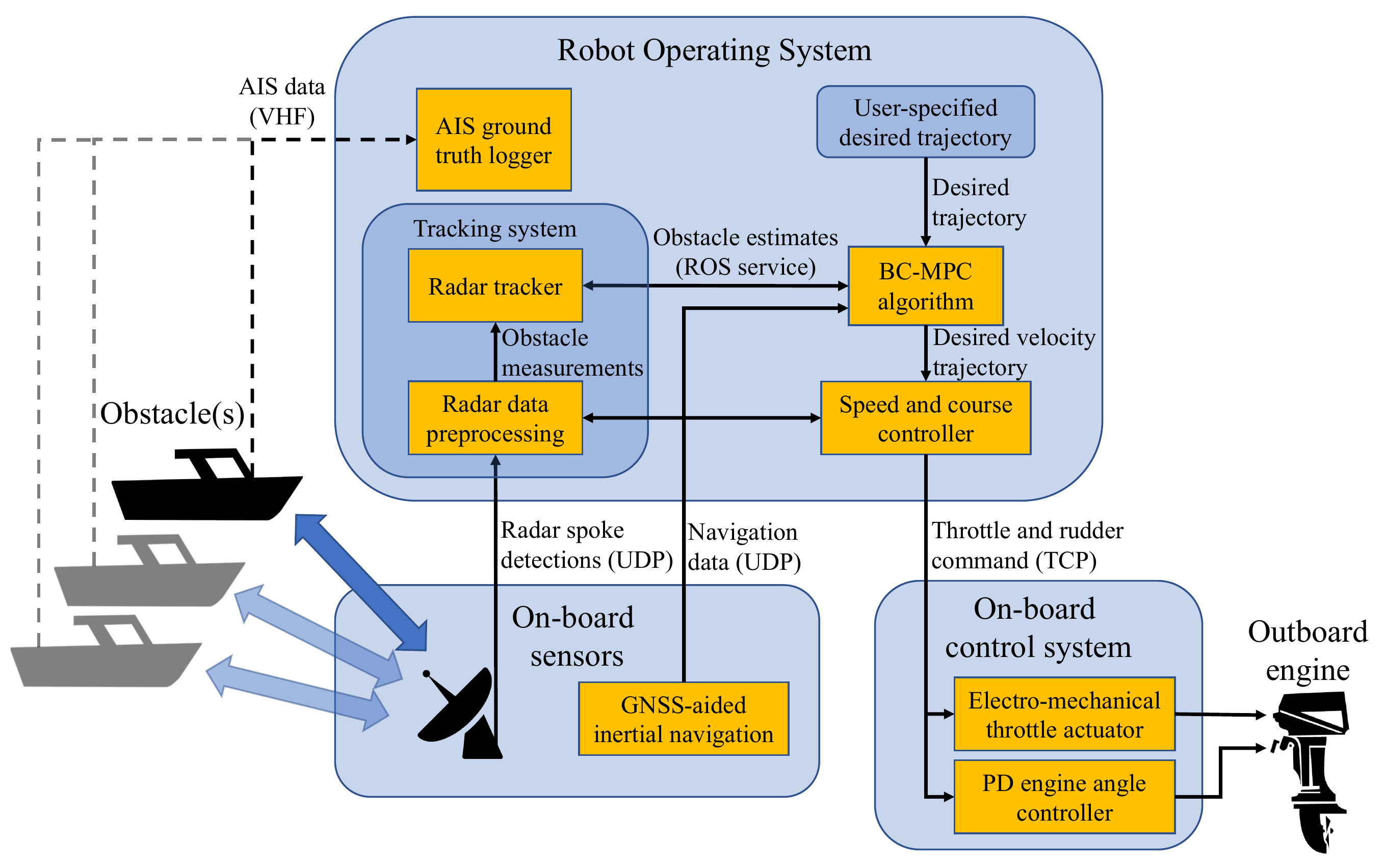}
  \caption{\label{fig:implementationArchitechture}Architecture of the COLAV implementation on the Telemetron \gls{asv}.}
\end{figure}
The implementation consists of a radar-based tracking system to provide obstacle estimates, the \gls{bc-mpc} algorithm and the model-based speed and course controller described in \cref{sec:ASVcontrol} for low-level vessel control.
The system was implemented on a processing platform with an Intel® i7 $3.4~\si{\giga\hertz}$ CPU running Ubuntu 16.04 Linux, using the \gls{ros} \cite{Quigley09}. \Cref{fig:implementationArchitechture} shows the implementation architecture.

The tracking system receives spoke detections from the radar through a UDP interface.
The detections are transformed to a local reference frame and clustered together to form one measurement per obstacle, which is a common assumption for many tracking algorithms.
The obstacle measurements are used by the radar tracker, which is based on a \gls{pdaf}.
See \cite{Wilthil2017} for more details on the tracking system.

The \gls{bc-mpc} algorithm interfaces the tracking system using a \gls{ros} service, which enables request-response functionality for providing obstacle estimates.
The \gls{bc-mpc} outputs a desired velocity trajectory to the model-based speed and course controller, which specifies a throttle and rudder command to the on-board control system through a TCP interface.
The on-board control system has an electro-mechanical actuator for controlling the motor throttle, while the rudder command is handled by steering the outboard engine angle to the desired angle using a PD controller and a hydraulic actuation system.

\begin{figure}
  \centering
  \begin{subfigure}[t]{.49\textwidth}
    \includegraphics[width=\textwidth]{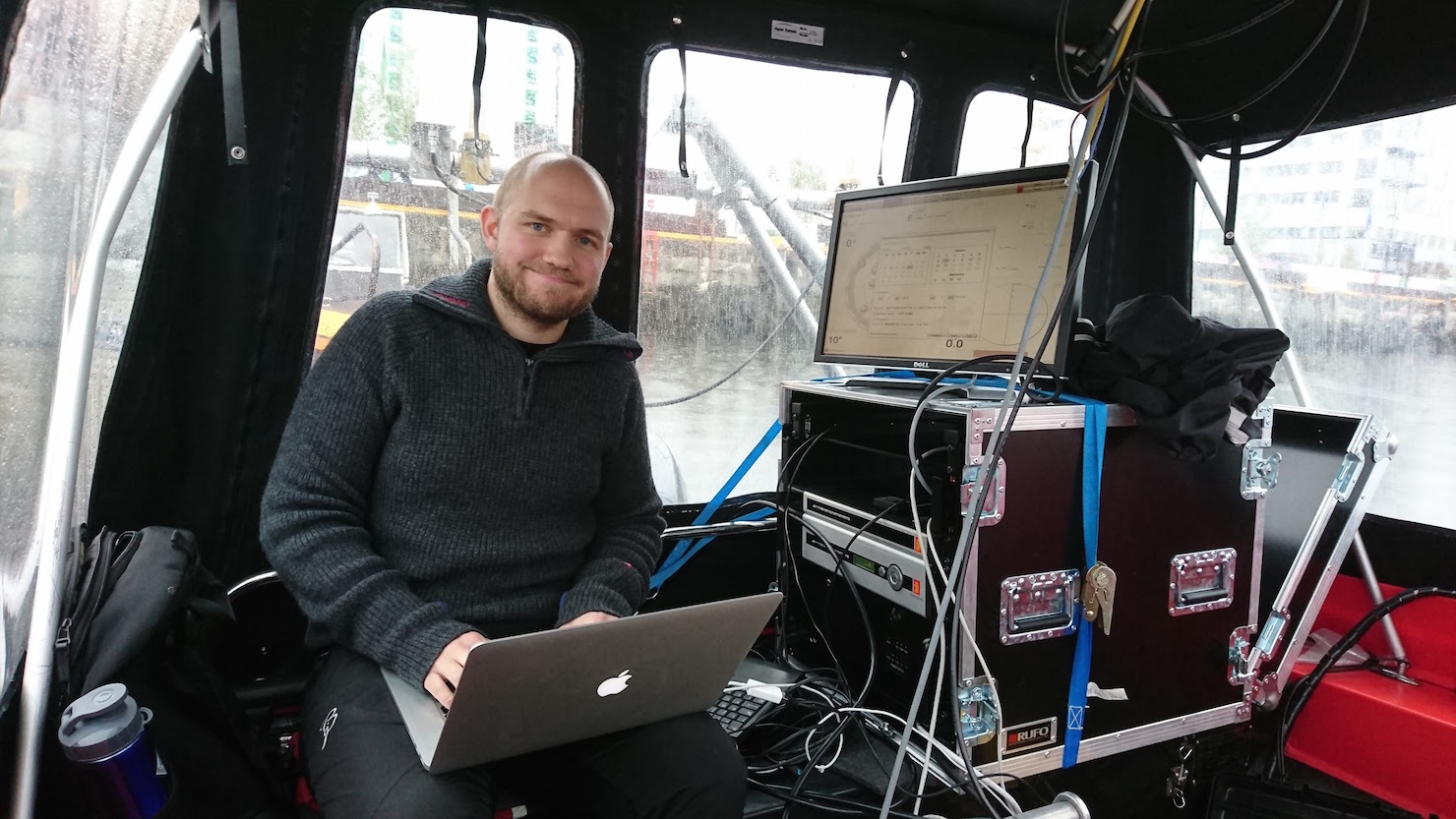}
    \caption{\label{fig:Erik_rack}Erik Wilthil in the back of the Telemetron \gls{asv} with the navigation system and the processing platform in the rack to the right.}
  \end{subfigure}\hfill
  \begin{subfigure}[t]{.49\textwidth}
    \includegraphics[width=\textwidth]{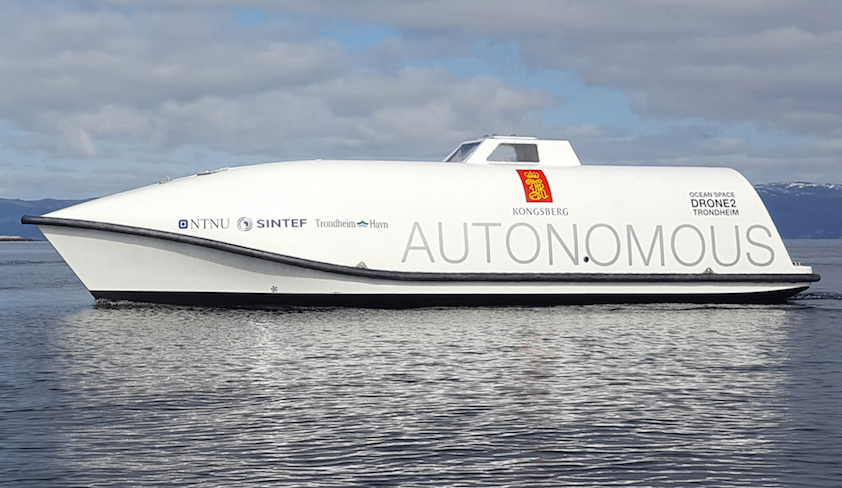}
    \caption{\label{fig:KSXOSD1}The Kongsberg Seatex Ocean Space Drone 2, which is identical to the \acrlong{osd1}.
Courtesy of Kongsberg Seatex.}
  \end{subfigure}
  \caption{\label{fig:WilthilAndOSD1}The inside of the Telemetron \gls{asv} (a) and the Kongsberg Seatex Ocean Space Drone 2 (b).}
\end{figure}

The system receives \gls{ais} messages over VHF to obtain ground-truth trajectories for the vessels involved in the experiments.
Notice that these are subject to the uncertainty of the navigation system providing the \gls{ais} data on the given vessels.
They are, however, expected to be much more precise than the estimates from the radar-based tracking system. \Cref{fig:Erik_rack} shows the inside of the Telemetron \gls{asv}, with the navigation system and processing platform.

The Kongsberg Seatex \gls{osd1} was used as the obstacle.
This was originally an offshore lifeboat, which has been fitted with a full control and navigation system for testing autonomous control systems, shown in \cref{fig:KSXOSD1}.
The vessel is $12~\si{\meter}$ long, and has a mass of approximately $10$ metric tons.

During the experiments, the \gls{osd1} was steered on constant course with a speed of approximately $2.5~\si{\meter\per\second}$ ($5$ knots) using an autopilot.
In addition to the \gls{osd1}, several commercial and leisure crafts were present in the area, affecting some of the scenarios.

We included four different scenarios in the experiments:
\begin{enumerate}
  \item Head on.
The ownship and \gls{osd1} approaches each other on reciprocal courses.
With respect to \gls{colregs}, both vessels are required to perform starboard maneuvers.
  \item Crossing from starboard.
The \gls{osd1} approaches from $90\si{\degree}$ on the ownship's starboard side.
In this case, \gls{colregs} requires the ownship to avoid collision, preferably by making a starboard maneuver and passing behind the \gls{osd1}.
  \item Overtaking.
The ownship approaches the \gls{osd1} from behind with a higher speed. \Gls{colregs} requires the ownship to avoid collision by passing on either side.
We prefer, however, to pass the \gls{osd1} on its port side by doing a port maneuver.
  \item Crossing from port.
Similar scenario as crossing from starboard, but here the \gls{osd1} approaches the ownship from the port side.
In this case, \gls{colregs} deems the ownship as the stand-on vessel, and the \gls{osd1} is supposed to avoid collision.
The \gls{osd1} will, however, keep its speed and course, requiring action with respect to rule $17$ where the give-way vessel fails to avoid collision.
The ownship must then avoid collision, preferably avoiding maneuvering to port.
\end{enumerate}
In the following sections, we present three head-on scenarios, two crossing from starboard scenarios, one overtaking scenario and one crossing from port scenario.

\subsection{Head on: Experiments 1.1--1.3}
The first experiments we performed were a number of head-on scenarios.
In these scenarios, the desired trajectory inputted to the \gls{bc-mpc} algorithm is a straight-line trajectory approaching the \gls{osd1} on a reciprocal course, resulting in a collision with a relative bearing of $0\si{\degree}$ if the desired trajectory is followed.
With respect to \gls{colregs}, both vessels should perform starboard maneuvers.
However, in our case, the \gls{osd1} violates \gls{colregs} by keeping its speed and course constant throughout the scenario.

To verify that the \gls{bc-mpc} algorithm worked as it was supposed to, we first used \gls{ais} for providing obstacle estimates in Experiment 1.1.
The \gls{osd1} is equipped with an \gls{ais} transceiver providing low-noise estimates of the position, speed and course, originating from a Kongsberg Seatex SeaNav 300 navigation system.
\begin{figure}
    \centering
    \includegraphics[width=.6\textwidth]{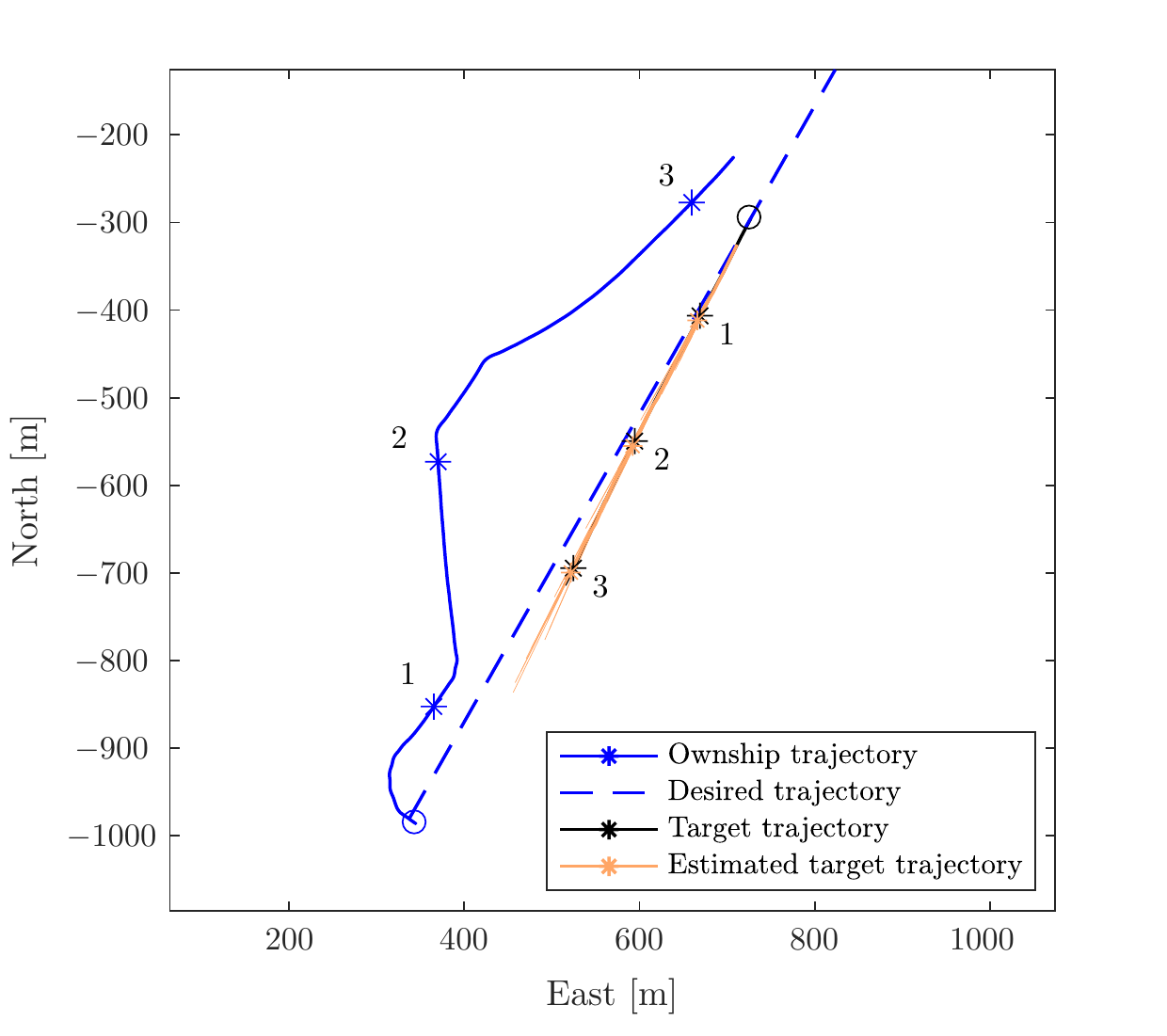}
    \caption{\label{fig:Head_on_AIS}Experiment 1.1: Head-on scenario using AIS for providing obstacle estimates.
The ownship and obstacle initial positions are marked with circles, the estimated obstacle trajectory is shown with the thick orange line, while predicted future trajectory for the obstacle at each timestep are shown as the thin orange lines.
The numbers represent time markers for each $60~\si{\second}$.
The obstacle trajectory in black is located behind the estimated obstacle trajectory in orange, since they both originate from the same AIS data in this experiment.}
\end{figure}
As shown in \cref{fig:Head_on_AIS}, we successfully avoid collision by performing a port maneuver in this scenario.
This violates the preferred \gls{colregs} action to turn to starboard in a head-on situation, and is most likely caused by the ownship approaching the obstacle on the port side of the desired trajectory, which together with the slightly angled obstacle trajectory makes a port maneuver attractive.
We do, however, pass the obstacle with a large clearance, which demonstrates the \gls{colregs} awareness of the \gls{bc-mpc} algorithm.
Moreover, the maneuver is smooth with a sufficient course change to be readily observable for other vessels.
\begin{figure}
    \centering
    \begin{subfigure}[b]{.49\textwidth}
      \includegraphics[width=\textwidth]{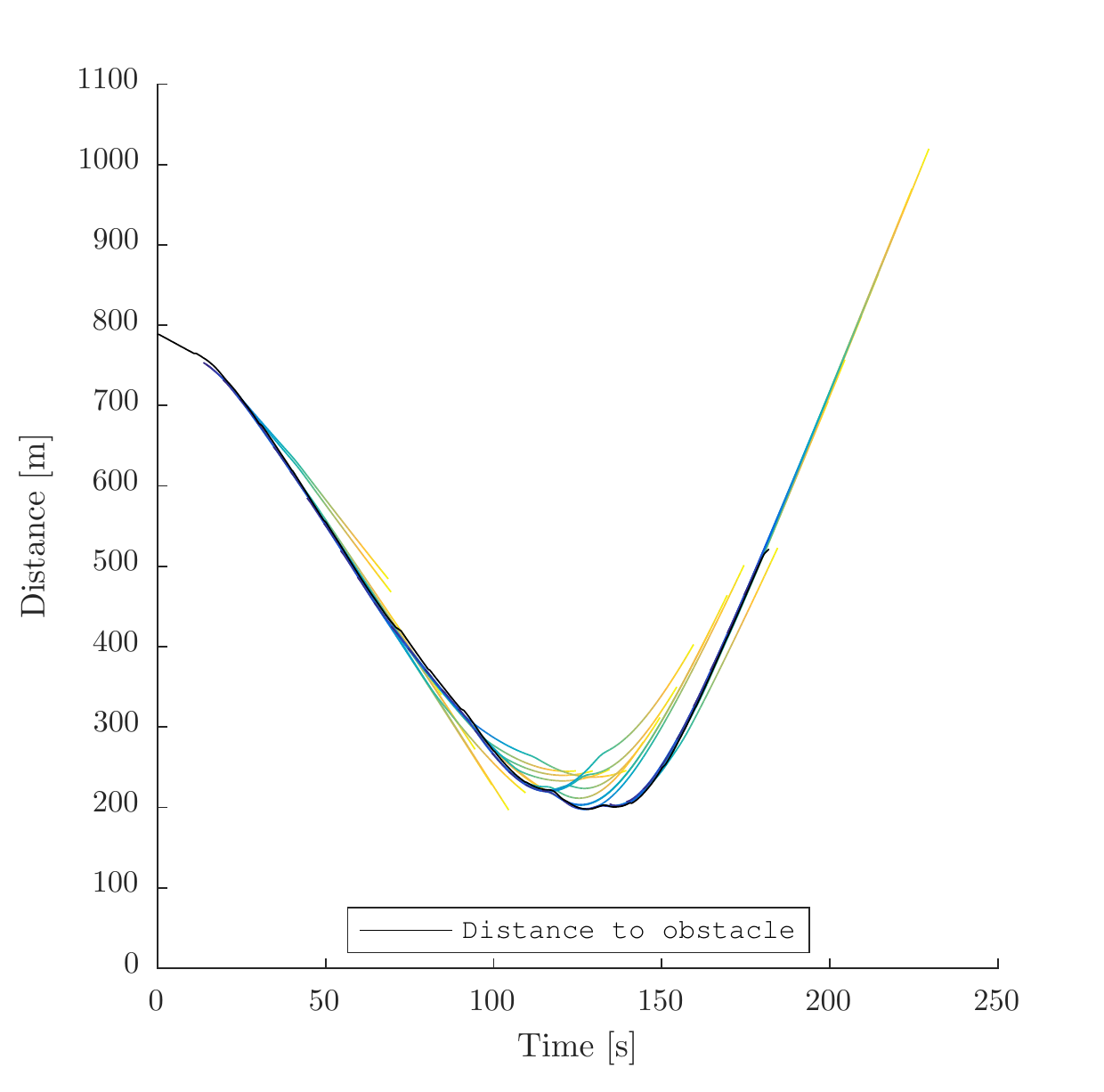}
      \caption{\label{fig:Head_on_AIS_estimates_dist}The black line shows the actual distance between the vessels, while the colored lines show the predicted future distance at each \gls{bc-mpc} iteration.
Blue represents the start of the predictions while yellow represents the end.}
    \end{subfigure}\hfill
    \begin{subfigure}[b]{.49\textwidth}
      \includegraphics[width=\textwidth]{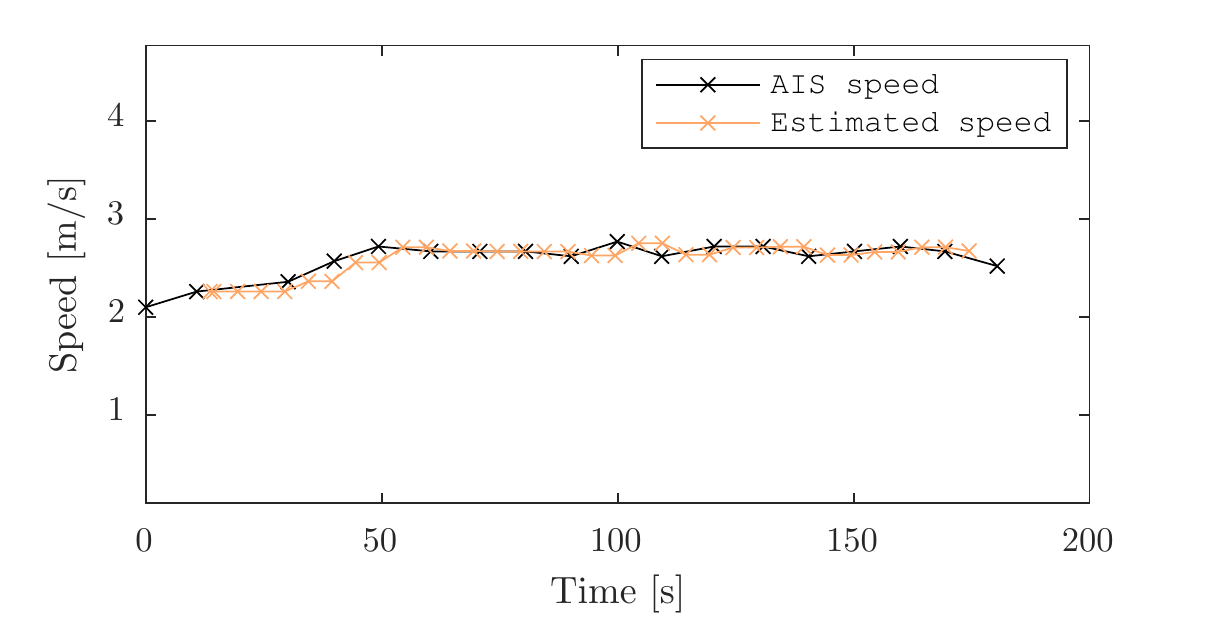} \\
      \includegraphics[width=\textwidth]{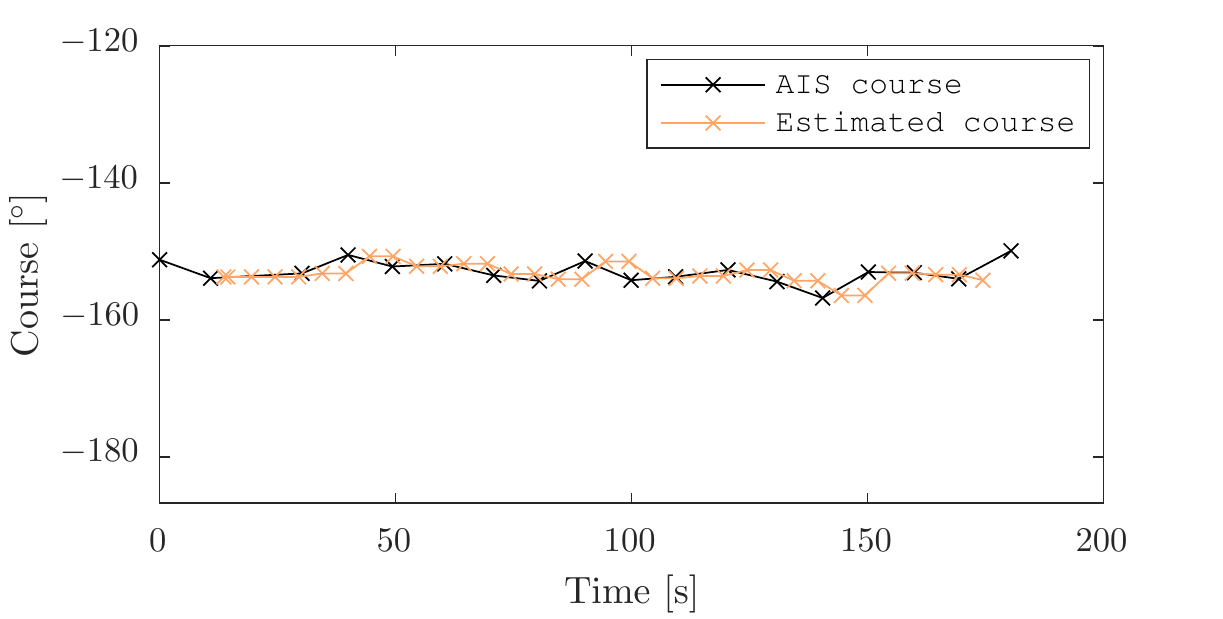}
      \caption{\label{fig:Head_on_AIS_estimates_est}Speed and course estimates.
The black crosses represent received \gls{ais} messages, while the orange crosses represent \gls{bc-mpc} iterations.}
    \end{subfigure}
    \caption{\label{fig:Head_on_AIS_estimates}Distance to the \gls{osd1} (a) and the estimated speed and course (b) during Experiment 1.1.}
\end{figure}
\Cref{fig:Head_on_AIS_estimates_dist} shows the distance between the \gls{osd1} and the ownship, and the predicted future distance given the trajectory the \gls{bc-mpc} algorithm chose at each iteration, while \Cref{fig:Head_on_AIS_estimates_est} shows the estimated and actual speed and course of the \gls{osd1}.
The estimated values are in this case based on \gls{ais}, and hence equal to the ground truth.
The \gls{osd1} does, however, transmit \gls{ais} messages quite seldom, introducing some delay in the estimated speed and course.

\begin{figure}
    \centering
    \begin{subfigure}[b]{.49\textwidth}
      \includegraphics[width=\textwidth]{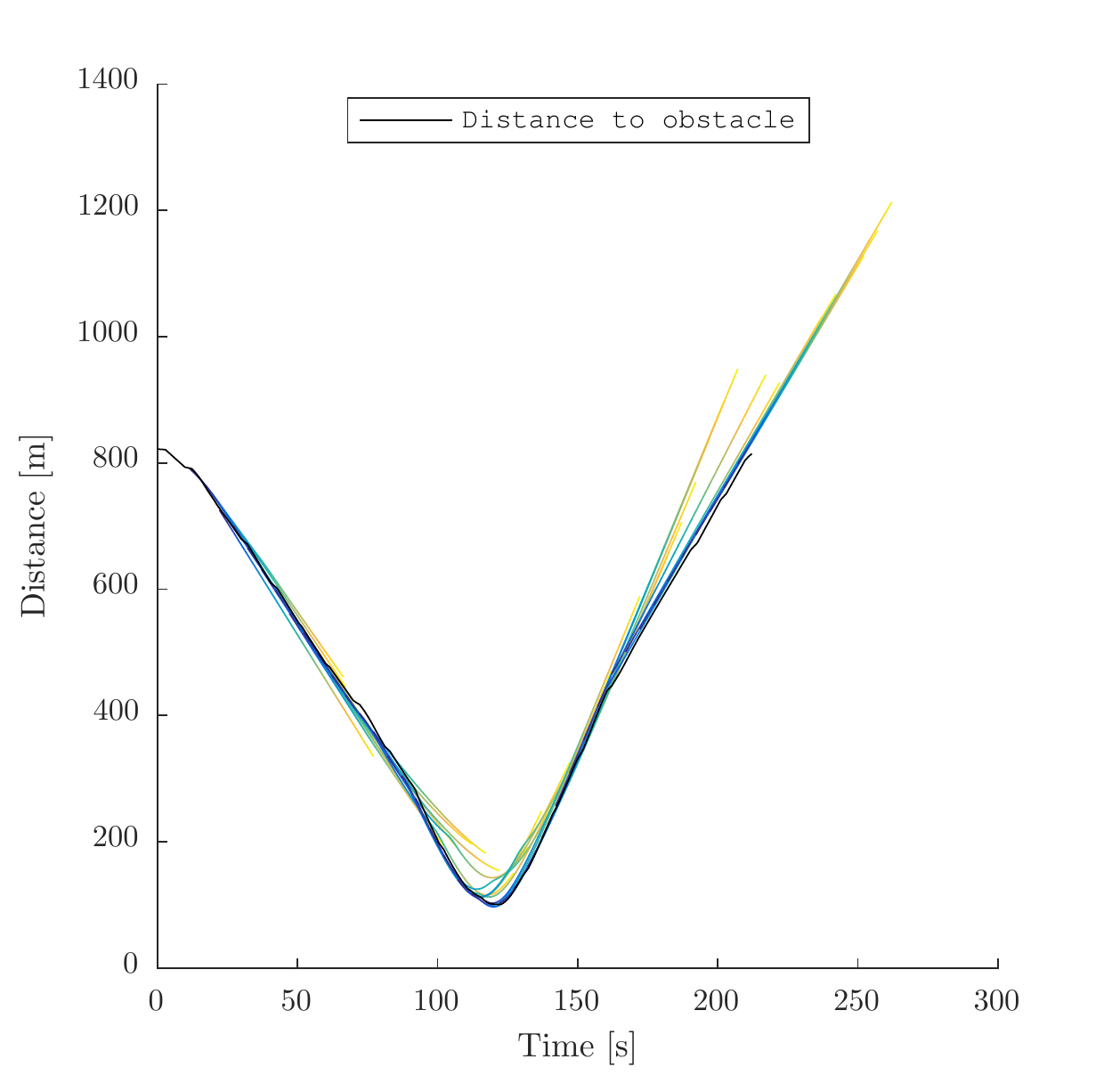}
      \caption{\label{fig:Head_on_radar_1_estimates_dist}The black line shows the actual distance between the vessels, while the colored lines show the predicted future distance at each \gls{bc-mpc} iteration.
Blue represents the start of the predictions while yellow represents the end.}
    \end{subfigure}\hfill
    \begin{subfigure}[b]{.49\textwidth}
      \includegraphics[width=\textwidth]{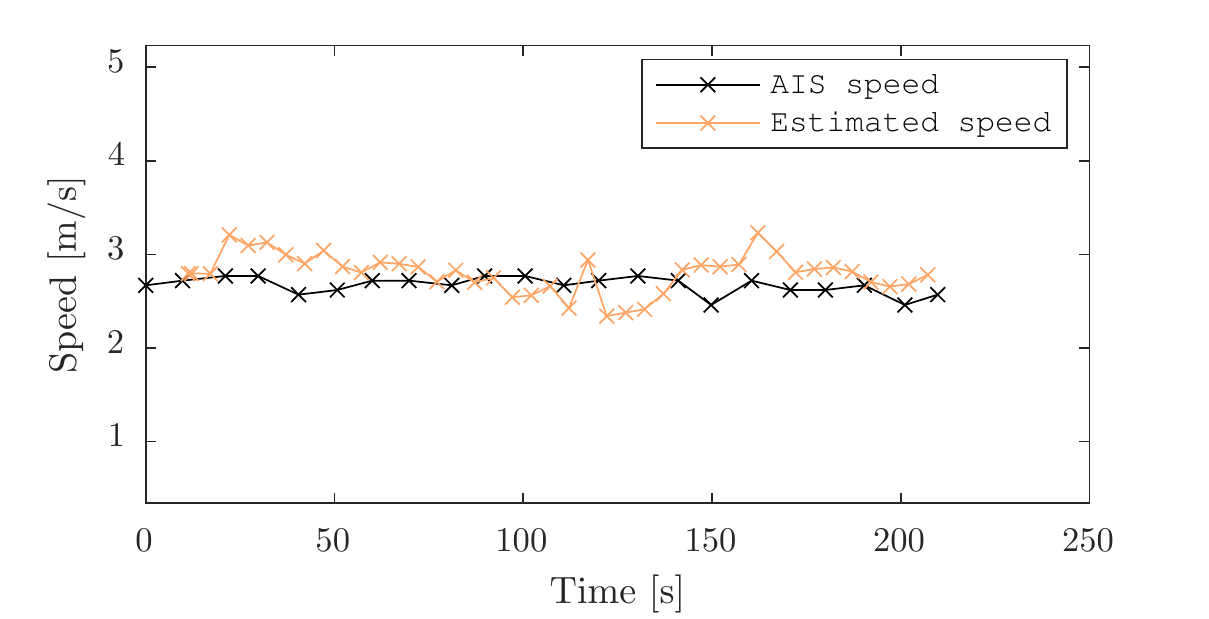} \\
      \includegraphics[width=\textwidth]{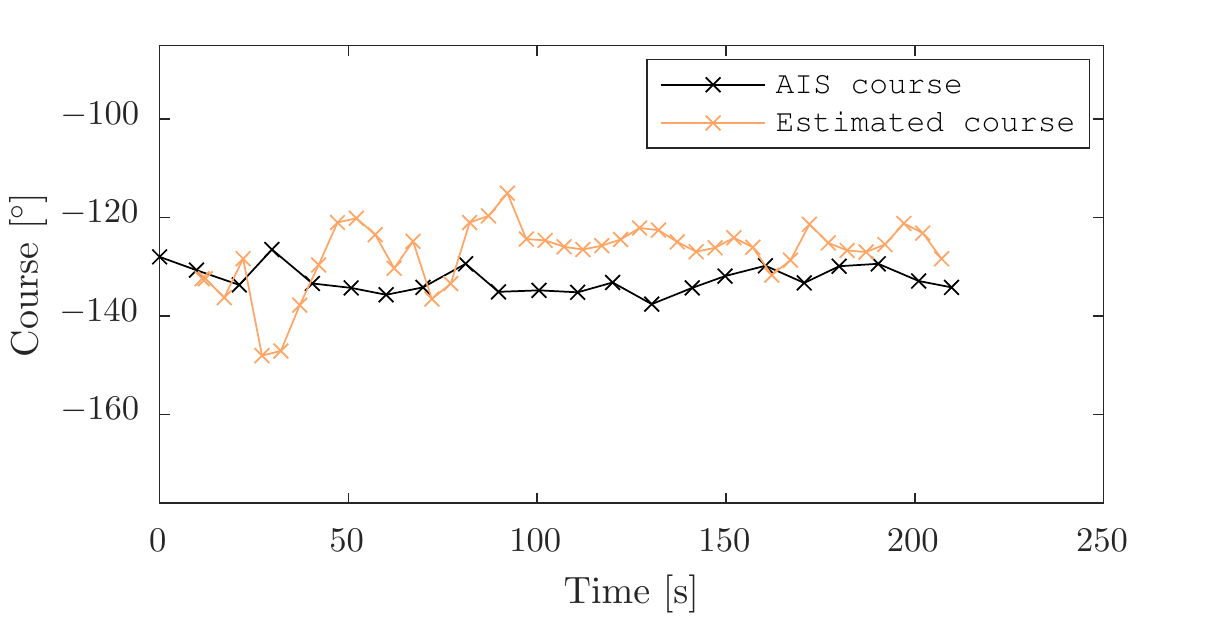}
      \caption{\label{fig:Head_on_radar_1_estimates_est}Speed and course estimates.
The black crosses represent received \gls{ais} messages, while the orange crosses represent \gls{bc-mpc} iterations.}
    \end{subfigure}
    \caption{\label{fig:Head_on_radar_1_estimates}Distance to the \gls{osd1} (a), and the estimated speed and course (b) during Experiment 1.2.}
\end{figure}

Following this experiment, we performed several experiments using the radar-based tracking system for providing obstacle estimates.
\begin{figure}
    \centering
    \includegraphics[width=.6\textwidth]{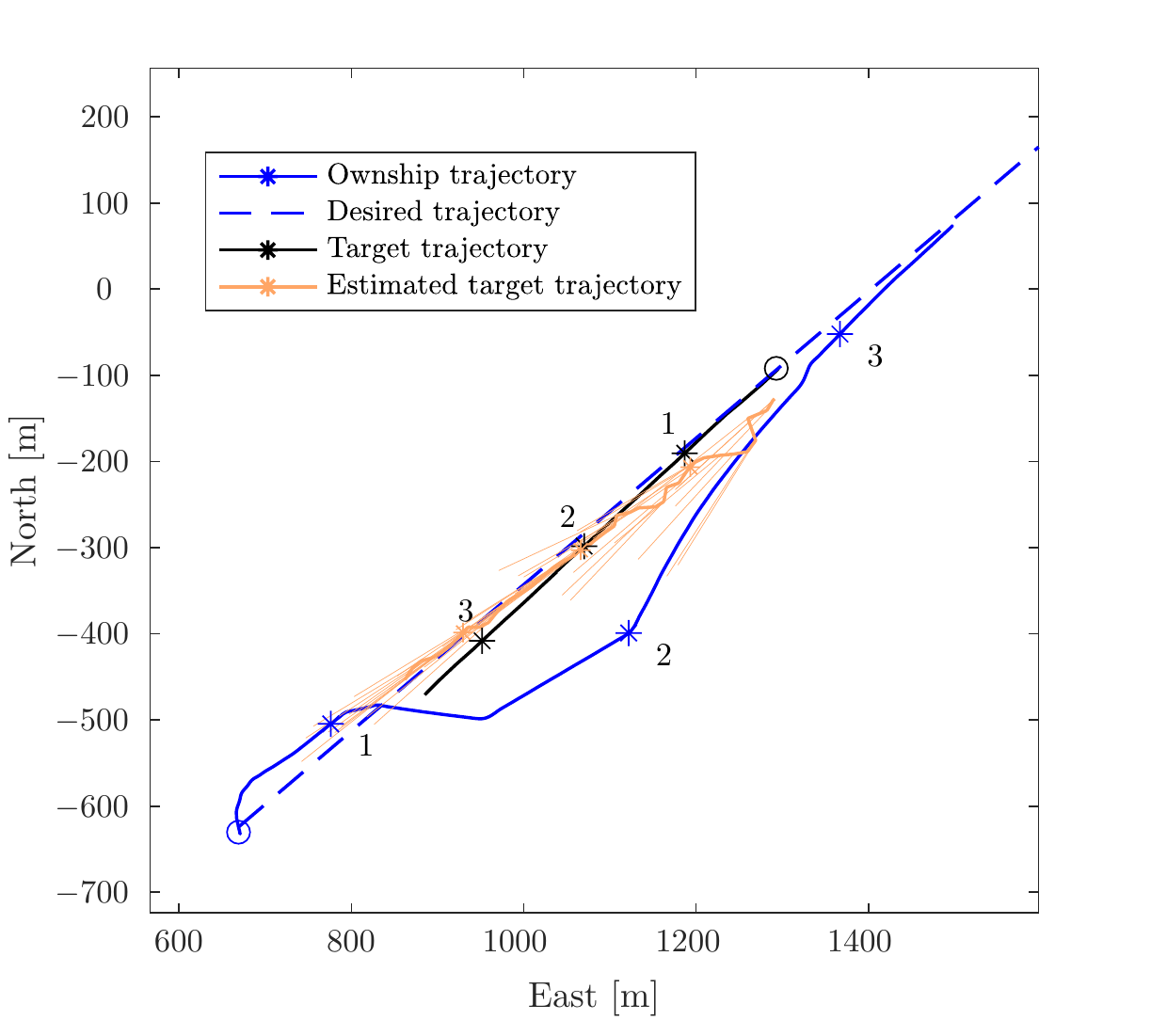}
    \caption{\label{fig:Head_on_radar_1}Experiment 1.2: Head-on scenario using the radar-based tracking system for providing obstacle estimates.
The ownship and obstacle initial positions are marked with circles, the estimated obstacle trajectory is shown with the thick orange line, while predicted future trajectory for the obstacle at each timestep are shown as the thin orange lines.
The numbers represent time markers for each $60~\si{\second}$.}
\end{figure}
\Cref{fig:Head_on_radar_1} shows the results from Experiment 1.2, a similar experiment as the one performed with \gls{ais}.
In this experiment, the ownship performs a starboard maneuver in order to avoid collision, as preferred by \gls{colregs}.
As shown in the figure, there is a fair amount of noise on the obstacle estimates, in particularly the course estimate.
This is confirmed by the course estimate shown in \cref{fig:Head_on_radar_1_estimates_est}, which shows course fluctuations often in excess of $20\si{\degree}$.
Despite this, the ownship performs a smooth maneuver, which demonstrates the \gls{bc-mpc} algorithm's robustness with respect to noise on the obstacle estimates.
This is also shown in \cref{fig:Head_on_radar_1_estimates_dist}, where the predicted distance to the obstacle varies quite much without making the algorithm decide on a new maneuver.

\begin{figure}
    \centering
    \includegraphics[width=.6\textwidth]{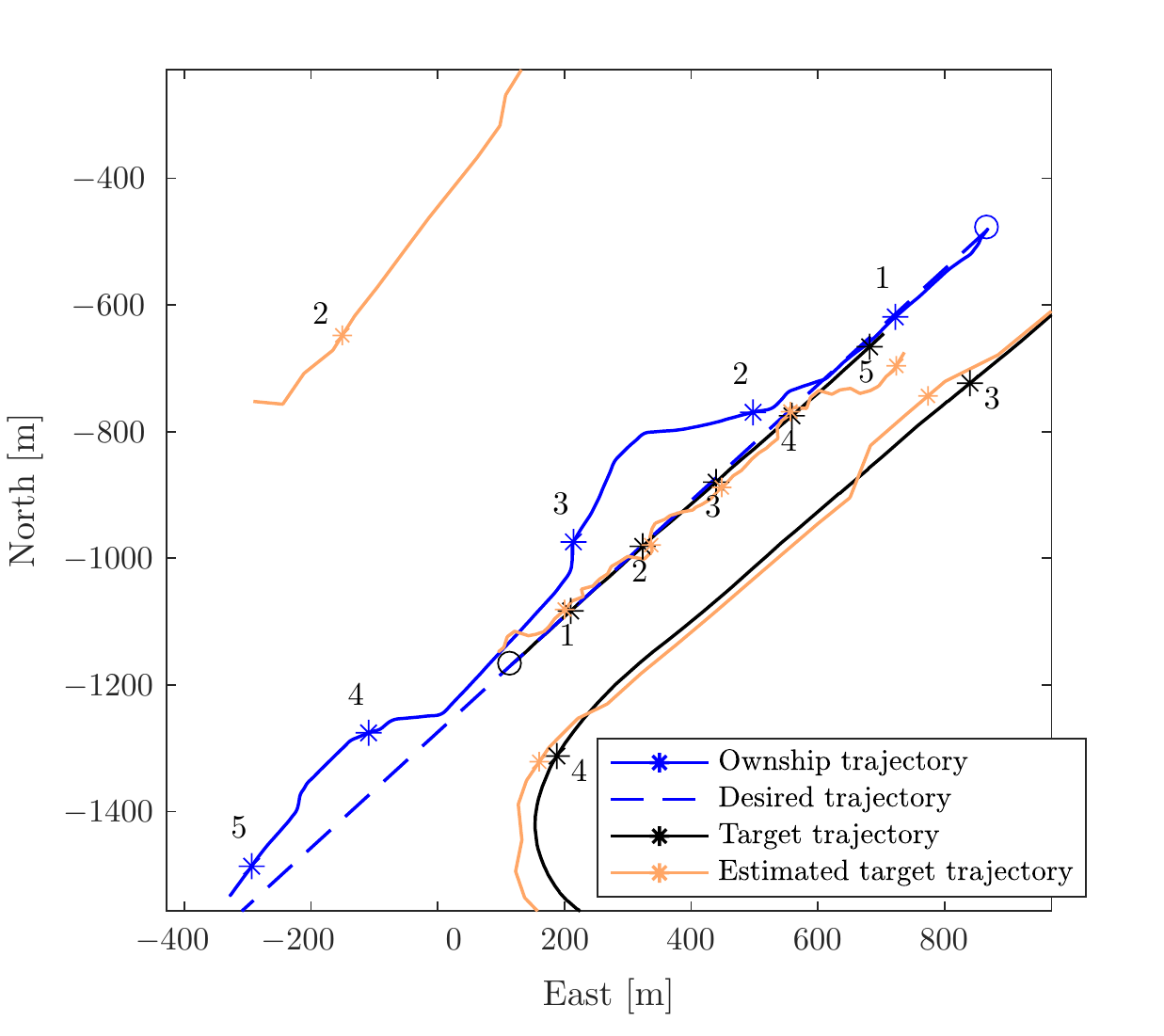}
    \caption{\label{fig:Head_on_radar_2}Experiment 1.3: Head-on scenario using the radar-based tracking system for providing obstacle estimates.
The ownship and obstacle initial positions are marked with circles, and the estimated obstacle trajectory is shown with the thick orange line.
The numbers represent time markers for each $60~\si{\second}$.
In this experiment, two vessels unexpectedly entered the scenario bringing the total vessels included up to three.
The leisure craft in the upper-left corner was traveling towards North-East and did not have \gls{ais}, so there is no ground truth trajectory for this vessel.}
\end{figure}
\begin{figure}
  \centering
  \includegraphics[width=.9\textwidth]{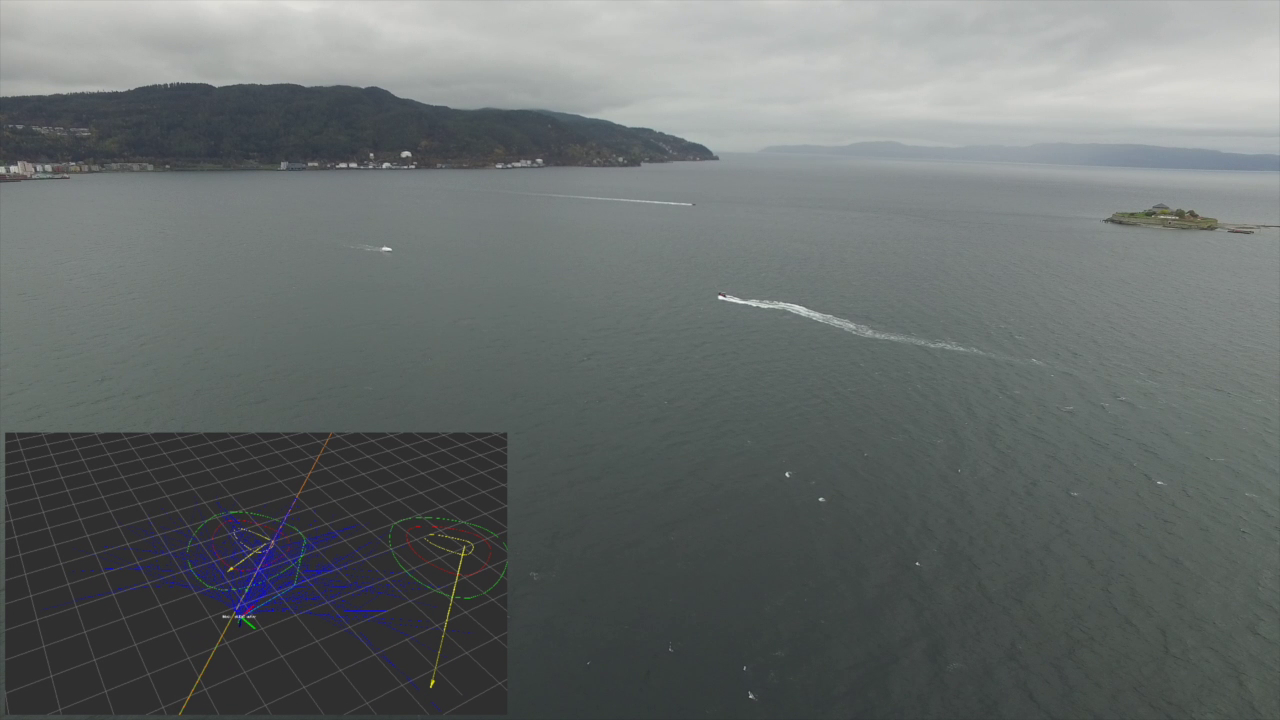}
  \caption{\label{fig:dronePictureWithROS}Drone picture during Experiment 1.3, approximately at the second time mark.
The ownship is located in the middle of the picture, with the \gls{osd1} to the left.
The vessel in the background is a high-speed leisure craft.
Yellow arrows in the visualization represent the estimated obstacle speed and course, while the orange line is the desired trajectory.
The blue lines are the feedback-corrected \gls{bc-mpc} pose trajectories, while the green line is the selected trajectory.
Notice that the estimated course of the \gls{osd1} deviates quite much from the actual vessel course, which was aligned by the orange line.}
\end{figure}
The last head-on scenario, Experiment 1.3, is shown in \cref{fig:Head_on_radar_2}, where we approach the \gls{osd1} from north-east.
The predicted future obstacle trajectories at each iteration are omitted from the following figures to improve the readability.
This scenario was slightly more complex, as two other vessels unexpectedly entered the scenario.
One of these was a high-speed leisure craft approaching from the west, while the other was a high-speed passenger ferry approaching from south-east, behind the ownship.
The leisure craft did not have \gls{ais}, and we do therefore not have a ground-truth trajectory for this vessel. \Cref{fig:dronePictureWithROS} shows an image captured by a drone during this experiment, with algorithm visualization embedded in the lower left corner.
As in the previous scenario, we avoid the \gls{osd1} by doing a starboard maneuver.
Following this, we approach the desired trajectory before the passenger ferry approaches from abaft.
With respect to \gls{colregs}, this is an overtaking situation where we are deemed the stand-on vessel, and the passenger ferry ``Trondheimfjord II'' is supposed to give way to us.
However, as mentioned earlier, the  algorithm is designed to also handle the situations where the give-way vessel does not adhere to its obligations, requiring action by the stand-on vessel.
Hence, the algorithm chooses to do a new starboard maneuver to let the passenger ferry pass.
Eventually, the ferry turns towards the Trondheim Harbor allowing the ownship to approach the desired trajectory once again.
There is some wobbling in the ownship trajectory which is most likely caused by obstacle estimate noise.
This could possibly be avoided by changing the tuning parameters of the \gls{bc-mpc} algorithm, namely increasing the transitional cost weight.
\begin{figure}
    \centering
    \begin{subfigure}[b]{.49\textwidth}
      \includegraphics[width=\textwidth]{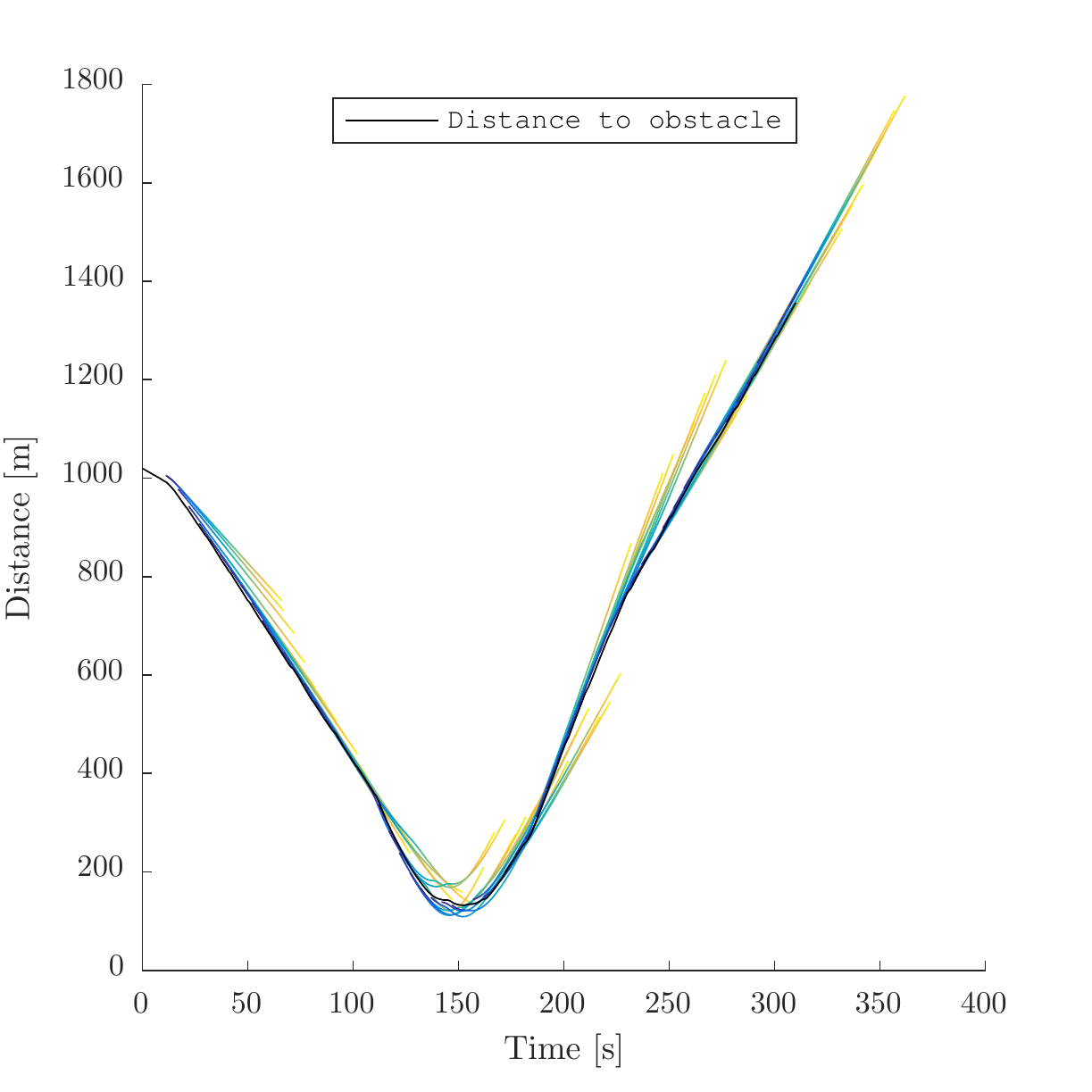}
      \caption{\label{fig:Head_on_radar_2_estimates_dist}The black line shows the actual distance between the vessels, while the colored lines show the predicted future distance at each \gls{bc-mpc} iteration.
Blue represents the start of the predictions while yellow represents the end.}
    \end{subfigure}\hfill
    \begin{subfigure}[b]{.49\textwidth}
      \includegraphics[width=\textwidth]{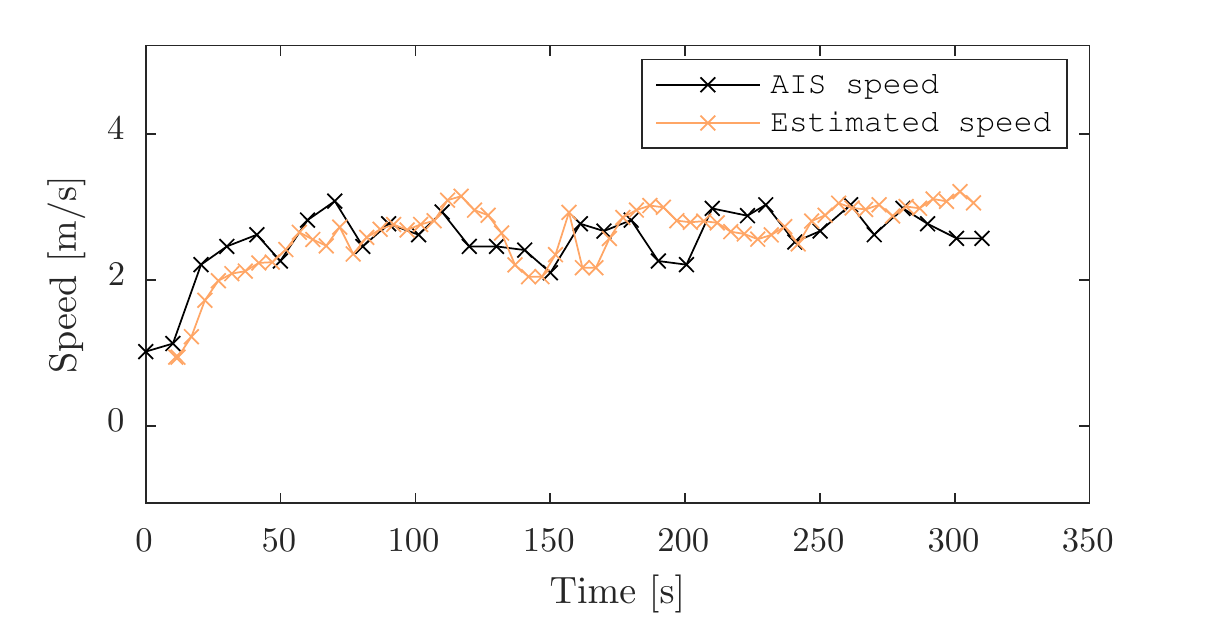} \\
      \includegraphics[width=\textwidth]{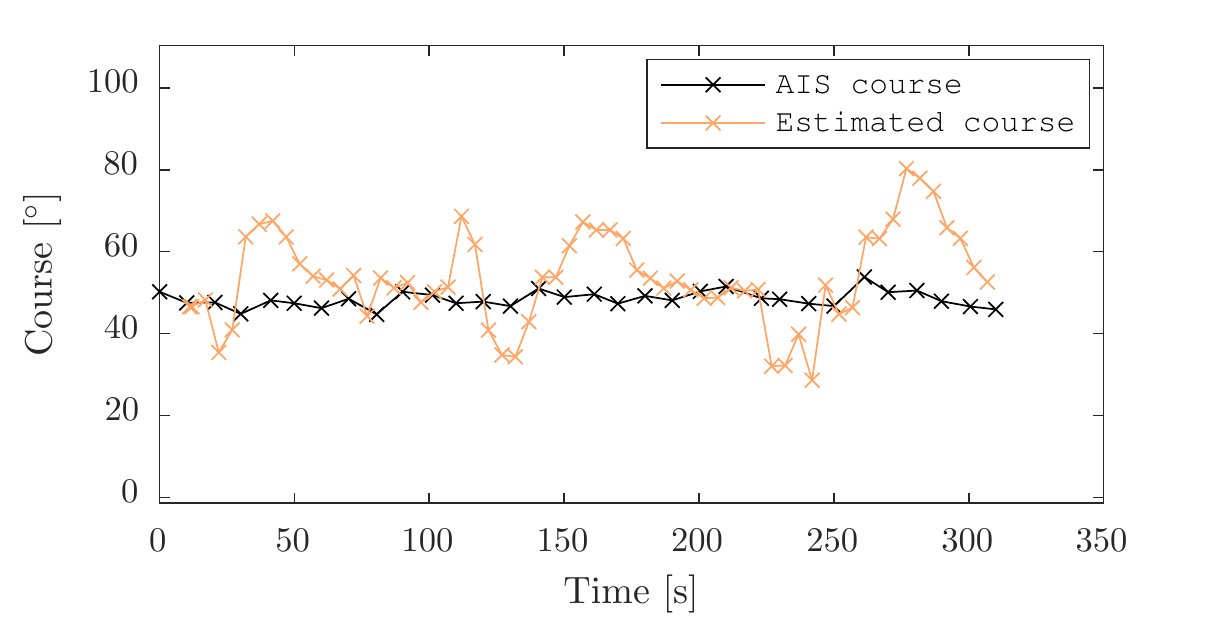}
      \caption{\label{fig:Head_on_radar_2_estimates_est}Speed and course estimates.
The black crosses represent received \gls{ais} messages, while the orange crosses represent \gls{bc-mpc} iterations.}
    \end{subfigure}
    \caption{\label{fig:Head_on_radar_2_estimates}Distance to the \gls{osd1} (a), and and the estimated speed and course (b) during Experiment 1.3.}
\end{figure}
\begin{figure}
    \centering
    \begin{subfigure}[b]{.49\textwidth}
      \includegraphics[width=\textwidth]{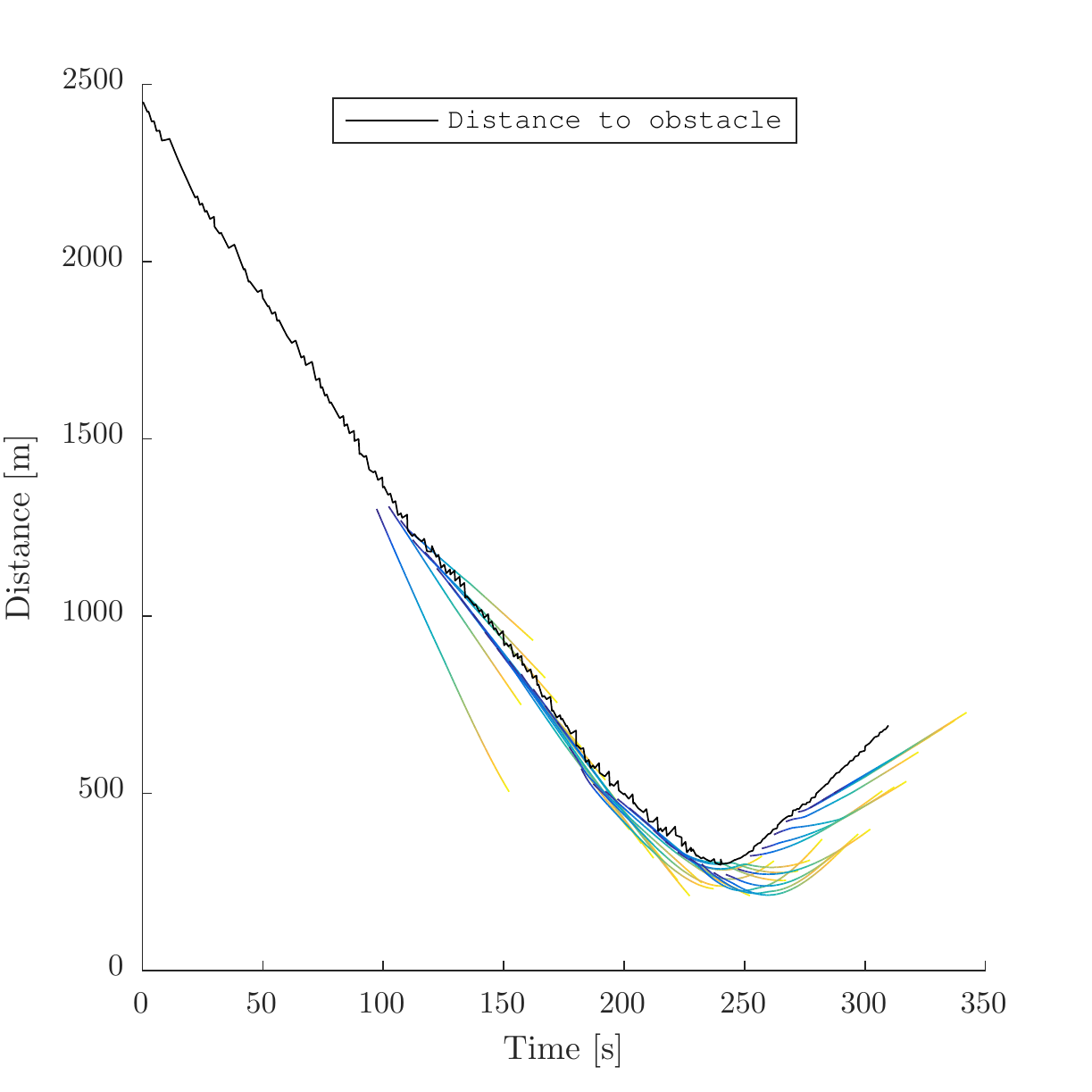}
      \caption{\label{fig:Head_on_radar_2_estimates_dist_TRONDHEIMSFJORD}The black line shows the actual distance between the vessels, while the colored lines show the predicted future distance at each \gls{bc-mpc} iteration.
Blue represents the start of the predictions while yellow represents the end.}
    \end{subfigure}\hfill
    \begin{subfigure}[b]{.49\textwidth}
      \includegraphics[width=\textwidth]{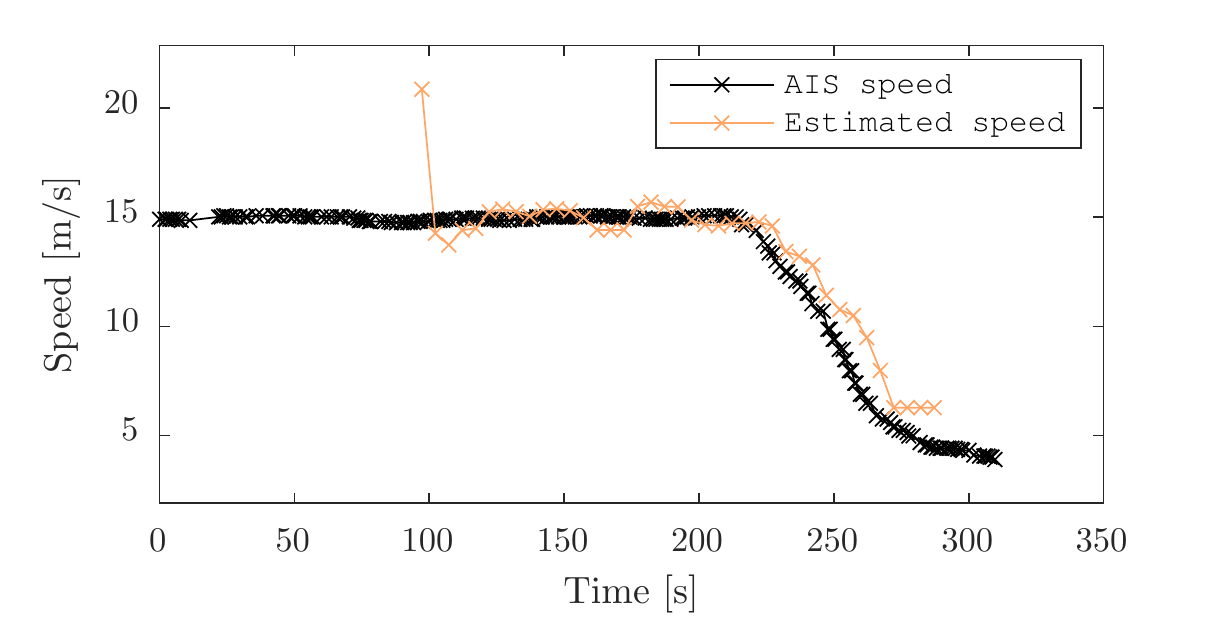} \\
      \includegraphics[width=\textwidth]{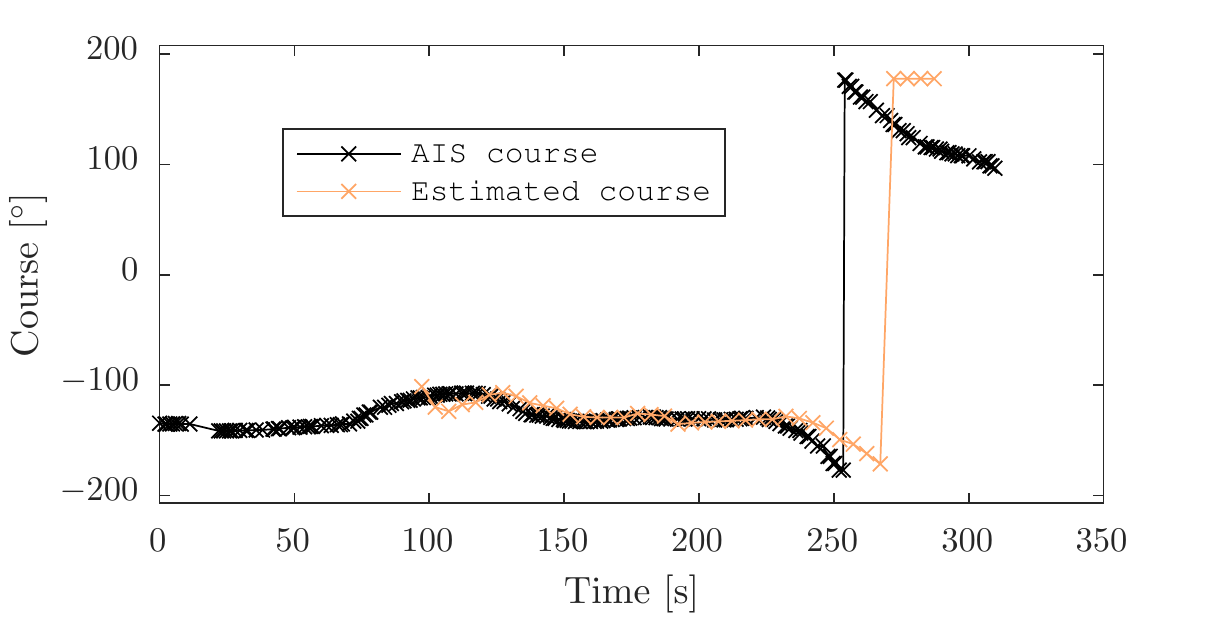}
      \caption{\label{fig:Head_on_radar_2_estimates_est_TRONDHEIMSFJORD}Speed and course estimates.
The black crosses represent received \gls{ais} messages, while the orange crosses represent \gls{bc-mpc} iterations.
The radar did not detect the Trondheimsfjord II after the course discontinuity at approximately $t=270~\si{\second}$, causing a large course error.}
    \end{subfigure}
    \caption{\label{fig:Head_on_radar_2_estimates_TRONDHEIMSFJORD}Distance to the Trondheimfjord II, and the estimated speed and course during Experiment 1.3.}
\end{figure}
In this experiment, the amount of estimate noise is even larger than in the previous experiment, with course fluctuations up to $40\si{\degree}$ as seen in \cref{fig:Head_on_radar_2_estimates_est}.
Still, as shown in \cref{fig:Head_on_radar_2_estimates_dist,fig:Head_on_radar_2}, the \gls{bc-mpc} manages to make quite smooth maneuvers, which again shows robustness with respect to obstacle estimate noise. \Cref{fig:Head_on_radar_2_estimates_TRONDHEIMSFJORD} shows similar plots for the Trondheimfjord II ferry.
Notice that it takes some time before the tracking system detects that the passenger ferry makes a maneuver, which is due to a limited sample rate on the radar combined with some latency in the \gls{pdaf} tracking system.

\subsection{Crossing from starboard: Experiments 2.1--2.2}
Crossing from starboard is a more complex scenario than the head-on scenario.
We performed two experiments with the \gls{osd1} approaching on collision course from starboard.
The scenarios were constructed such that the desired trajectory coincides with the obstacle trajectory, resulting in a collision with a relative bearing of $-90\si{\degree}$ if the desired trajectory is followed.
In such a scenario, the ownship is deemed the give-way vessel and should avoid collision by preferably maneuvering to starboard and passing abaft of the stand-on vessel.

\begin{figure}
    \centering
    \includegraphics[width=.6\textwidth]{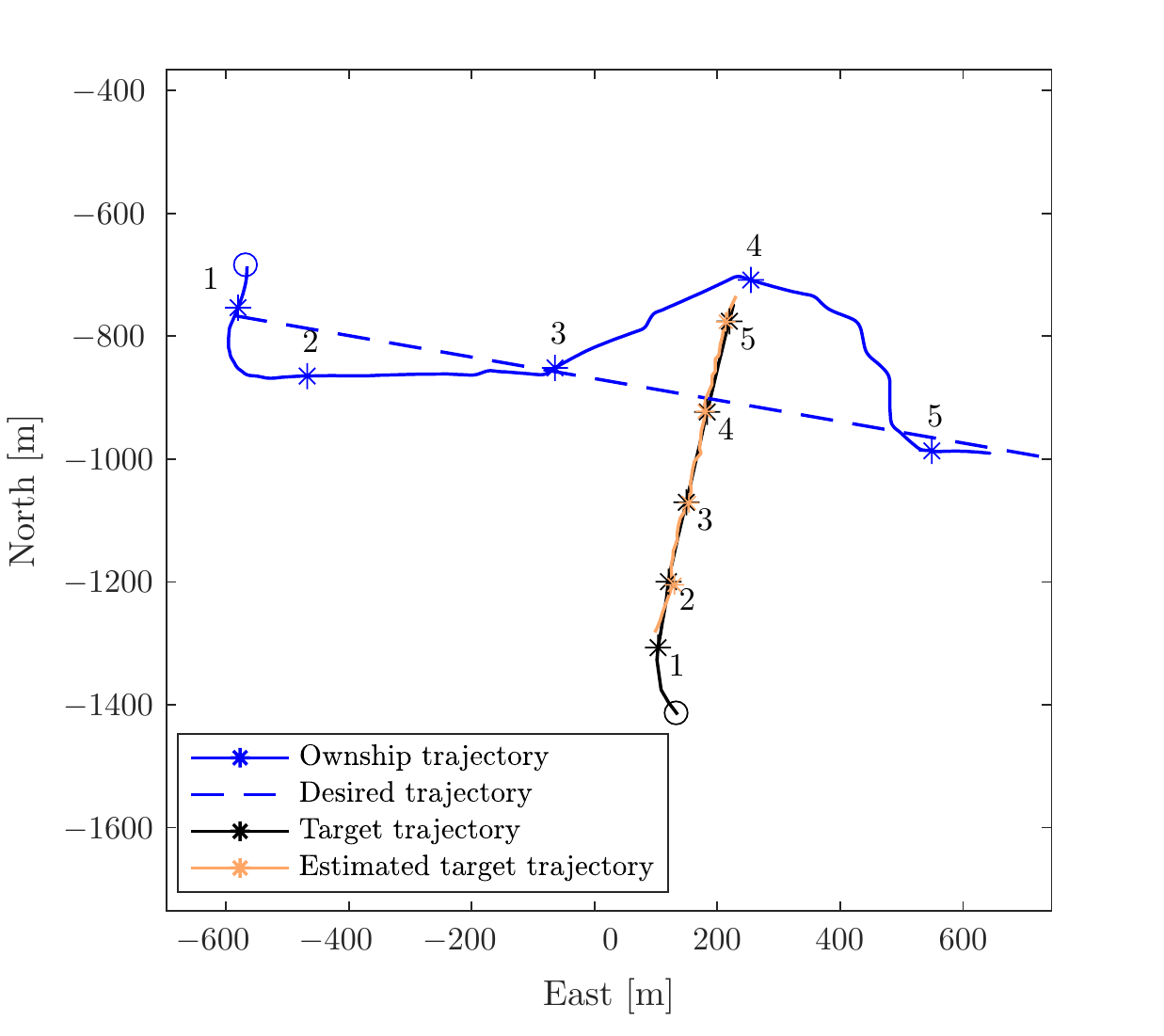}
    \caption{\label{fig:Crossing_starboard_radar_1}Experiment 2.1: Crossing from starboard scenario using the radar-based tracking system for providing obstacle estimates.
The ownship and obstacle initial positions are marked with circles, and the estimated obstacle trajectory is shown with the thick orange line.
The numbers represent time markers for each $60~\si{\second}$.}
\end{figure}
In Experiment 2.1, shown in \Cref{fig:Crossing_starboard_radar_1}, we avoided collision with the \gls{osd1} by maneuvering to port and passing in front of the obstacle.
This can be considered as suboptimal with respect to the preferred action being passing abaft of the obstacle.
However, the minimum distance to the obstacle is $214.0~\si{\meter}$, meaning that the obstacle is only slightly inside the margin region.
With this in mind, the maneuver is considered to be safe.
This is, once again, an example on how the algorithm demonstrates \gls{colregs} awareness.

\begin{figure}
    \centering
    \includegraphics[width=.6\textwidth]{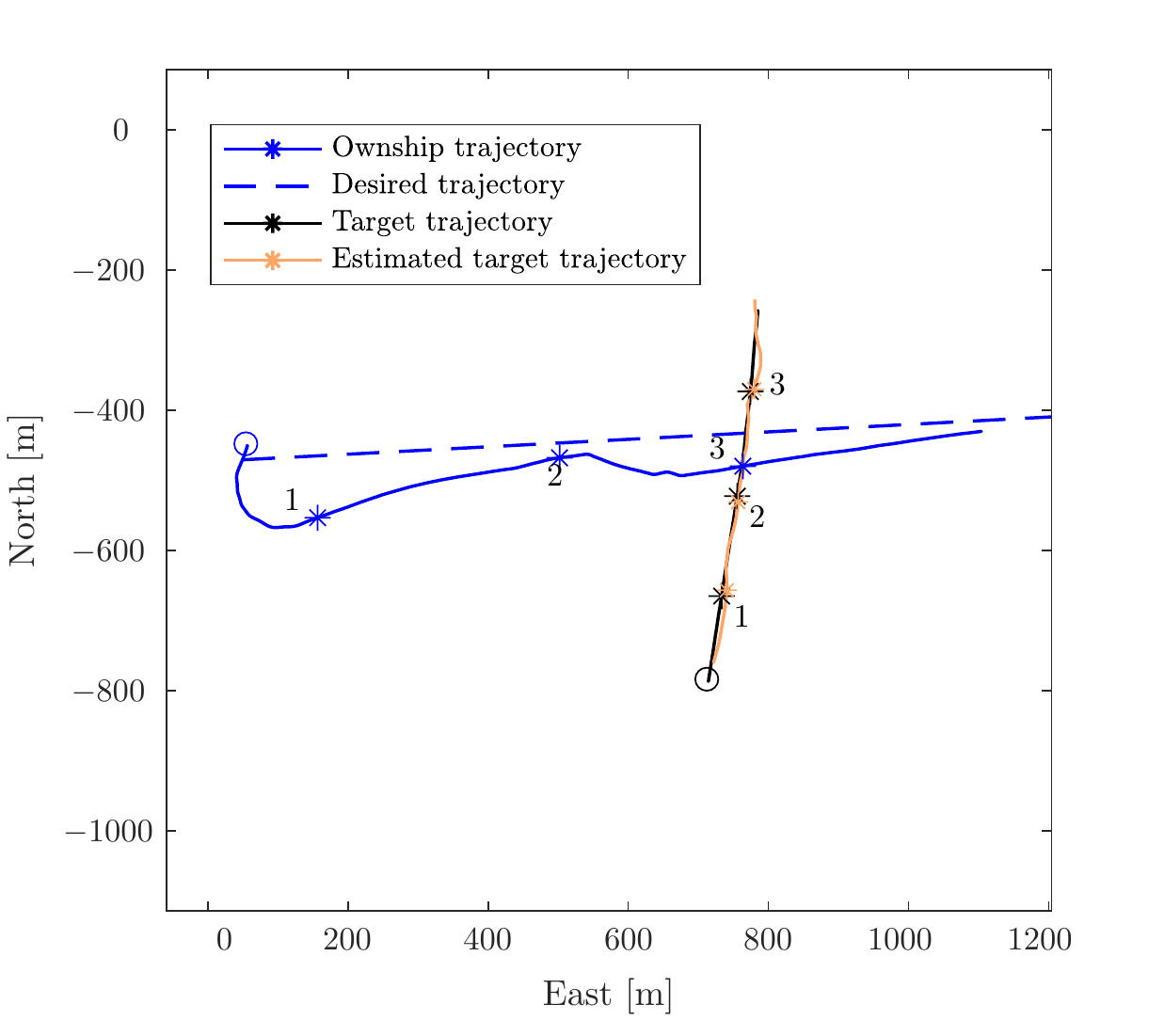}
    \caption{\label{fig:Crossing_starboard_radar_2}Experiment 2.2: Crossing from starboard scenario using the radar-based tracking system for providing obstacle estimates.
The ownship and obstacle initial positions are marked with circles, and the estimated obstacle trajectory is shown with the thick orange line.
The numbers represent time markers for each $60~\si{\second}$.}
\end{figure}
In Experiment 2.2, shown in \Cref{fig:Crossing_starboard_radar_2}, we avoided collision by passing abaft of the \gls{osd1}, as preferred by \gls{colregs}.
In this experiment, the minimum distance to the obstacle was $106.2~\si{\meter}$, significantly closer than when we passed in front of the obstacle.
This is still only slightly inside the margin region, remembering that the elliptical \gls{colregs} penalty function is smaller abaft the obstacle than in front of the obstacle.

\subsection{Overtaking: Experiment 3}
\begin{figure}
    \centering
    \includegraphics[width=.6\textwidth]{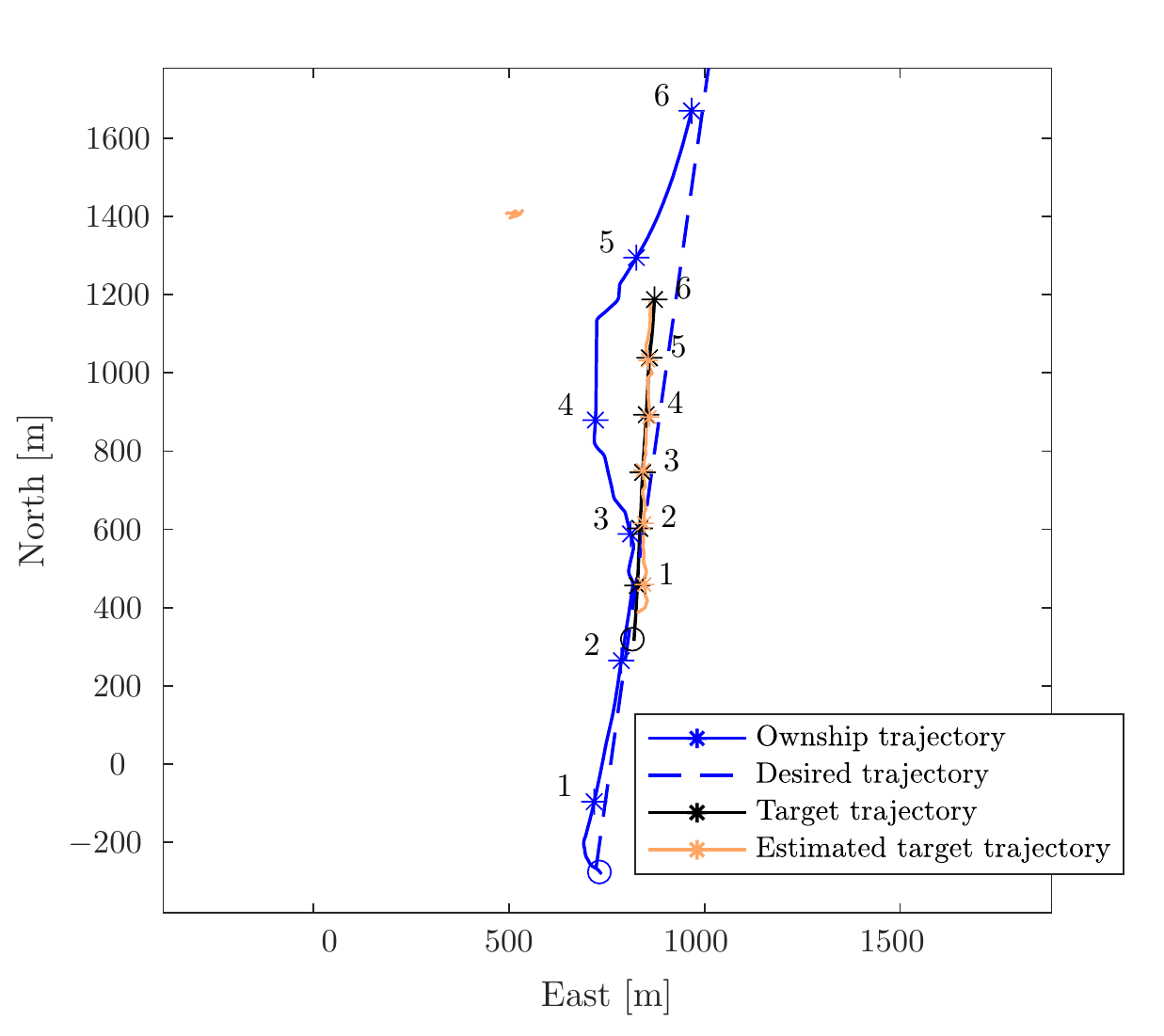}
    \caption{\label{fig:Overtaking_radar}Experiment 3: Overtaking scenario using the radar-based tracking system for providing obstacle estimates.
The ownship and obstacle initial positions are marked with circles, and the estimated obstacle trajectory is shown with the thick orange line.
The numbers represent time markers for each $60~\si{\second}$.
The trajectory located at approximately $(1450,500)$ originates from a navigational aid, which was detected approximately $130~\si{\second}$ into the experiment.}
\end{figure}
Another distinct situation is when the ownship approaches an obstacle from behind, overtaking it.
With respect to \gls{colregs}, the overtaking vessel is deemed the give-way vessel, while the overtaken vessel is deemed the stand-on vessel.
There is no strict rules on whether the give-way vessel should pass the stand-on vessel on the port or starboard side, but we prefer to pass on the port side of the stand-on vessel, as this does not block the stand-on vessel's possibilities in maneuvering to starboard if it finds itself in a head-on or crossing situation while being overtaken.

\Cref{fig:Overtaking_radar} shows Experiment 3, where the ownship overtakes the \gls{osd1}.
The ownship maneuvers to port, passing the \gls{osd1} on her port side.
The ownship trajectory is quite smooth, but turns towards the desired trajectory a bit early.
This was caused by the radar tracking system detecting a navigational aid in front of the ownship on the port side, which made maneuvering closer to the desired trajectory preferable.
The ownship was approximately $200~\si{\meter}$ in front of the obstacle when doing this maneuver.
Notice that we currently do not distinguish between dynamic and static objects in the tracking system, hence this navigational aid was considered as a moving vessel.
The closest distance to the obstacle during the overtaking maneuver was $127.3~\si{\meter}$, approximately equal to the size of the margin region on the port side of the obstacle.

\subsection{Crossing from port: Experiment 4}
\begin{figure}
    \centering
    \includegraphics[width=.6\textwidth]{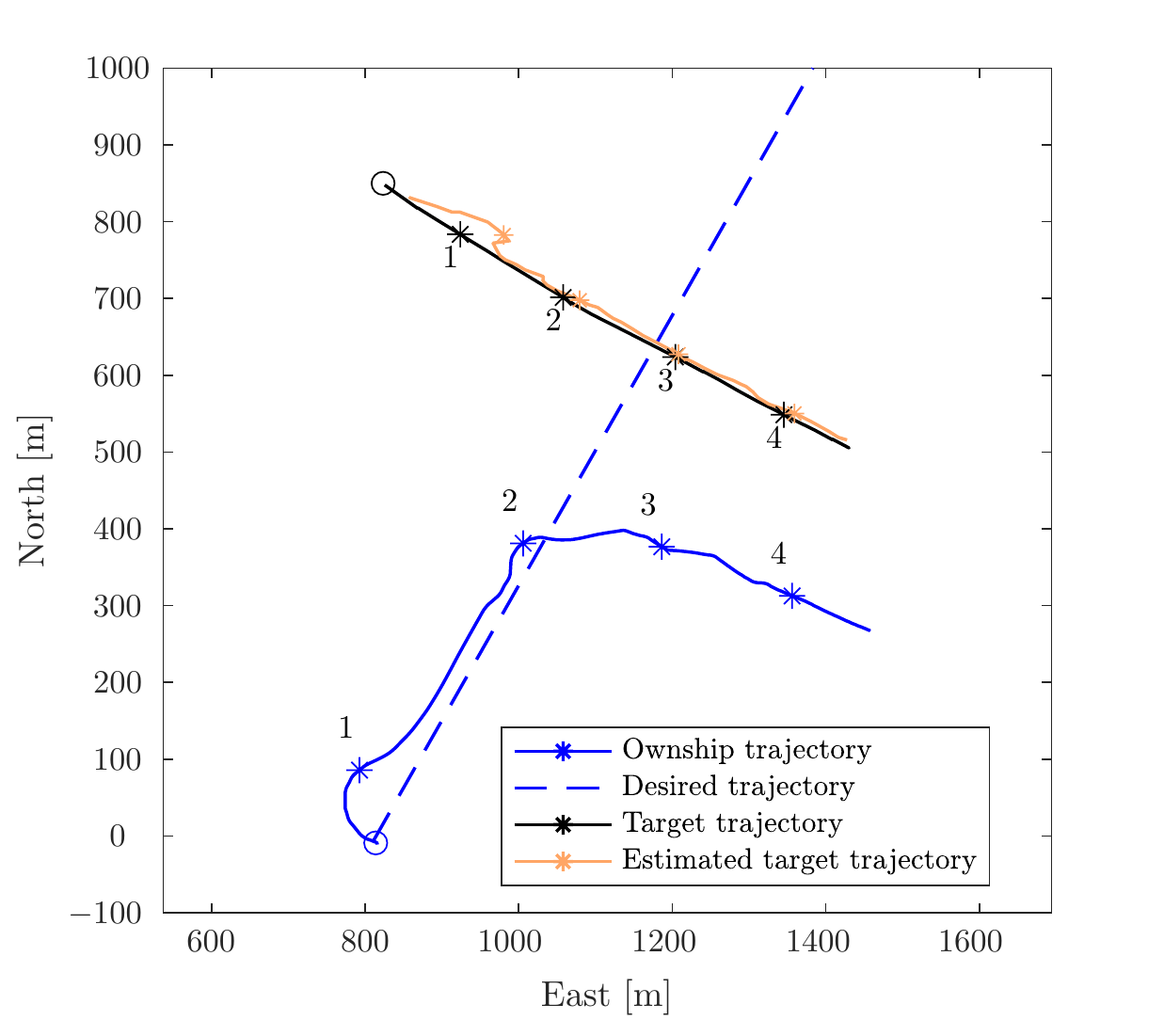}
    \caption{\label{fig:Crossing_port_radar}Experiment 4: Crossing from port scenario using the radar-based tracking system for providing obstacle estimates.
The ownship and obstacle initial positions are marked with circles, and the estimated obstacle trajectory is shown with the thick orange line.
The numbers represent time markers for each $60~\si{\second}$.}
\end{figure}
The last scenario we tested was a crossing from port, which may be the most complex scenario of the ones presented in this article.
This situation was generated similarly as the crossing from starboard situation, but with a relative bearing of $90\si{\degree}$ instead of $-90\si{\degree}$.
Here, \gls{colregs} deems the ownship as the stand-on vessel, while the \gls{osd1} is deemed the give-way vessel.
However, the \gls{osd1} keeps its speed and course, requiring the ownship to avoid collision.
In such a situation \gls{colregs} recommends the ownship to avoid maneuvering to port, favoring a starboard maneuver.

\Cref{fig:Crossing_port_radar} shows the results from Experiment 4, where the \gls{bc-mpc} algorithm maneuvers the ownship to starboard, following the recommendations in \gls{colregs} regarding this situation.
The algorithm chose to maneuver the ownship at the minimum speed in order to minimize the distance to the desired trajectory.
This minimum speed ensures maneuverability of the ownship, and was by coincidence similar as the speed of the \gls{osd1} during the experiment, resulting in the ownship trajectory following parallel to the obstacle trajectory.
Obviously, the ownship could increase the speed and pass in front of the obstacle, but this is not apparent to the \gls{bc-mpc} algorithm due to the limited prediction horizon.
In a hybrid \gls{colav} architecture, this situation would be solved by the mid-level \gls{colav} algorithm, designed with a longer prediction horizon than the \gls{bc-mpc} algorithm.

\subsection{Experiment summary}
The \gls{bc-mpc} algorithm has been tested in four different scenarios, each with different desirable behavior.
A total of 7 experiments is presented, and the key points and numbers of the experiments are given in \cref{tab:expSummary}.

In the head-on experiments, both \gls{ais} and radar tracking was used for providing obstacle estimates.
In Experiment 1.1, where we used \gls{ais} for providing obstacle estimates, the ownship avoided collision by passing the obstacle on its starboard side, violating the desired behavior of \gls{colregs}.
The ownship did, however, maneuver with increased distance to the obstacle compared to experiments 1.2 and 1.3 where we passed the obstacle on its port side in accordance with the desired behavior of \gls{colregs}.
In experiments 1.2 and 1.3, we used the radar tracking system for obtaining obstacle estimates, which provided estimates with a large amount of noise compared to Experiment 1.1.
The \gls{bc-mpc} algorithm did, however, not seem to be significantly affected by this noise.

In the crossing from starboard experiments we only used radar tracking for providing obstacle estimates.
In Experiment 2.1, we passed in front of the obstacle.
This is not strictly forbidden by \gls{colregs}, but is neither desirable.
The ownship did, however, have a large clearance to the obstacle, demonstrating the \gls{colregs} awareness of the \gls{bc-mpc} algorithm.
In Experiment 2.2, we passed behind the obstacle, complying with the desirable behavior of \gls{colregs}.

In Experiment 3.1 the ownship overtook the obstacle on its port side. \Gls{colregs} does not dictate which side the obstacle should be passed on, but by maneuvering to port the obstacle is free to maneuver to starboard if it finds itself in a separate collision situation.

Experiment 4.1 is a crossing situation where the ownship is deemed the stand-on vessel, and the \gls{osd1} is required to avoid collision.
The \gls{osd1} did, however, not fulfill her obligation to avoid collision, requiring that the ownship avoided collision in accordance with rule 17 of \gls{colregs}.
The ownship avoided collision by performing a starboard maneuver, as suggested by \gls{colregs}.

In some of the experiments, the \gls{bc-mpc} algorithm chose to ignore rules 13--15 of \gls{colregs}.
This is because the algorithm is designed to comply with rule 17 of \gls{colregs}, which requires that the ownship avoids collision in cases where it is deemed the stand-on vessel if the other vessel does not avoid collision, ignoring rules 13-15 if required.
In cases where the desired rule 13--15 behavior is ignored, the obstacle is passed with extra clearance.

\begin{table}
\caption{\label{tab:expSummary}Key points and numbers from the experiments.
*In Experiment 2.1 we passed in front of the obstacle, while COLREGs prefers that the ownship pass behind the obstacle.
Passing in front is, however, not strictly forbidden.}
\centering
\begin{tabular}{llll}
\toprule
\textbf{\begin{tabular}{@{}l@{}}Experiment type \\ and number \end{tabular}} & \textbf{\begin{tabular}{@{}l@{}}Obstacle \\ sensor\end{tabular}} &\textbf{\begin{tabular}{@{}l@{}}Rule 13--15 \\ compliance\end{tabular}} & \textbf{\begin{tabular}{@{}l@{}}Minimum distance \\to obstacle\end{tabular}} \\
\midrule
Head on\\
\hspace{1cm}1.1                 & AIS & No                          & $197.8~\si{\meter}$         \\
\hspace{1cm}1.2                 & Radar & Yes                          & $100.8~\si{\meter}$       \\
\hspace{1cm}1.3                 & Radar & Yes                          & $132.5~\si{\meter}$      \\
\cmidrule{1-4}
Crossing from starboard\\
\hspace{1cm}2.1                 & Radar & Yes*                     & $214.0~\si{\meter}$        \\
\hspace{1cm}2.2                 & Radar & Yes                          & $106.2~\si{\meter}$        \\
\cmidrule{1-4}
Overtaking\\
\hspace{1cm}3.1                 & Radar & Yes                          & $127.3~\si{\meter}$        \\
\cmidrule{1-4}
Crossing from port\\
\hspace{1cm}4.1                 & Radar & N/A                          & $231.7~\si{\meter}$        \\
\bottomrule
\end{tabular}
\end{table}

\section{Conclusion and further work}\label{sec:conclusions}
We have presented a new algorithm called the \acrfull{bc-mpc} algorithm for \gls{asv} \acrfull{colav}.
The algorithm has been thoroughly tested in closed-loop full-scale experiments in the Trondheimsfjord in October 2017, using a radar-based system for obstacle detection and tracking.
The algorithm performs well and displays good robustness with respect to noise on obstacle estimates, which is a significant source of disturbance when using tracking systems based on exteroceptive sensors to provide estimates of obstacle position, course and speed.
During the experiments, leisure and commercial vessels entered some of the scenarios by coincidences and were successfully avoided by the \gls{bc-mpc} algorithm without human intervention.

The \gls{bc-mpc} algorithm is intended for use as a short-term \gls{colav} algorithm, and should therefore always be able to find a feasible solution to avoiding collision.
As required by \gls{colregs}, this means that the normal \gls{colregs} rules dictating the vessel behavior in head-on, crossing and overtaking scenarios sometimes needs to be ignored.
However, the algorithm is motivated to follow the normal behavior described by \gls{colregs} when possible.
The algorithm is compliant with rules 8 and 17 of \gls{colregs}, and motivated to follow rules 13--15 if possible, which is why the algorithm is denoted as \gls{colregs}-aware.

In the future, we would like to do an extensive simulation study, analyzing the algorithm performance to a greater detail than what is possible through full-scale experiments.
This can prove valuable in order to further develop the algorithm, and to tune the objective function to e.g. obtain smoother trajectories, which would improve the indirect communication of intention between the \gls{asv} and other vessels at sea based on their maneuvers.
Furthermore, we would also like to distinguish between static and moving obstacles in the algorithm, and combine it with a long-term \gls{colav} algorithm in a hybrid architecture.

\subsubsection*{Acknowledgments}
This work was supported by the Research Council of Norway through project number 244116 and the Centres of Excellence funding scheme with project number 223254. The authors would like to express great gratitude to Kongsberg Maritime and Maritime Robotics for providing high-grade navigation technology, the Ocean Space Drone 1 and the Telemetron \gls{asv} at our disposal for the experiments.

\bibliographystyle{apalike}
\bibliography{root}
\end{document}